\documentclass[preprint,3p,12pt,times]{elsarticle}

\usepackage{graphicx}
\usepackage{epsfig}
\usepackage{epstopdf}
\usepackage{amssymb}
\usepackage{amsthm}
\usepackage{bm}
\usepackage{placeins}

\usepackage[colorlinks=true,citecolor=blue,linkcolor=black]{hyperref}
\usepackage{color}
\usepackage{amsmath}
\usepackage{multirow}
\usepackage[justification=centering]{caption}
\usepackage{subcaption}
\usepackage{color}

\usepackage{booktabs} 
\usepackage{lineno}
\journal{Elsevier}

\begin{document}

\title{SPH-based framework for modelling fluid-structure interaction problems with finite deformation and fracturing}

\author[iitd]{Md Rushdie Ibne Islam\corref{coraut}}
\ead{rushdie@am.iitd.ac.in}

\cortext[coraut]{Corresponding Author}
\address[iitd]{Department of Applied Mechanics, Indian Institute of Technology Delhi, New Delhi 110016, India} 

\begin{abstract}
Understanding crack propagation in structures subjected to fluid loads is crucial in various engineering applications, ranging from underwater pipelines to aircraft components. This study investigates the dynamic response of structures, including their damage and fracture behaviour under hydrodynamic load, emphasizing the fluid-structure interaction (FSI) phenomena by applying Smoothed Particle Hydrodynamics (SPH). The developed framework employs weakly compressible SPH (WCSPH) to model the fluid flow and a pseudo-spring-based SPH solver for modelling the structural response. For improved accuracy in FSI modelling, the $\delta$-SPH technique is implemented to enhance pressure calculations within the fluid phase. The pseudo-spring analogy is employed for modelling material damage, where particle interactions are confined to their immediate neighbours. These particles are linked by springs, which don't contribute to system stiffness but determine the interaction strength between connected pairs. It is assumed that a crack propagates through a spring connecting a particle pair when the damage indicator of that spring exceeds a predefined threshold. The developed framework is extensively validated through a dam break case, oscillation of a deformable solid beam, dam break through a deformable elastic solid, and breaking dam impact on a deformable solid obstacle. Numerical outcomes are subsequently compared with the findings from existing literature. The ability of the framework to accurately depict material damage and fracture is showcased through a simulation of water impact on a deformable solid obstacle with an initial notch.

\end{abstract}


\maketitle

\section{Introduction}
In recent years, there has been a growing focus on fluid-structure interaction (FSI), which plays a prominent role in numerous engineering and industrial contexts. Examples of these applications include coastal engineering, the shipbuilding industry, and aviation. Understanding the crack propagation in structures under fluid load is critical for enhancing safety, preventing environmental disasters, reducing economic losses, and advancing engineering innovation in complex fluid-structure interaction scenarios. The intricacies of fluid-structure interaction (FSI) problems often render them beyond the reach of analytical solutions. Furthermore, the high cost and logistical difficulties associated with experimental studies in FSI have spurred the adoption of numerical modelling as an attractive alternative.

Over recent decades, various methodologies have been developed to tackle the complex challenges of fluid-structure interaction (FSI) problems. Although mesh-based methods such as finite difference method (FDM), finite volume method (FVM), and finite element method (FEM) \cite{hu2009two, slone2002dynamic, heil2004efficient} have achieved a degree of success for FSI simulations, they often require additional computationally intensive numerical schemes (e.g., interface tracking or re-meshing etc.) when dealing with free surfaces, moving boundaries, and deformable structures. The computation becomes even more complex with propagating cracks and material separation as field variables exhibit discontinuities. Traditional mesh-based methods such as FEM are unsuitable \cite{belytschko1996difficulty}, and while discontinuous enrichment helps in modelling the cracks \cite{fries2010extended}, implementing these additional numerical strategies is not only intricate but also computationally intensive, and they can frequently introduce instability issues. As a promising alternative, Lagrangian particle-based methods have gained increasing favour in FSI modelling. These methods are becoming more popular due to their meshless and Lagrangian nature, which makes them well-suited for representing free-surface fluid flow and the substantial deformation of solid structures. Moreover, Lagrangian particle-based meshless approaches offer a natural and efficient means of capturing moving interfaces and finite deformation in structures encountered in FSI problems.

Smoothed Particle Hydrodynamics (SPH), initially designed to address challenges in astrophysical contexts, has gained widespread acclaim as a leading meshless method \cite{gingold1977smoothed, lucy1977numerical}. SPH operates as a truly meshless and Lagrangian particle-based approach, where individual particles represent the material points and carry the field variables\cite{libersky2005smooth, liu2010smoothed}. Each particle exclusively engages with its neighbouring counterparts in this method through a kernel function. The extent of this interaction is governed by the smoothing length, which defines the dimensions of the local neighbourhood, also referred to as the influence domain. Notably, the kernel function exhibits a characteristic bell-shaped profile designed to maximize interaction strength with immediate neighbours and progressively diminish it as the distance between interacting particles increases. SPH offers distinct advantages in handling scenarios involving free-surface flow, finite material deformation, moving interfaces and boundaries. Its wide range of applications can be found in dynamic fluid flow \cite{monaghan1994simulating, adami2012generalized, fraga2019implementation}, geotechnical simulations \cite{bui2008lagrangian, chen2011practical, peng2015sph, peng2016unified}, explosive and impact events \cite{liu2003smoothed, islam2017computational, islam2020pseudo, islam2023comparison}. SPH also plays a prominent role in FSI simulations \cite{antoci2007numerical, rafiee2009sph, salehizadeh2022coupled}.    

SPH methodologies can be of different types, such as weakly compressible SPH (WCSPH), incompressible SPH (ISPH), total Lagrangian SPH (TLSPH) etc. WCSPH and ISPH are the most prominent techniques for fluid flow, whereas standard SPH and TLSPH are used for modelling the deformation of solids. In WCSPH, the time step size used for numerical integration is quite small, whereas ISPH allows a larger time step size for integration. Another advantage of ISPH is that it produces smooth pressure fields compared to WCSPH simulations \cite{marrone2015prediction, meringolo2017filtering, pahar2016modeling, pahar2017coupled}. However, the computational cost per step is relatively low in WCSPH. For large-scale simulations, implementing parallelized techniques for ISPH \cite{trask2015highly, guoa2015developing, guo2018new} is more challenging than its WCSPH counterpart. Nonetheless, the conventional WCSPH method is hindered by substantial pressure fluctuations. While these fluctuations have a limited impact on flow kinematics, they pose significant challenges in Fluid-Structure Interaction (FSI) modelling. This is because pressure fluctuations can lead to inaccuracies in assessing the interacting forces at the fluid-structure interface. To address this issue, several numerical schemes (e.g. $\delta$-SPH method \cite{molteni2009simple, marrone2011delta}, Rusanov flux \cite{ferrari2009new} etc.) have been introduced to get smooth pressure fields.

As for the solid phase, two methods are mainly used, i.e., the conventional SPH based on the Eulerian kernel and TLSPH. In the conventional SPH approach, particle positions are updated at each computational step, and kernel functions are computed based on these updated positions \cite{libersky2005smooth}. Consequently, the traditional SPH kernel function is often called the Eulerian because particles can enter and exit its influence domain. This Eulerian kernel function is known to introduce a well-recognized issue called tensile instability \cite{swegle1995smoothed}, leading to local particle clustering and the formation of unphysical numerical fractures. To mitigate the tensile instability, a commonly used correction method is the artificial pressure/stress technique \cite{monaghan2000sph, gray2001sph}, which can effectively alleviate the issue. The tensile instability can be circumvented by calculating kernel functions using reference particle positions \cite{vignjevic2006sph}. The kernel function based on the reference configuration is the Lagrangian kernel, and the corresponding SPH formulation is called TLSPH. TLSPH eliminates tensile instability if computations are consistently based on the initial configuration \cite{belytschko2000unified, vidal2007stabilized}. However, the original TLSPH method faces limitations in modelling scenarios involving significant material distortion and separation due to negative Jacobians \cite{islam2023comparison}. Moreover, the traditional SPH-based frameworks provide better agreements than TLSPH when compared with the experimental and other numerical results for finite deformation and failure of materials \cite{islam2022equivalence, islam2022large, islam2023comparison}.  

Most SPH-based FSI modelling uses different combinations of these methods \cite{antoci2007numerical, rafiee2009sph, salehizadeh2022coupled}. Despite the success of SPH in modelling FSI problems, however, important issues on material damage and fracture are yet to be addressed. Limited studies can be found in the literature on a stable, accurate, efficient SPH framework for modelling FSI problems involving material damage and fracture \cite{rahimi2023sph}. In this context, the research community increasingly embraces SPH and its extensions, primarily due to their innate ability to model crack propagation \cite{vignjevic2006sph, rahimi2022modeling, zhao2023simulation}. Among these extensions, the General Particle Dynamics (GPD) framework, built upon SPH, has gained widespread adoption for simulating progressive failure in slopes and the fracturing of rocks \cite{zhou2015novel, zhai2016effects, zhou20173d}. Another notable SPH extension is the pseudo-spring augmented SPH \cite{chakraborty2013pseudo}, which establishes connections between each particle and its immediate neighbours through pseudo-springs. The advancement of a crack front occurs when the pseudo-spring linking any two immediate neighbours is disrupted. The crack paths can be modelled in this approach without refinement, enrichment or visibility criteria. A slightly adapted version of this methodology has proven effective in simulating failures in both brittle and ductile materials \cite{islam2017computational, islam2020pseudo}.

This work presents a coupled WCSPH framework for fluid-structure interaction with deformable structures undergoing damage and fracture. The WCSPH enhanced with numerical schemes to improve accuracy and stability is used for simulating the fluid flow. A pseudo-spring analogy in traditional SPH has been adopted for modelling crack initiation and propagation in structures. The effectiveness of the proposed approach is demonstrated through several numerical illustrations. The paper is organized as follows. Section 2 discusses the governing equations, their discretization, boundary conditions and stability schemes for fluid simulation using WCSPH. Section 3 discusses the SPH method for solid deformation and the pseudo-spring analogy for modelling material damage and fracture. Section 4 discusses the coupling strategy of WCSPH and pseudo-spring-based SPH. Section 5 presents some numerical examples for verification and validation purposes. The crack initiation and propagation in elastic obstacles with a notch due to water impact is also demonstrated through an example. Finally, some conclusions are drawn in section 6.  

\section{Weakly compressible smoothed particle hydrodynamics (WCSPH) for fluid flow}\label{wcsph}
The foundational equations governing the dynamic fluid flow encompass the principles of mass conservation and momentum preservation principles. These equations encompass:

\begin{equation}\label{con}
    %\frac{d \rho^f}{dt} = - \rho^f \frac{\partial v^f_\beta}{\partial x_\beta}
    \dfrac{d\rho}{dt} = -\rho \dfrac{\partial v^\beta}{\partial x^\beta},
\end{equation}

\begin{equation}
    %\frac{dv^f_\alpha}{dt} = - \frac{1}{\rho^f} \frac{\partial p}{\partial x_\alpha} + \frac{1}{\rho^f} \frac{\partial \tau_{\alpha \beta}}{\partial x_\alpha} + g_\alpha
    \frac{d v^\alpha}{dt} = - \frac{1}{\rho} \frac{\partial p}{\partial x^\alpha} + \frac{1}{\rho} \frac{\partial \tau^{\alpha \beta}}{\partial x^\alpha} + g^\alpha,
\end{equation}
where $\rho$ is the material density. In the moving Lagrangian frame, we represent the time derivative as $\dfrac{d}{dt}$. At an individual material point, we describe the spatial coordinates using the notation $x^\alpha$ for the component indexed as $\alpha$, while the velocity at this point is denoted as $v^\alpha$. $\tau^{\alpha\beta}$ denotes the $\alpha$ and $\beta$ elements of the viscous stress:

\begin{equation}
    \tau^{\alpha \beta} = \mu_f \left( \frac{\partial v^\alpha}{\partial x^\beta} + \frac{\partial v^\beta}{\partial x^\alpha} \right),
\end{equation}
where $\mu_f$ is the dynamic viscosity of the fluid. $g^\alpha$ is the $\alpha$ component of the body force. $p$ is the pressure, and we derive the value of $p$ by employing a weakly compressible equation of state model \cite{monaghan1994simulating}, formulated as follows in this study:

\begin{equation}
    p = p_0 \left[ \left( \frac{\rho}{\rho_0} \right)^\gamma - 1 \right],
\end{equation}
where $\gamma=7$, $\rho_0$ represents the reference material density and $p_0=\frac{c^2_0\rho_0}{\gamma}$ with $c_0$ representing the speed of sound.

SPH, classified as a collocation method, divides the domain into a collection of material points, commonly known as particles, whether they are distributed regularly or irregularly. The partial differential equations associated with the conservation equations are converted into a set of equivalent ordinary differential equations. These transformed equations are then solved using one of the numerous numerical integration techniques. Field variables pertaining to a particle situated at the material point $x_i^\alpha$ are determined by considering neighboring particles positioned at $x_j^\alpha$. This computation relies on the utilization of a kernel function denoted as $W(q, h)$. Notably, this kernel function serves as an approximation of the Dirac-delta function. The parameter $q$ is defined as the normalized distance between $x_i^\alpha$ and $x_j^\alpha$, with this normalization being relative to the smoothing length $h$ i.e., $q = (||x_i^\alpha - x_j^\alpha||)/h$. In this work, we use the Wendland C2 kernel function \cite{wendland1995piecewise}:

\begin{equation}\label{kernel}
    W(q, h)=\alpha_d 
\begin{cases}
    (q + 0.5) (2 - q)^4, & \text{if } q\le 2\\
    0,                                & \text{otherwise}
\end{cases}
\end{equation}
where $\alpha_d=7/(32 \pi h^2)$ in 2D. To maintain simplicity in our discussion, from this point onward, we will denote the kernel function $W(q, h)$ for a particle pair $i$ and $j$ as $W_{ij}$, and its derivative as $W_{ij,\beta}$. The discretized forms of the conservation equations are as follows:

\begin{equation}\label{con_dis}
   \frac{d\rho_i}{dt} = \sum_j m_j v_{ij}^\beta {W}_{ij,\beta} + \delta h c_0 \sum_j 2 \frac{m_j}{\rho_j}(\rho_i - \rho_j) \frac{x_{ij}^\beta}{||x_i^\beta - x_j^\beta||^2 + 0.01h^2} {W}_{ij,\beta},
 \end{equation}

\begin{equation}\label{mom_f}
    \frac{d v_i^\alpha}{dt} =  \sum_j m_{j}\left(\frac{\tau_i^{\alpha\beta}}{\rho^2_i}+\frac{\tau_j^{\alpha\beta}}{\rho^2_j}-\pi_{ij}\delta^{\alpha\beta} \right) {W}_{ij,\beta} - \sum_j m_{j}\left(\frac{p_i}{\rho^2_i}+\frac{p_j}{\rho^2_j}\right) {W}_{ij,\beta} + g^\alpha, 
\end{equation}
where $v^\beta_{ij}=v^\beta_i - v^\beta_j$. There are two distinct terms on the right-hand side of the equation \ref{con_dis}. The first term signifies the SPH discretization of equation \ref{con}, while the second term introduces an additional numerical diffusion component referred to as $\delta$-SPH \cite{molteni2009simple}. We use $\delta=0.1$ in this paper. The $\delta$-SPH term ensures a smooth pressure field in WCSPH simulations. $\pi_{ij}$ is the artificial correction term and maintains numerical stability in the presence of shock. The following form is used in this work \cite{monaghan1983shock}: 

\begin{equation}\label{eq8}
    \pi_{ij}= 
\begin{cases}
    \dfrac{-\beta_1 \bar{c}_{ij}\mu_{ij} + \beta_2 \mu^2_{ij}}{\bar{\rho}_{ij}},& \text{if } v^\alpha_{ij} x^\alpha_{ij}\le 0\\
    0,              & \text{otherwise}
\end{cases}
\end{equation} 
where,
\begin{equation}
\mu_{ij}=\dfrac{hv^\alpha_{ij} x^\alpha_{ij}}{||x_i^\alpha - x_j^\alpha||^2+ 0.01h^2},
\end{equation} 
where $x^\alpha_{ij}=x^\alpha_i-x^\alpha_j$, $\bar{c}_{ij}$ represents the average sound speed calculated across particles $i$ and $j$ and $\bar{\rho}_{ij} = 0.5(\rho_i + \rho_j)$.

\subsection{No-slip boundary condition}\label{bc}

\begin{figure}[hbtp!] %trim={<left> <lower> <right> <upper>}
\centering
\begin{subfigure}[t]{0.8\textwidth}
\includegraphics[width=\textwidth]{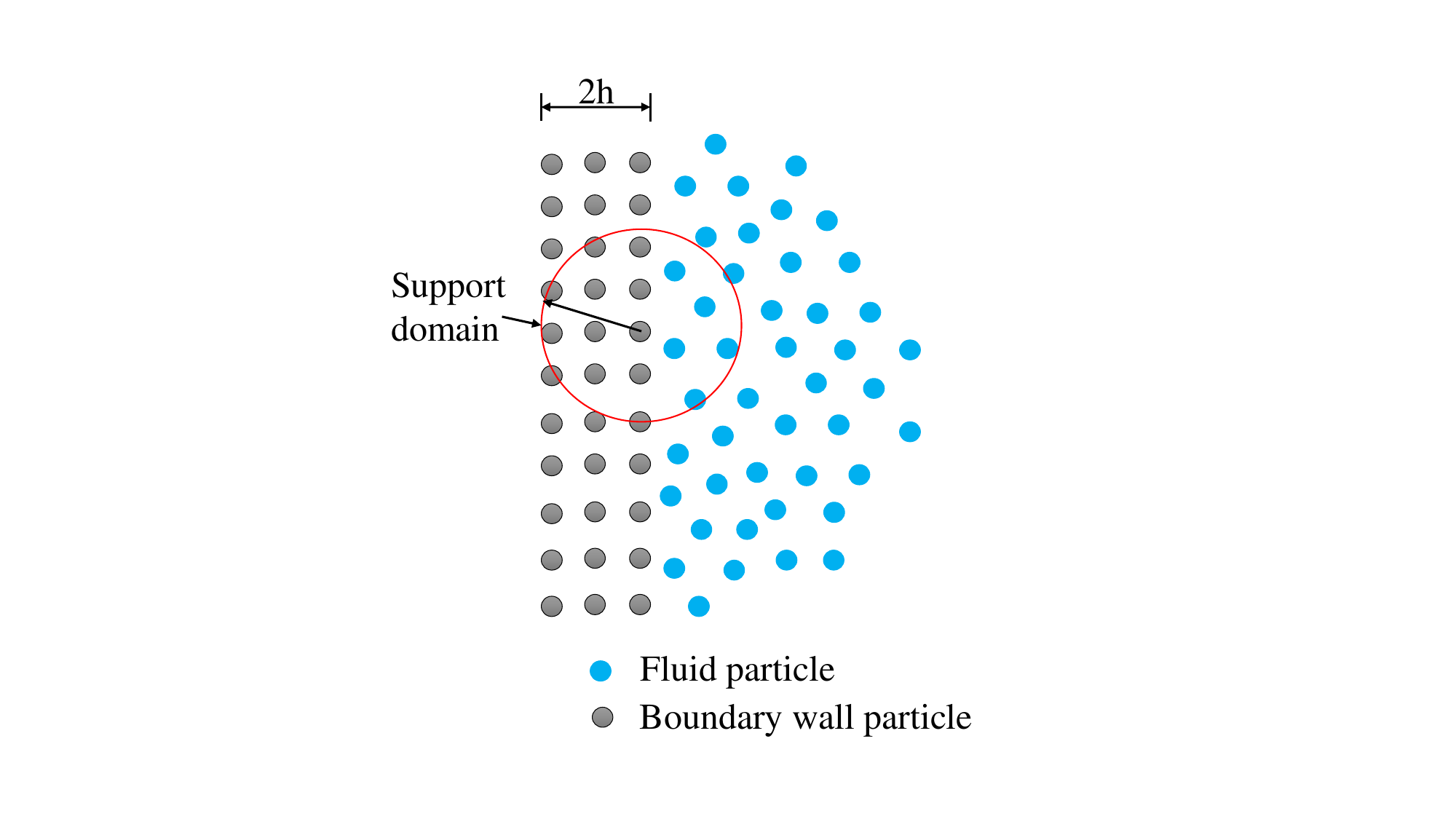}
\end{subfigure}
\caption{Diagram of the boundary treatment at the solid wall}\label{boundary}
\end{figure}

To maintain the no-slip solid boundary condition along solid walls, we introduce boundary particles that contain extrapolated information about velocity and pressure \cite{adami2012generalized}. We represent the solid wall boundaries using distinct boundary particles spanning $2h$ (Fig. \ref{boundary}). The particles at the walls are assigned the same initial properties (inter-particle distance, particle mass and density) as the fluid particles. The field variables of the solid boundary wall particles are extrapolated from the adjacent fluid particles and remain fixed in their initial positions. These solid boundary wall particles participate like regular particles for field variable computation of fluid particles. The pressure values at the solid boundary wall particles are obtained using the information from neighbouring fluid particles in the following form:    

\begin{equation}
    p_w = \frac{\sum_f p_f W(x_{wf}) + (g^\beta - a^\beta_w) \sum_f \rho_f x^\beta_{wf} W(x_{wf})}{\sum_f W(x_{wf})},
\end{equation}
where the solid boundary wall particles are denoted by subscript $w$ and $f$ represents the fluid particles. $a^\beta_w$ represents the $\beta$ component of the specified acceleration of the solid boundary wall particles. The following equation is used to calculate the density of the solid boundary wall particles: 

\begin{equation}
    \rho_w = \rho_0 \left( \frac{p_w}{p_0} + 1\right)^\frac{1}{\gamma}.
\end{equation}

\section{Pseudo-spring based SPH for solid deformation}\label{sph}
The conservation equations for material deformation due to external loading are

\begin{equation}
    \frac{d\rho}{dt} = -\rho \frac{\partial v^\beta}{\partial x^\beta},
\end{equation}

\begin{equation}
    \dfrac{dv^\alpha}{dt} =  \dfrac{1}{\rho} \dfrac{\partial \sigma^{\alpha\beta}}{\partial x^\beta},
\end{equation}
where the Cauchy stress tensor's component corresponding to indices $\alpha$ and $\beta$ is symbolized as $\sigma^{\alpha\beta}$. The above conservation equations are discretized as: 

\begin{equation}
    \dfrac{d\rho_i}{dt} = \sum_j m_{j}v^\beta_{ij} {W}_{ij,\beta},
\end{equation}

\begin{equation}\label{mom_solid}
    \dfrac{dv^\alpha_i}{dt} = \sum_j m_{j}\left(\dfrac{\sigma^{\alpha\beta}_i}{\rho^2_i}+\dfrac{\sigma^{\alpha\beta}_j}{\rho^2_j}-\pi_{ij}\delta^{\alpha\beta}-P^{a}_{ij}\delta^{\alpha\beta} \right) {W}_{ij,\beta},
\end{equation}
where $P^{a}_{ij}$ is the artificial pressure correction term, and its purpose is to prevent the occurrence of tensile instability, which occurs when particles tend to cluster together, leading to the development of unrealistic cracks. This adjustment introduces a short-range repulsive force using the given expression \cite{monaghan2000sph}:

\begin{equation}\label{eq_ap}
   P^{a}_{ij}=\gamma \left(\dfrac{|P_{r_i}|}{\rho^2_i}+\dfrac{|P_{r_j}|}{\rho^2_j}\right) \left[\dfrac{W(d_{ij})}{W(\Delta p)}\right]^{\bar{n}},
\end{equation}
where $\gamma$ symbolizes the adjustment parameter, and $\bar{n}$ is defined as $W(0)/W(\Delta p)$, where $\Delta p$ signifies the average particle spacing in the initial configuration.

\subsection{Constitutive model for elastic structure}
The Cauchy stress tensor, denoted as $\sigma^{\alpha \beta}$, consists of two main components: the hydrostatic pressure $p$ and the deviatoric stress $S^{\alpha \beta}$ ($\sigma^{\alpha\beta}= S^{\alpha\beta}-p\delta^{\alpha\beta}$). We have employed a linear equation of state to calculate the hydrostatic pressure in deformable solids \cite{eliezer1986introduction} as $p=K\left(\frac{\rho}{\rho_0}-1\right)$, $K$ being the bulk modulus. The rate of change of the deviatoric stress $S^{\alpha \beta}$ is calculated is determined by the following equation:    

\begin{equation}\label{eq11}
    \dot{S}^{\alpha\beta}=2\mu \left(\dot{\epsilon}_{\alpha\beta}-\dfrac{1}{3}\delta^{\alpha\beta}\dot{\epsilon}^{\gamma\gamma}\right)+S^{\alpha\gamma}\omega^{\beta \gamma} +S^{\gamma\beta}\omega^{\alpha\gamma},
\end{equation}
Here, $\mu$ is the shear modulus. The above Jaumann stress rate is used to ensure frame independence. The strain rate tensor and spin tensor are represented as $\dot{\epsilon}^{\alpha \beta}$ and $\omega^{\alpha \beta}$, respectively. These tensors can be calculated as follows:

\begin{equation}\label{eq12}
\begin{array}{rcl}
    \dot{\epsilon}^{\alpha \beta} = \dfrac{1}{2} \left(l^{\alpha \beta} + l^{\beta \alpha} \right) &,& 
    \omega^{\alpha \beta} = \dfrac{1}{2} \left(l^{\alpha \beta} - l^{\beta \alpha} \right),
\end{array}
\end{equation} 
On the other hand, the velocity gradient tensor $l^{\alpha \beta}$ is determined by:
\begin{equation}\label{esph_l}
l^{\alpha \beta}_{ESPH} = -\sum_j (v^{\alpha}_i - v^{\alpha}_j) {W}_{ij,\beta} \dfrac{m_j}{\rho_j}.
\end{equation} 

\subsection{Definition of immediate neighbour particles for approximation}
The kernel functions employed in SPH exhibit their maximum values near the centre particle of their compact support. As one moves away from this centre particle, the magnitude of these functions rapidly diminishes. Consequently, particles located close to a reference particle denoted as $i$ have a significantly greater impact on the approximation than those positioned near the outer boundary of the kernel support. Considering this behaviour, the field variables at particle $i$ are estimated by summing the contributions solely from its immediate neighbouring particles in our work. Therefore, only those particles that can be directly connected to particle $i$ through straight lines, without intersecting any other particles within the domain, are considered for the approximation \cite{chakraborty2013pseudo, islam2017computational}. When employing a rectangular particle distribution, this criterion implies that any interior particle is influenced by its eight nearest neighbours, a boundary particle is affected by five nearest neighbours, and three nearest neighbours influence a corner particle. Here, a gradient correction method \cite{chen1999corrective} is employed to mitigate the truncation errors emerging due to the incomplete or partial support domain. In SPH, the inclusion of a gradient correction method serves the purpose of attaining both zeroth-order consistency ($C^0$ consistency near boundary particles) and first-order consistency ($C^1$ consistency in interior particles). In this study, we substitute $W_{ij,\beta}$ with $\hat{W}{ij,\beta}$ in order to accomplish this, with $\hat{W}{ij,\beta}$ being computed as follows::

\begin{equation}\label{csph_app_1}
    \hat{W}_{ij,\beta}=B_i^{\beta\alpha}W_{ij,\alpha} ~~\text{with}~~ \mathbf{B_i=A_i^{-1}}~~~\text{and}~~~A_i^{\beta\alpha}=-\sum_j \frac{m_j}{\rho_j}x^{\beta}_{ij}W_{ij,\alpha}.
\end{equation}
We checked the error caused by this reduction in the number of interacting particles by approximating the function $sin \frac{\pi x}{2}, 0<x<1$ and its derivatives. This exercise revealed that the use of only neighbouring particles introduced negligible error.

\subsection{Pesudo-spring analogy in SPH}
In our framework, it is presumed that the closest neighbouring particles are linked to the $i^{th}$ particle through what we refer to as pseudo springs. These pseudo springs are introduced solely for the purpose of modelling interactions between connecting particles and do not impart any additional stiffness to the system. This technique is more detailed in \cite{chakraborty2013pseudo} and \cite{islam2017computational}. These pseudo springs are responsible for defining the level of interaction denoted as $f_{ij}$ between the connected particles. Specifically, when the material connecting particles $i$ and $j$ is undamaged, the value of $f_{ij}$ is set to 1. However, it is assumed that these pseudo springs will fail when certain predetermined criteria are met, such as reaching a critical axial stress or strain along the $ij$ line or some other relevant parameter. When such failure occurs, we set the value of $f_{ij}$ to 0, and this change from 1 to 0 is considered permanent. Consequently, the presence of these permanently damaged or failed pseudo springs allows for tracking the crack path within the domain. To accommodate the evolving interactions between particles as a result of these pseudo springs, kernel functions utilized in SPH, denoted as $\hat{W}_{ij}$, along with their respective derivatives $\hat{W}_{ij,\beta}$ used in approximating field variables, are replaced by modified versions incorporating the interaction level $f_{ij}$. These modified functions are expressed as $f_{ij} \hat{W}_{ij}$ and $f_{ij} \hat{W}_{ij,\beta}$ reflecting the changing influence of particle connections due to the presence of damaged or failed pseudo springs. To visualize the path of a crack, we employ fringe plots of a damage variable denoted as $D$. This variable is defined for a given particle, say particle $i$, as the ratio of the count of pseudo springs for which $f_{ij} = 0$ (indicating failure) to the total count of initial pseudo springs connected to that particle. When $D = 1$, it signifies that all the pseudo springs linked to particle $i$ have experienced permanent failure or damage, essentially representing the complete failure of particle $i$. On the other hand, values of $D$ within the range $0 < D < 1$ imply that the material associated with particle $i$ has suffered partial damage. It is important to note that a crack can propagate even when $D_i < 1$, underscoring the idea that damage can extend beyond individual particles, affecting their connections and interactions.

\section{Coupling of WCSPH and Pseudo-spring based SPH}\label{fsi}
This section discusses the coupling strategy of the WCSPH and pseudo-spring SPH. In this coupling methodology, we tackle the fluid flow problem by applying WCSPH, bolstered by incorporating $\delta$-SPH techniques, as comprehensively discussed in Section \ref{wcsph}. Concurrently, our approach to solving structural deformation and failure leverages the pseudo-spring SPH, elaborated in Section \ref{sph}. We employ particles with identical initial spacing for the discretisation to maintain a seamless and harmonious treatment of fluid and solid phases. 

An explicit contact force algorithm is essential for accurately simulating the complex multi-body interactions between the fluid and deformable structure. In this study, we employ a soft repulsive particle contact model \cite{zhang2018meshfree}. This model incorporates a distance-dependent repulsive force, characterized by a finite magnitude, acting upon particles (both fluid and deformable structure) as they approach each other. This force is mathematically expressed as follows:

\begin{equation} \label{contact}
    F^\alpha_{ij} = 0.01 c^2 \zeta f(\eta) \frac{x_{ij}^\alpha}{r^2_{ij}}
\end{equation}

\begin{equation}
    \eta = \frac{r_{ij}}{0.75h_{ij}}
\end{equation}

\begin{equation}
    \zeta = 1 - \frac{r_{ij}}{\Delta d}, 0 < r_{ij} < \Delta p
\end{equation}

\begin{equation}\label{kernel_test}
    f(\eta)=
\begin{cases}
    2/3, & \text{if } 0 < \eta \le 2/3, \\
    (2 \eta - 1.5 \eta^2),    & \text{if } 2/3< \eta \le 1, \\
    0.5 (2-\eta^2), & \text{if } 1 < \eta \le 2, \\ 
    0, & \text{otherwise}.
\end{cases}
\end{equation}
The distance between two particles at the fluid-structure boundary, i.e., one fluid particle and another particle from the deformable structure, is denoted by $r$. This softer repulsive force effectively mitigates non-physical particle penetration while simultaneously reducing the occurrence of pressure disturbances \cite{zhang2018meshfree, liu2019smoothed}. While calculating the contact force and modelling the fluid-structure interactions, the SPH particles from the deformable structural domain interact with the fluid particles only through equations \ref{contact} - \ref{kernel_test} and vice-versa. Finally, we add the interaction forces to the discrete fluid and structure momentum conservation equations \ref{mom_f} and \ref{mom_solid}. A predictor-corrector integration method is utilized to solve the discretized equations governing the fluid-structure interaction (FSI) problem, with the time step determined through the Courant–Friedrichs–Lewy condition. 

\section{Numerical examples}
Within this section, we present several numerical examples. We utilize the proposed coupled WCSPH-pseudo-spring SPH method to simulate scenarios involving free-surface flow, elastic solids undergoing significant deformation, and fluid-structure interactions with deformable structures. To assess the accuracy of our simulations, we compare the numerical results with analytical solutions and experimental and numerical data available in the existing literature. In the last example, we simulate the material damage and fracture in the structure due to water wave interaction. For all the simulations, we utilize the WCSPH technique in conjunction with $\delta$-SPH correction to handle the fluid phase, while we adopt the pseudo-spring SPH approach to represent the deformable structure. 

\subsection{Dam break: collapse of a water column}
\begin{figure}[hbtp!] %trim={<left> <lower> <right> <upper>}
\centering
\begin{subfigure}[t]{0.8\textwidth}
\includegraphics[width=\textwidth,trim={100 100 100 100}, clip]{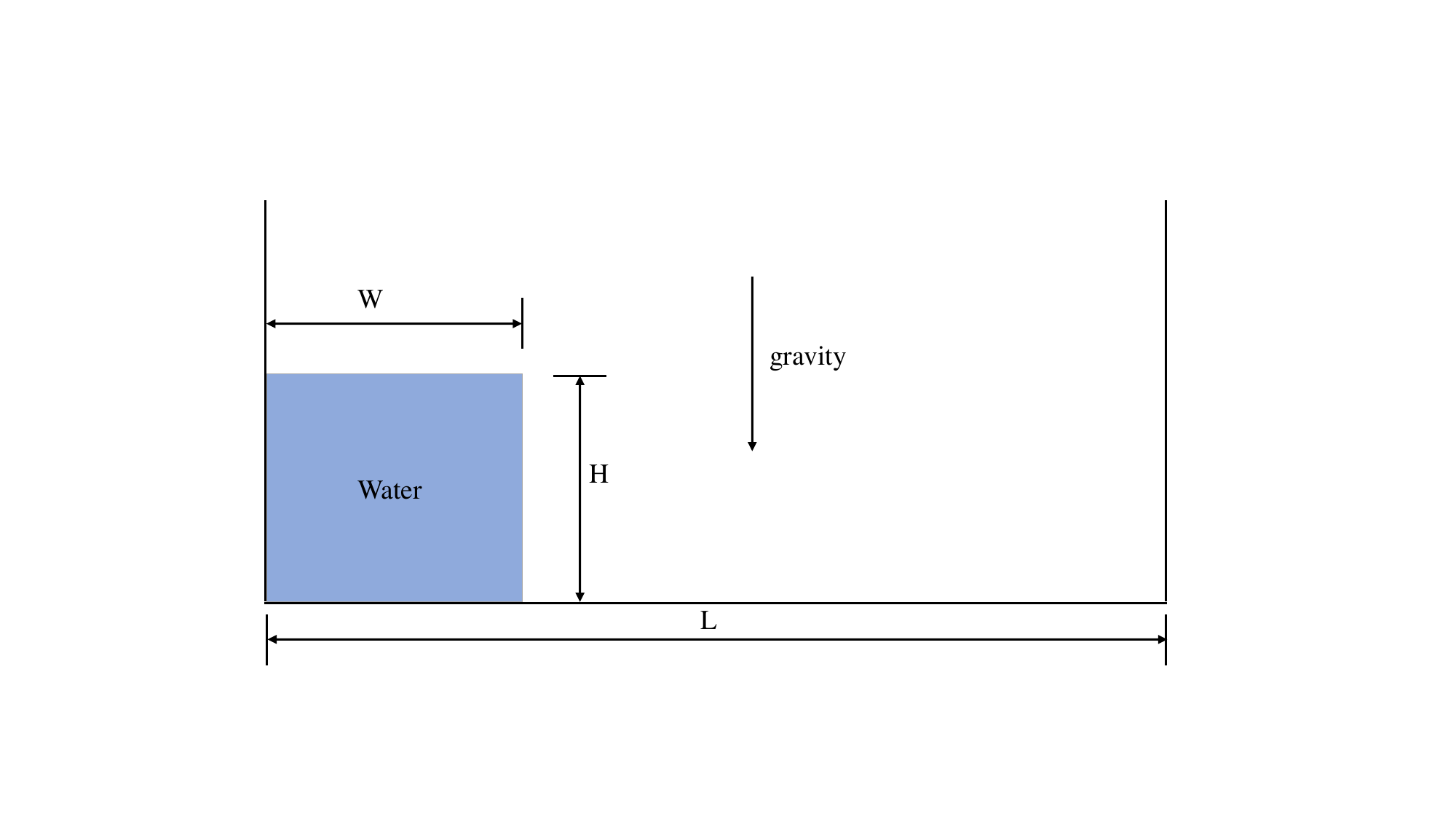}
\end{subfigure}
\caption{Setup for the dam break test}\label{dam_break}
\end{figure}

The phenomenon involving the collapse of a water column was initially explored in \cite{martin1952part} and has since been extensively examined through numerical simulations. Additionally, an analytical investigation of this problem was conducted in \cite{ritter1892reproduction}. These studies have become the benchmark tests routinely utilized to validate various computational frameworks simulating the free surface flow of water. The test setup is shown in Fig. \ref{dam_break} with $W=H=0.057$ m and $L=4H$. The water density is considered to be $1000~Kg/m^3$. The dynamic viscosity coefficient ($\mu_f$) is 0.05 Pa s. Three sizes of inter-particle spacing are used in the simulations with $\Delta p =0.00057$ m, $0.0014$ m and $0.0029$ m. The rigid wall and water interaction is modelled through the boundary condition described in section \ref{bc}.

\begin{figure}[hbtp!] %trim={<left> <lower> <right> <upper>}
\centering
\begin{subfigure}[t]{0.49\textwidth}
\includegraphics[width=\textwidth]{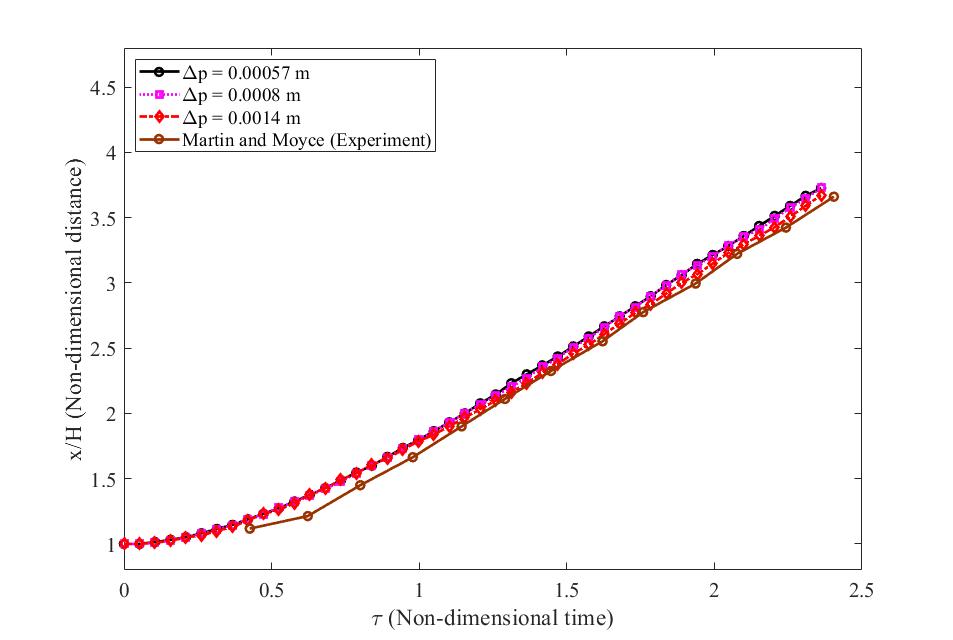}\caption{Effect of $\Delta p$}\label{dam_dp}
\end{subfigure}
\begin{subfigure}[t]{0.49\textwidth}
\includegraphics[width=\textwidth]{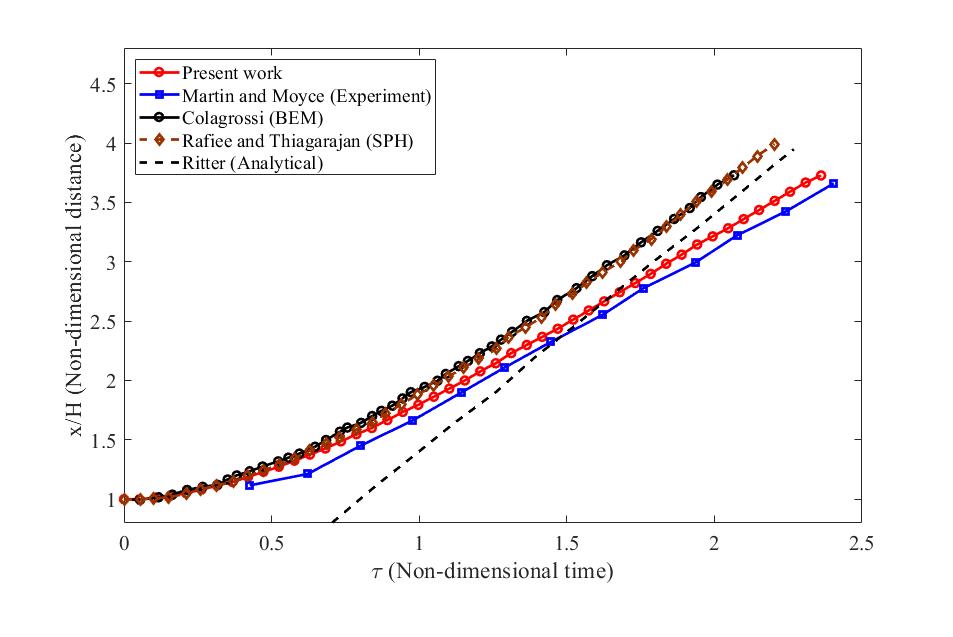}\caption{Comparison with other results with $\Delta p = 0.00057$ m}\label{dam_break_validation_2}
\end{subfigure}
\caption{Time history of the water-front toe (measured from the left wall) and the effect of inter-particle distance on the time history of the water-front toe in the dam break test ($\tau=\frac{t}{\sqrt{H/g}}$).}\label{dam_break_validation}
\end{figure}

The position of the water-front toe measured from the left wall is shown in Fig. \ref{dam_dp} at different time steps with the inter-particle resolutions and compared with the experimental results \cite{martin1952part}. The non-dimensional time coefficient is calculated as $\tau = \frac{1}{\sqrt{H/g}}$ with $g$ being the gravity force and the non-dimensional distance is $\frac{x}{H}$ with x being the current position of the water-front toe measured from the left wall. It can be observed that the present simulations agree well with the experimental result. The positional time history of the water-front toe is compared with other results available in the open literature in Fig. \ref{dam_break_validation_2} with $\Delta p = 0.00057$ m.  

In Fig. \ref{dam_break_contour}, we present the contours illustrating the velocity and pressure distribution of the water at different time steps. The simulation effectively captures the behaviour of free-surface flow influenced by the gravitational force of the dam. The simulation notably depicts the dam breaking, leading to water flow along the dry bed, culminating in an impact against a vertical rigid wall. Subsequently, the water rises, falls, and overturns backwards onto the underlying water. These flow patterns and pressure distributions closely mirror findings from previous research \cite{colagrossi2005meshless, rafiee2009sph, sun2020smoothed}.

\begin{figure}[hbtp!] %trim={<left> <lower> <right> <upper>}
\centering
\begin{subfigure}[t]{0.49\textwidth}
\includegraphics[width=\textwidth,trim={500 250 450 300}, clip]{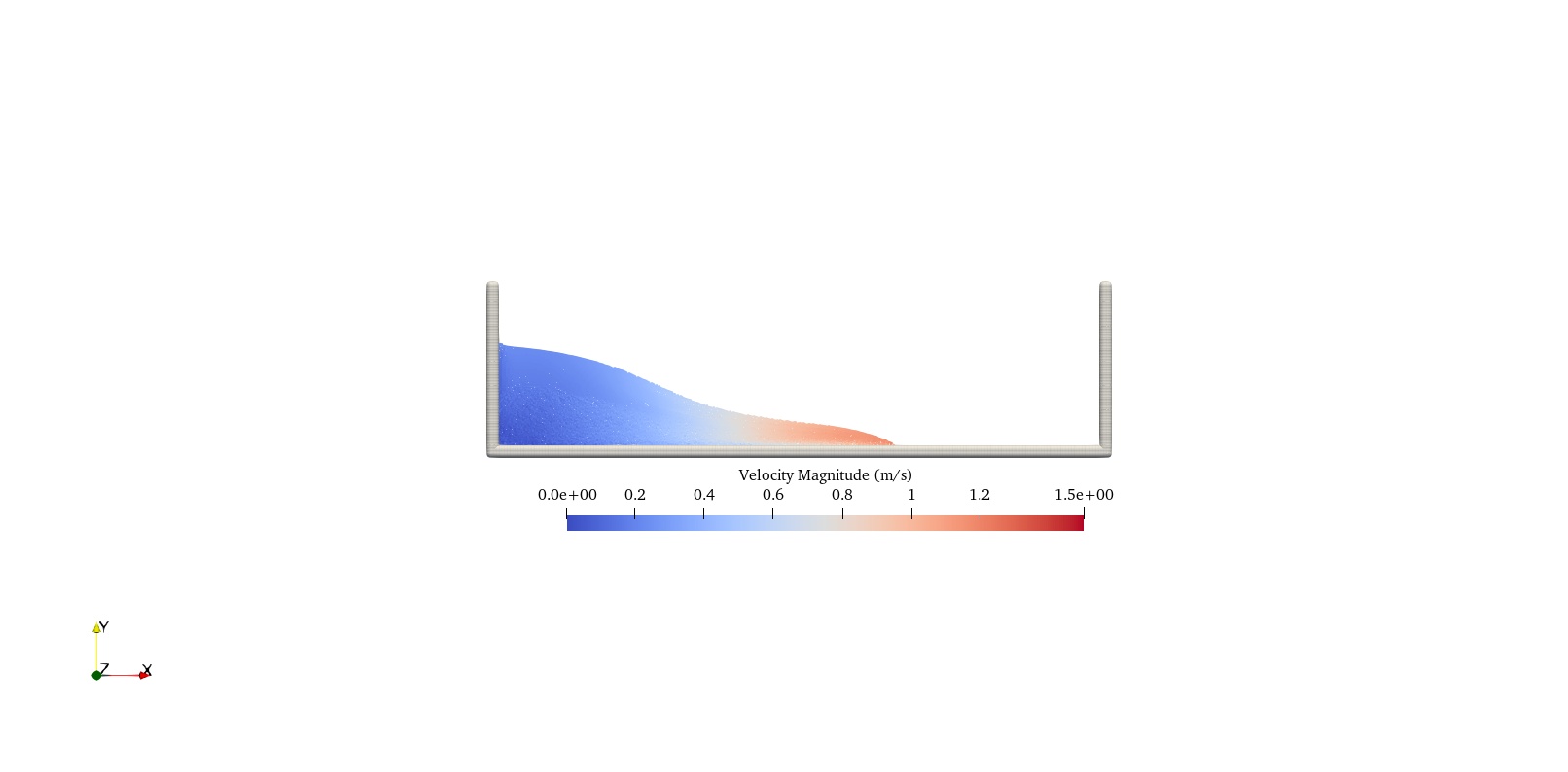}\caption{Time = 0.12 s}
\end{subfigure}
\begin{subfigure}[t]{0.49\textwidth}
\includegraphics[width=\textwidth,trim={500 250 450 300}, clip]{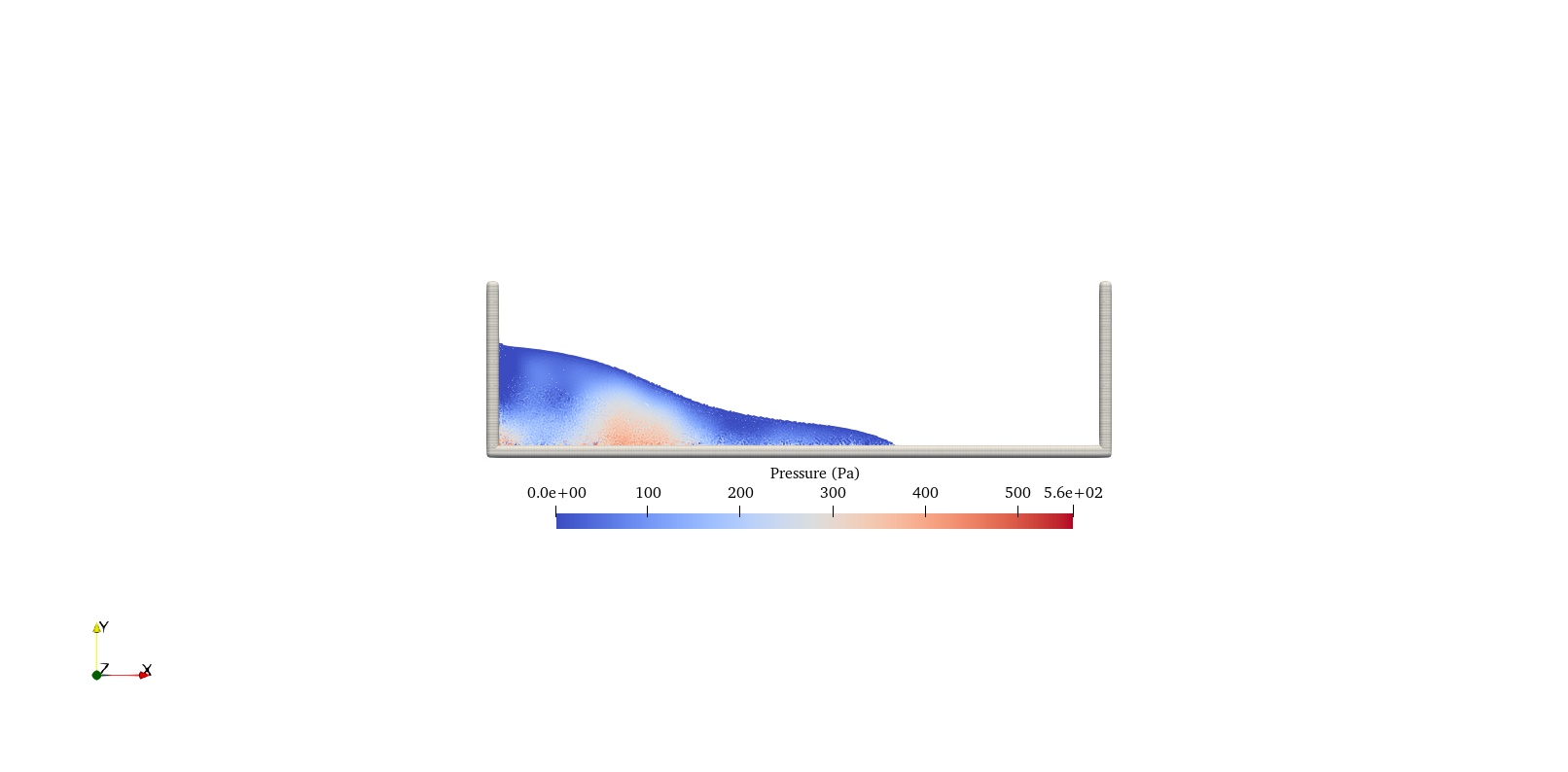}\caption{Time = 0.12 s}
\end{subfigure}
\begin{subfigure}[t]{0.49\textwidth}
\includegraphics[width=\textwidth,trim={500 250 450 300}, clip]{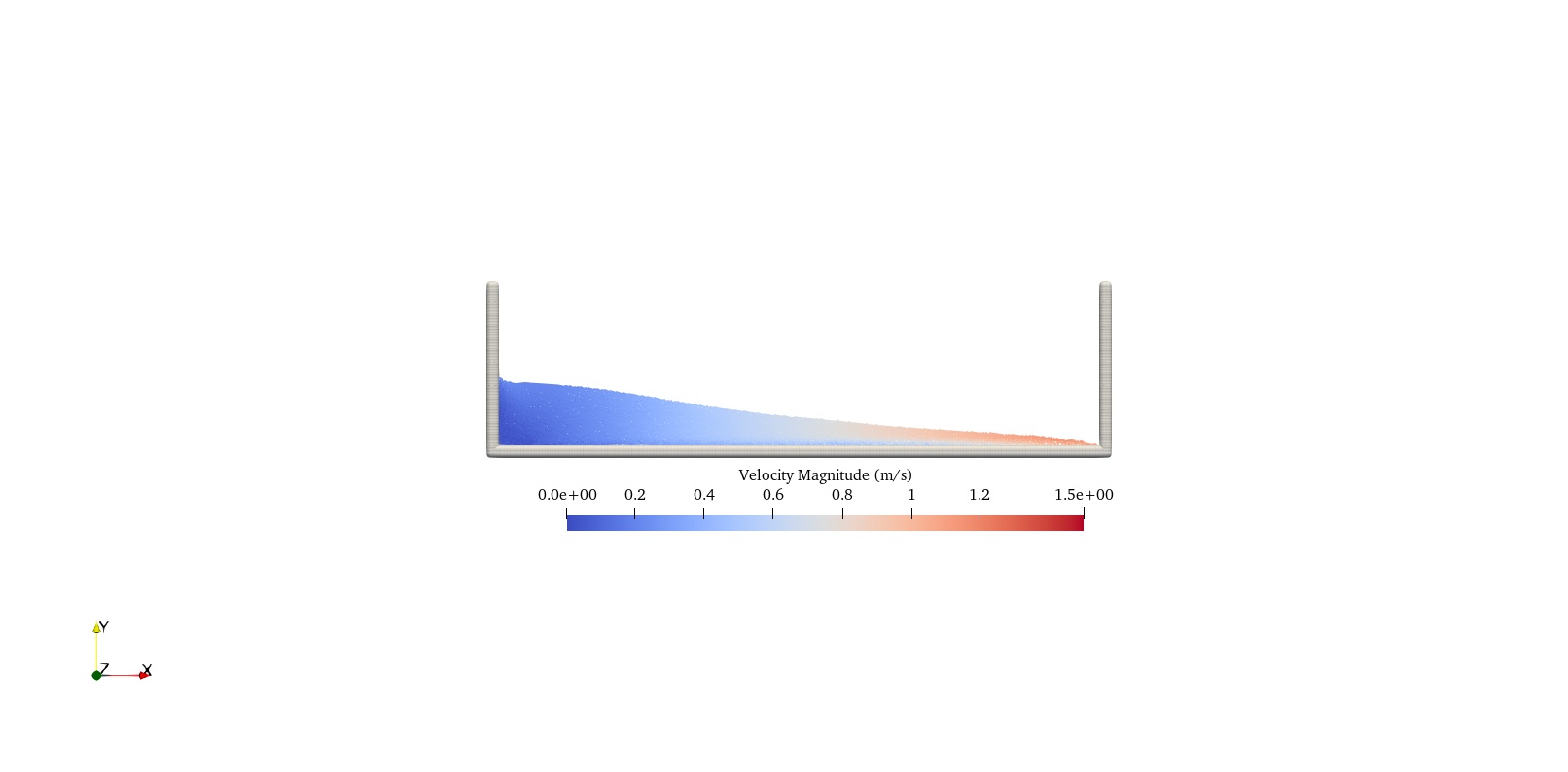}\caption{Time = 0.19 s}
\end{subfigure}
\begin{subfigure}[t]{0.49\textwidth}
\includegraphics[width=\textwidth,trim={500 250 450 300}, clip]{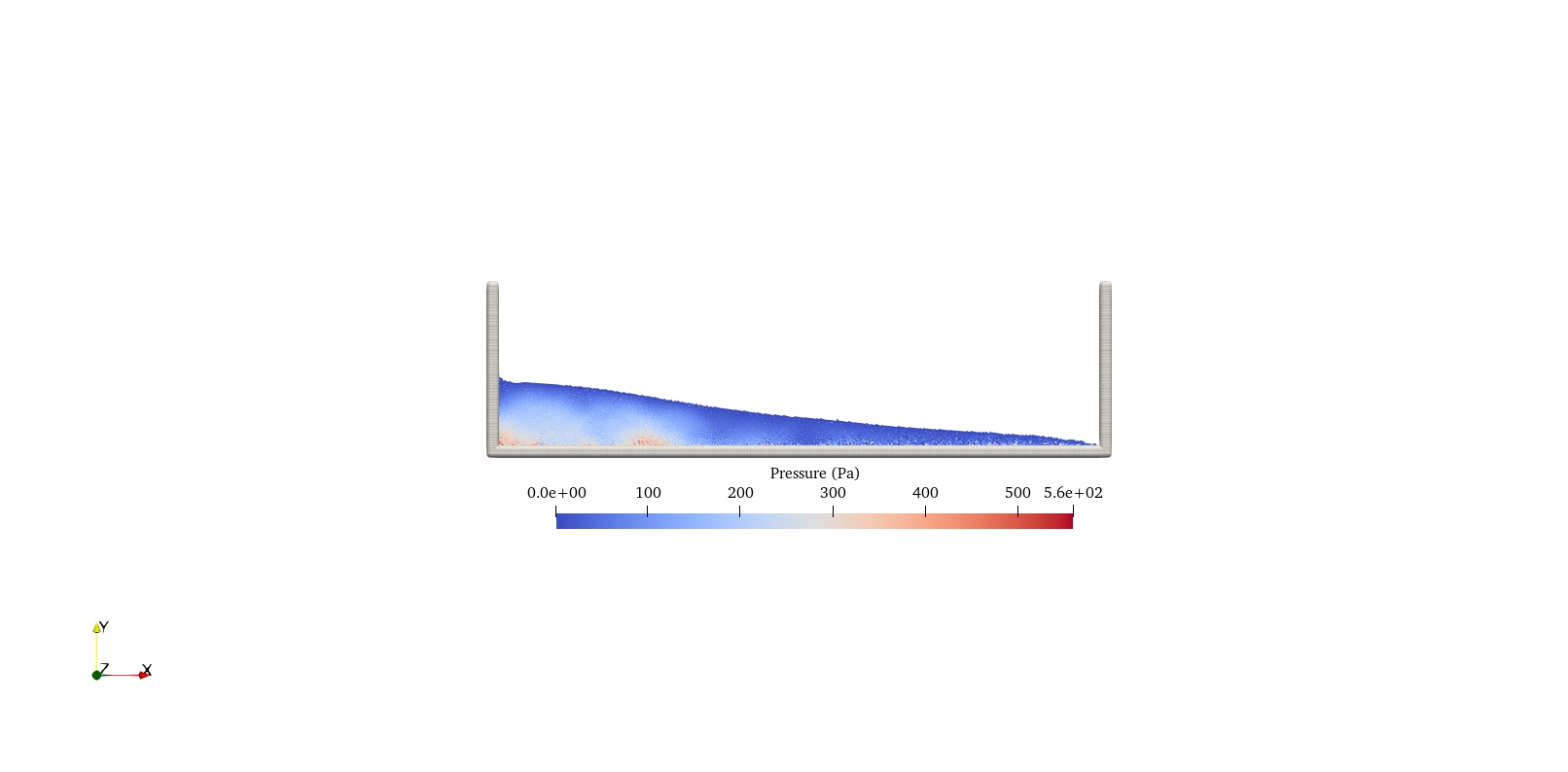}\caption{Time = 0.19 s}
\end{subfigure}
\begin{subfigure}[t]{0.49\textwidth}
\includegraphics[width=\textwidth,trim={500 250 450 300}, clip]{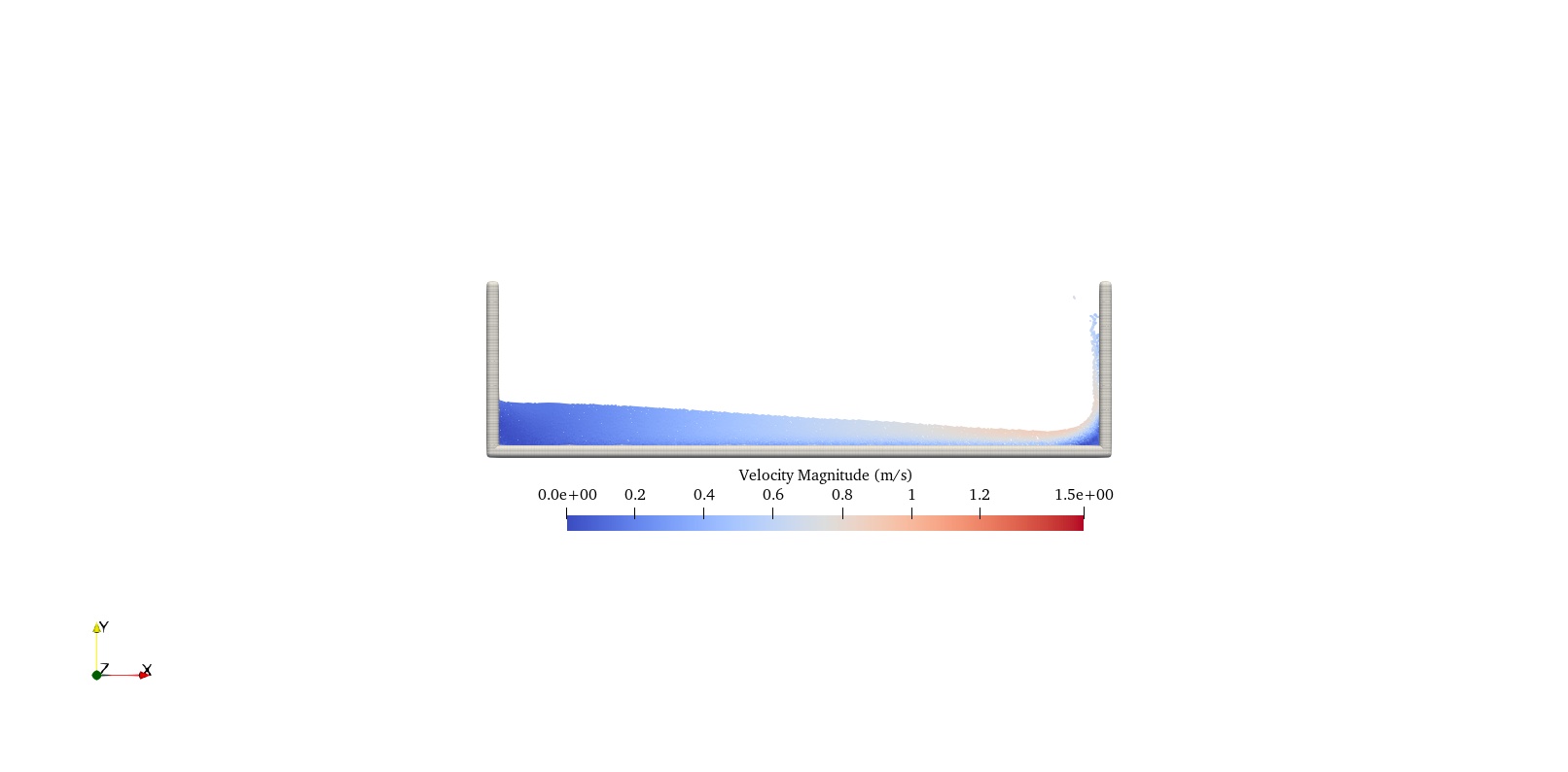}\caption{Time = 0.25 s}
\end{subfigure}
\begin{subfigure}[t]{0.49\textwidth}
\includegraphics[width=\textwidth,trim={500 250 450 300}, clip]{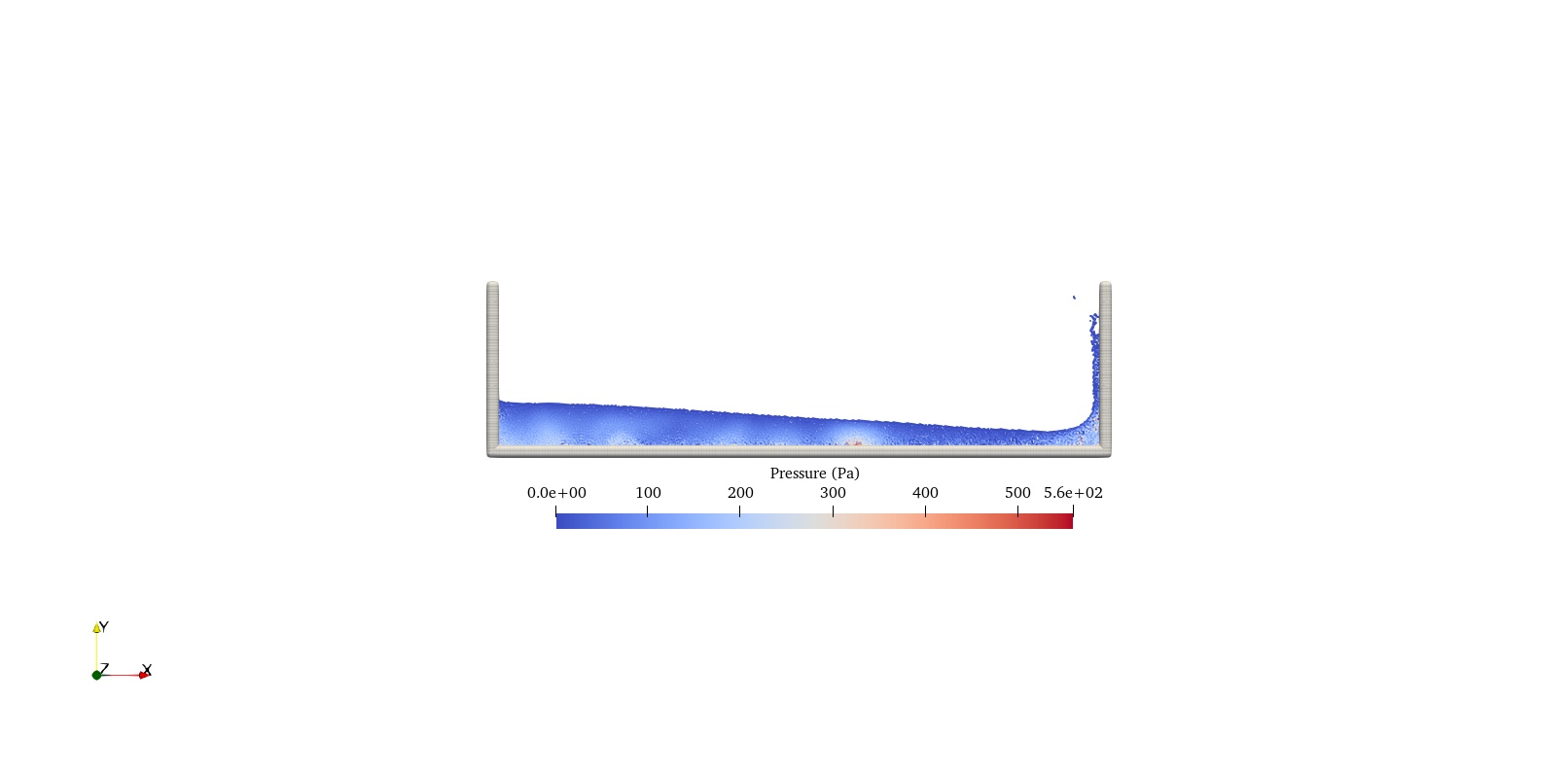}\caption{Time = 0.25 s}
\end{subfigure}
%\begin{subfigure}[t]{0.49\textwidth}
%\includegraphics[width=\textwidth,trim={500 250 450 300}, clip]{dam_break_0_40s_velocity.jpeg}\caption{Time = 0.40 s}
%\end{subfigure}
%\begin{subfigure}[t]{0.49\textwidth}
%\includegraphics[width=\textwidth,trim={500 250 450 300}, clip]{dam_break_0_40s_pressure.jpeg}\caption{Time = 0.40 s}
%\end{subfigure}
\caption{Contours of velocity magnitude and pressure distribution at 0.12, 0.19 and 0.25 s in dam break test}\label{dam_break_contour}
\end{figure}

\subsection{Oscillation of beam}
\begin{figure}[hbtp!] %trim={<left> <lower> <right> <upper>}
\centering
\begin{subfigure}[t]{0.7\textwidth}
\includegraphics[width=\textwidth,trim={150 250 75 150}, clip]{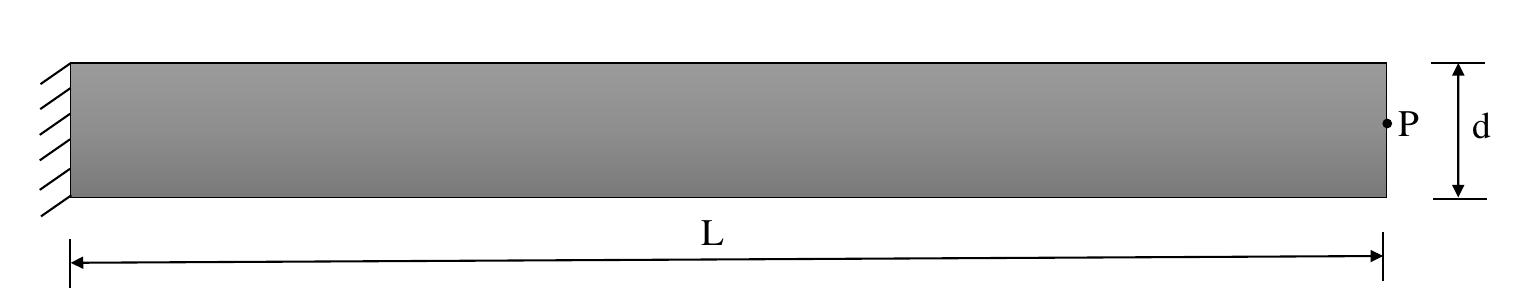}
\end{subfigure}
\caption{Schematic sketch of the beam under transverse oscillation}\label{beam}
\end{figure}

In this example, we show through a transverse oscillation of a beam that the present formulation is able to capture the deformation of deformable solids accurately. We consider an elastic cantilever beam (Fig. \ref{beam}) of length $L=10$ m and thickness $d=1$ m. 

The frequency of the oscillation is computed as $\omega^2=\frac{E d^2 k^4}{12 \rho (1 - \nu^2)}$ \citep{gray2001sph}; where, $\rho=7850 kg/m^3$ is the material density, $E=211 GPa$ is the elastic modulus and $\nu=0.3$ is the Poisson's ratio. Wave number $k$ is computed from the condition $cos(kL)sin(kL) = -1$; for the first mode, $kL=1.875$. Initially, the beam is set in motion with the following velocity function.

\begin{equation}
   \frac{v_y}{c_0} = V_f\frac{M\left\lbrace\cos (kx)-\cosh(kx)\right\rbrace-N\left\lbrace\sin(kx)-\sinh(kx)\right\rbrace}{Q}.
\end{equation}
where, $c_0$ is the sound speed in the medium, $V_f$ is the transverse velocity set as $V_f=0.05$, $M = \sin (kL) + \sinh (kL)$, $N = \cos (kL) + \cosh (kL)$ and $Q = 2(\cos (kL) \sinh(kL) - \sin (kL) \cosh (kL))$.

Simulations are performed with inter-particle spacing $\Delta p = 0.05$ m and $\frac{h}{\Delta p}=1.5$. $\gamma=0.3$ has been used to suppress the tensile instability in the present problem. The numerically computed time periods differ from the theoretical time period (0.114 s) by $7.2\%$ (in the case of SPH with $\gamma=0.3$, the time period found to be 0.122 s). It can be concluded that the present SPH formulation with $\gamma=0.3$ yields results close to the analytical solutions.

\subsection{Dam break - large deformation of an elastic gate}
\begin{figure}[hbtp!] %trim={<left> <lower> <right> <upper>}
\centering
\begin{subfigure}[t]{0.8\textwidth}
\includegraphics[width=\textwidth,trim={100 100 100 100}, clip]{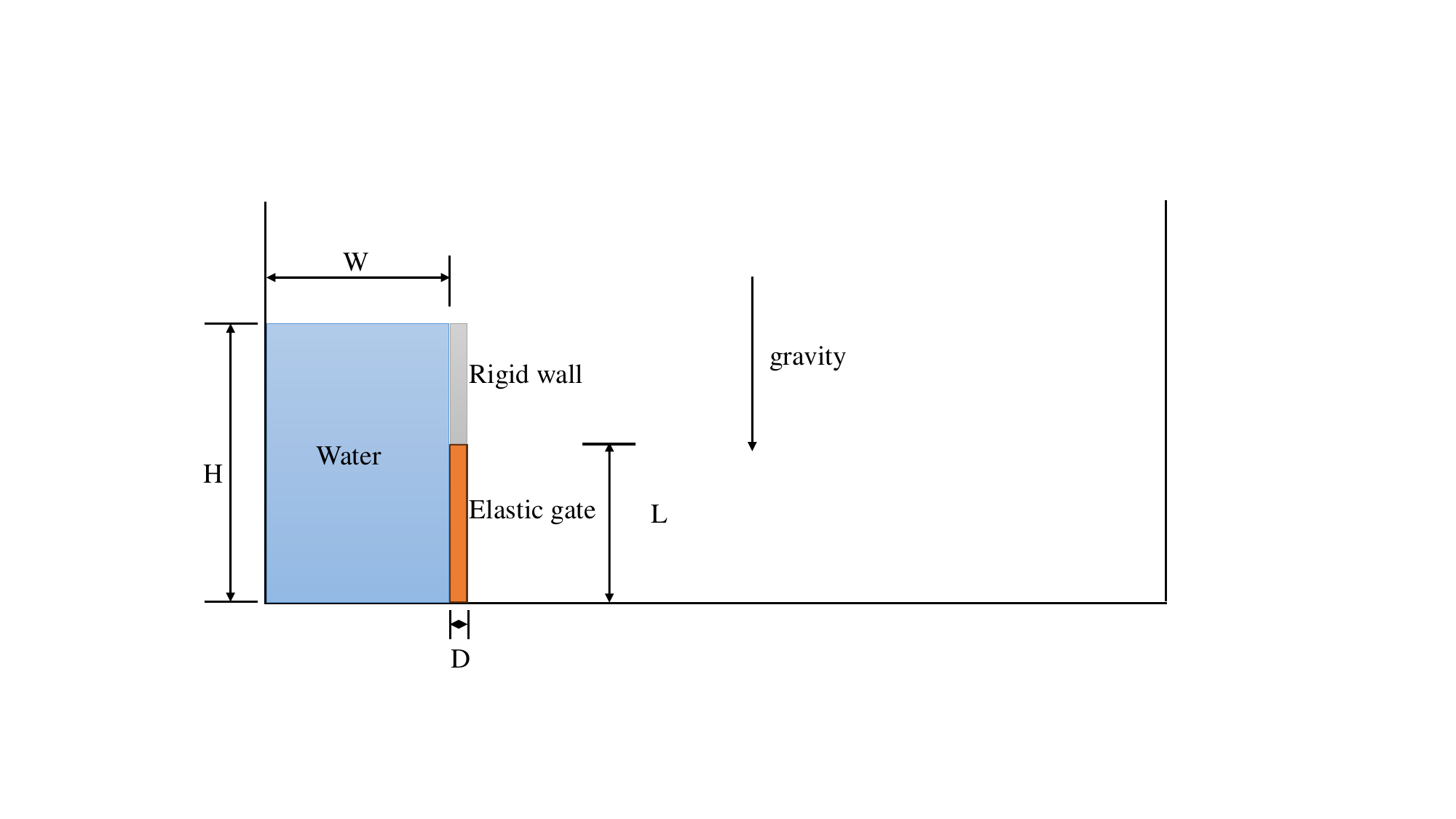}
\end{subfigure}
\caption{Setup of the dam break flow through an elastic gate}\label{dam_break_gate}
\end{figure}

The investigation into the deformation of an elastic gate due to water pressure was undertaken in \cite{antoci2007numerical}, employing both experimental and numerical methods. This problem is further modelled using different numerical techniques in \cite{rafiee2009sph, sun2020smoothed, salehizadeh2022coupled}. The set-up of the dam break flow through an elastic rubber gate is shown in Fig. \ref{dam_break_gate}. Initially, the water column is at rest, measuring $0.14$ m in height ($H$) and $0.1$ m in width ($W$). On the other hand, the elastic rubber gate has dimensions of $0.079$ m in length ($L$) and $0.005$ m in thickness ($D$). In this example, the water density is $1000~Kg/m^3$, and the dynamic viscosity coefficient ($\mu_f$) is taken as 0.05 Pa s. The density of the elastic rubber gate is $1100~Kg/m^3$. The elastic modulus is $12$ MPa, and the Poisson's ratio is $0.45$. The initial computational domain is discretized with $\Delta p = 0.0008$ m.

\begin{figure}[hbtp!] %trim={<left> <lower> <right> <upper>}
\centering
\begin{subfigure}[t]{0.49\textwidth}
\includegraphics[width=\textwidth]{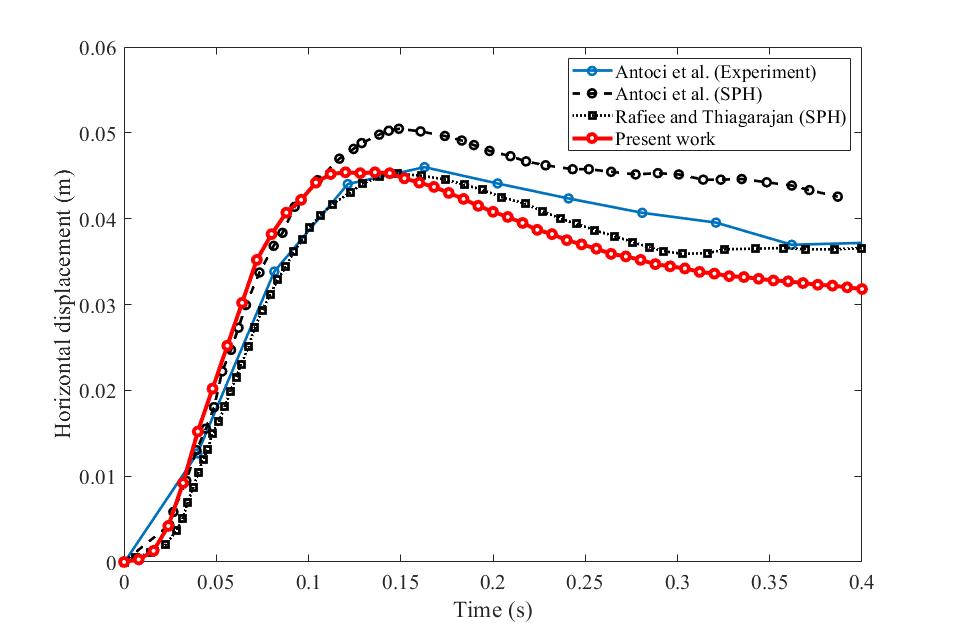}
\end{subfigure}
\begin{subfigure}[t]{0.49\textwidth}
\includegraphics[width=\textwidth]{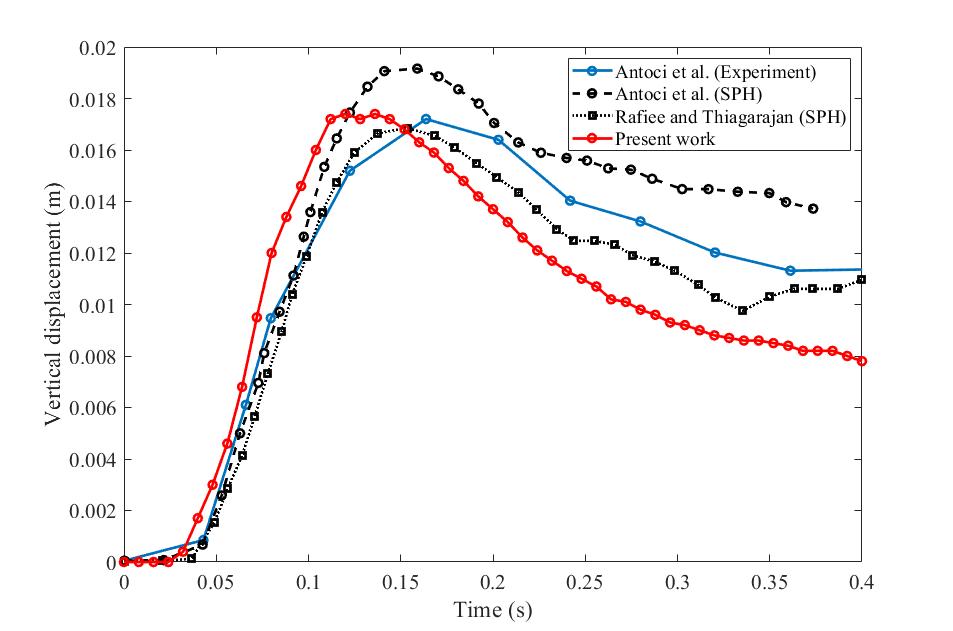}
\end{subfigure}
\caption{Comparison of time histories of the horizontal and vertical displacements of the free end of the gate}\label{dam_break_gate_hori}
\end{figure}

The water column applies force to the deformable rubber gate securely clamped to a rigid wall from above  (see Fig. \ref{dam_break_gate}). After releasing the elastic rubber gate, the water initiates contact with the gate and exits the tank through the gap between the gate and the unyielding bottom wall. The contact between the elastic rubber gate and the water is modelled through the soft repulsive particle contact model discussed in section \ref{fsi}, whereas the interaction between water and rigid wall is modelled using the boundary condition described in section \ref{bc}. Fig. \ref{dam_break_gate_hori} illustrates the temporal evolution of horizontal and vertical displacements observed at the free end of the gate. Our results have been compared with the experimental data \cite{antoci2007numerical} and other numerical results \cite{antoci2007numerical, rafiee2009sph}. The process unfolds in the following manner: initially, the water's pressure pushes the elastic gate aside, allowing the water to flow out. During the early stages, the horizontal displacement of the elastic gate increases rapidly. As the water depth in the enclosure decreases, the pressure force acting on the elastic gate diminishes, causing the gate to return towards its initial position slowly. Overall, our simulation closely aligns with the experimental and numerical data, indicating a significant level of agreement. We achieve a more favourable agreement in the early stage of the gate opening, but our simulation tends to underestimate the displacements slightly as the gate progresses toward closure.

\begin{figure}[hbtp!] %trim={<left> <lower> <right> <upper>}
\centering
%\begin{subfigure}[t]{0.4\textwidth}
%\includegraphics[width=\textwidth,trim={240 240 700 100}, clip]{dam_break_gate_16s_stress_xx.jpeg}\caption{Time = 0.16 s}
%\end{subfigure}
\begin{subfigure}[t]{0.4\textwidth}
\includegraphics[width=\textwidth,trim={240 240 700 100}, clip]{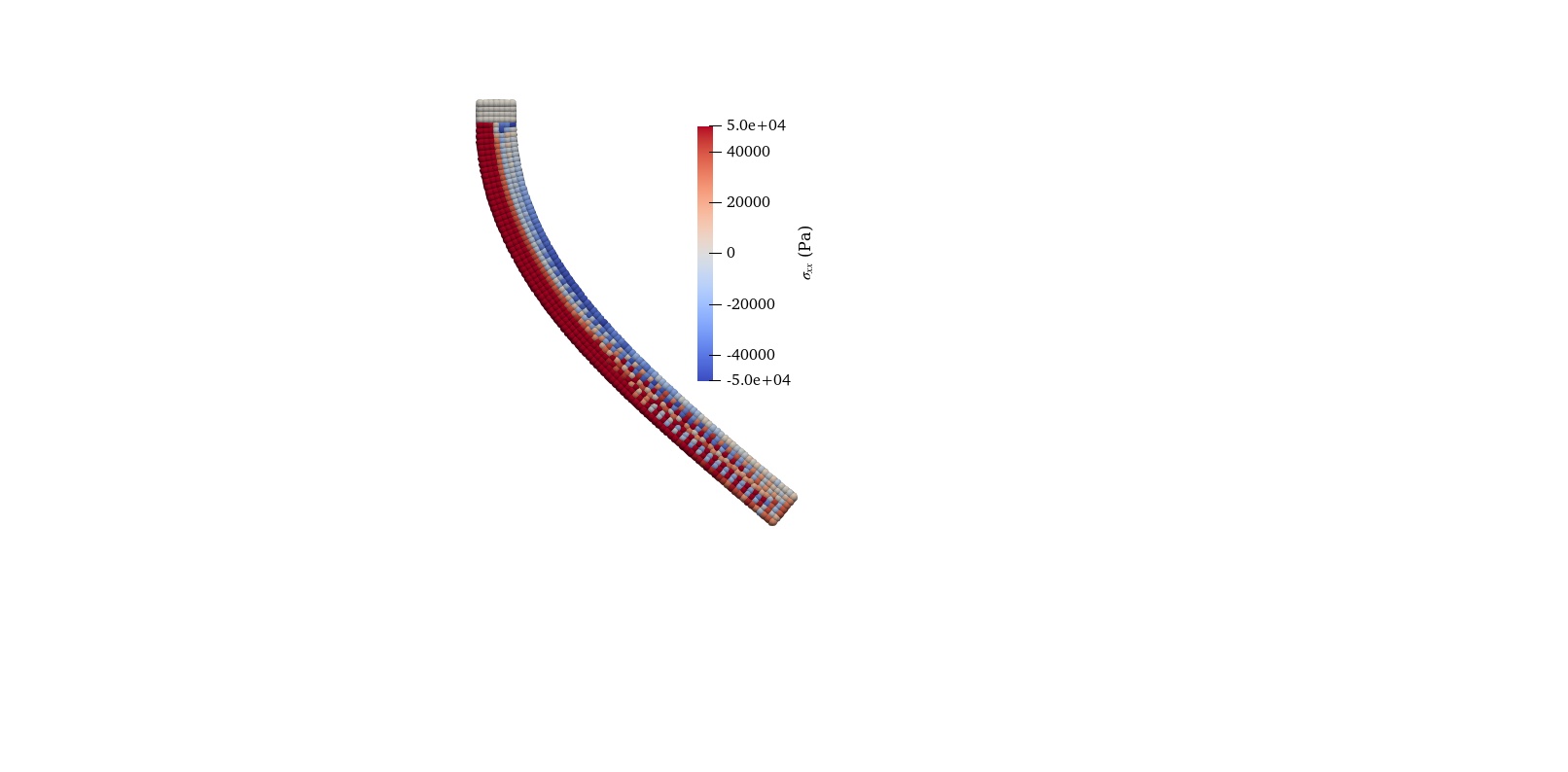}\caption{Time = 0.16 s}
\end{subfigure}
%\begin{subfigure}[t]{0.4\textwidth}
%\includegraphics[width=\textwidth,trim={240 240 700 100}, clip]{dam_break_gate_16s_stress_yy.jpeg}\caption{Time = 0.16 s}
%\end{subfigure}
\begin{subfigure}[t]{0.4\textwidth}
\includegraphics[width=\textwidth,trim={240 240 700 100}, clip]{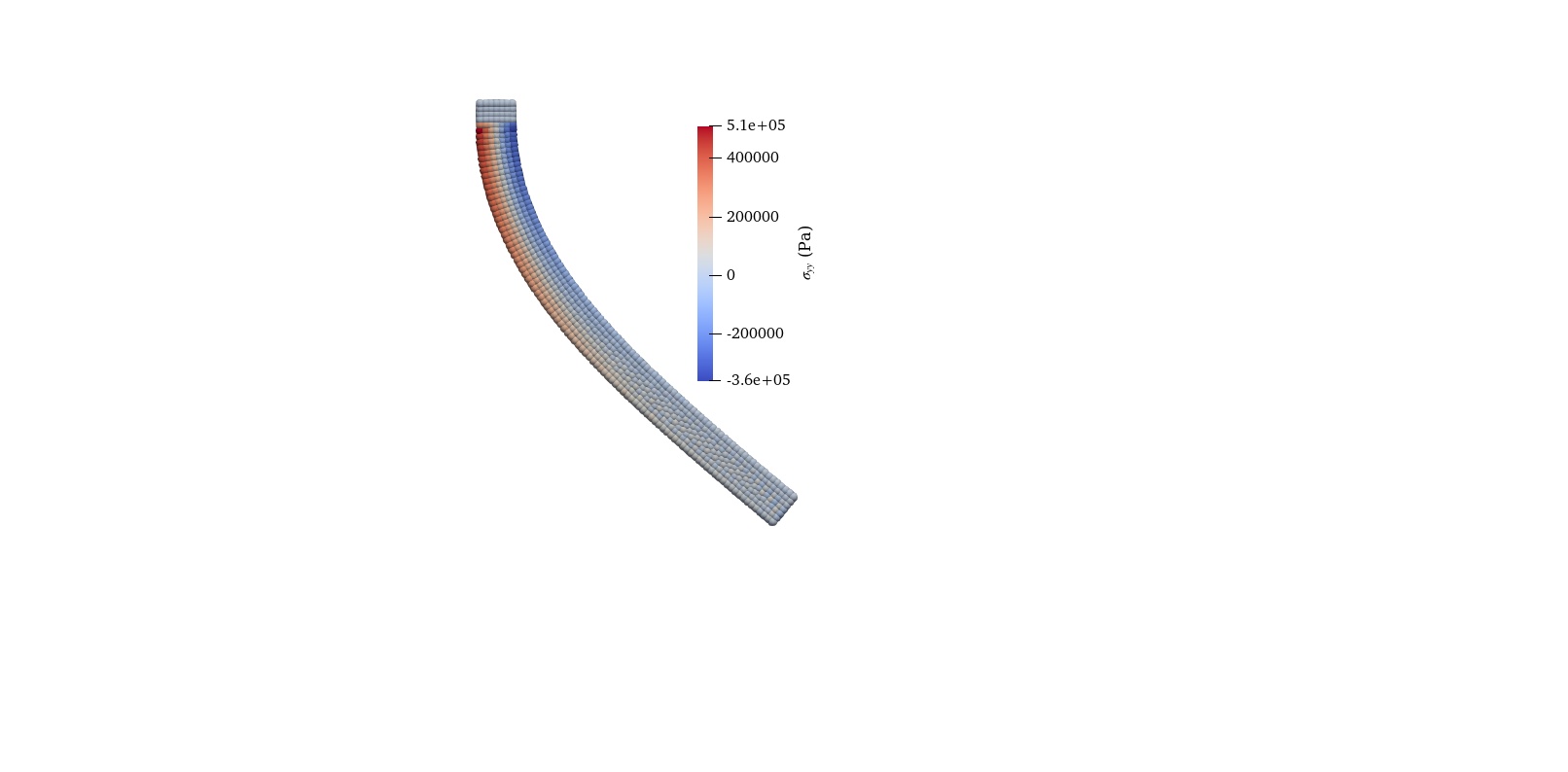}\caption{Time = 0.16 s}
\end{subfigure}
\caption{Pressure and stress distribution in the water and elastic gate}\label{dam_break_contour_stress}
\end{figure}

The simulation frames with the present framework at specific time points are presented in Fig. \ref{dam_break_contour} and compared with the experimental snapshots. It can be noted that the FSI coupling process with nonlinear characteristics is effectively replicated. The pressure distribution is also shown, and a consistent pressure distribution is observed. The simulation as a whole maintains stability, with no occurrences of instability or simulation failure observed. The maximum stress is observed on the inner side of the anchored gate's end, where the maximum bending moment is concentrated. (see Fig. \ref{dam_break_contour_stress}).

\begin{figure}[hbtp!] %trim={<left> <lower> <right> <upper>}
\centering
\begin{subfigure}[t]{0.35\textwidth}
\includegraphics[width=\textwidth,trim={0 0 0 0}, clip]{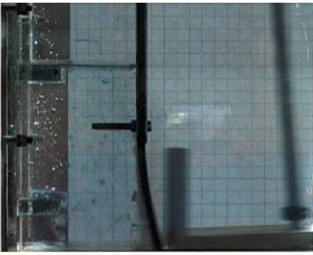}\caption{Time = 0.04 s}
\end{subfigure}
\begin{subfigure}[t]{0.37\textwidth}
\includegraphics[width=\textwidth,trim={250 180 700 125}, clip]{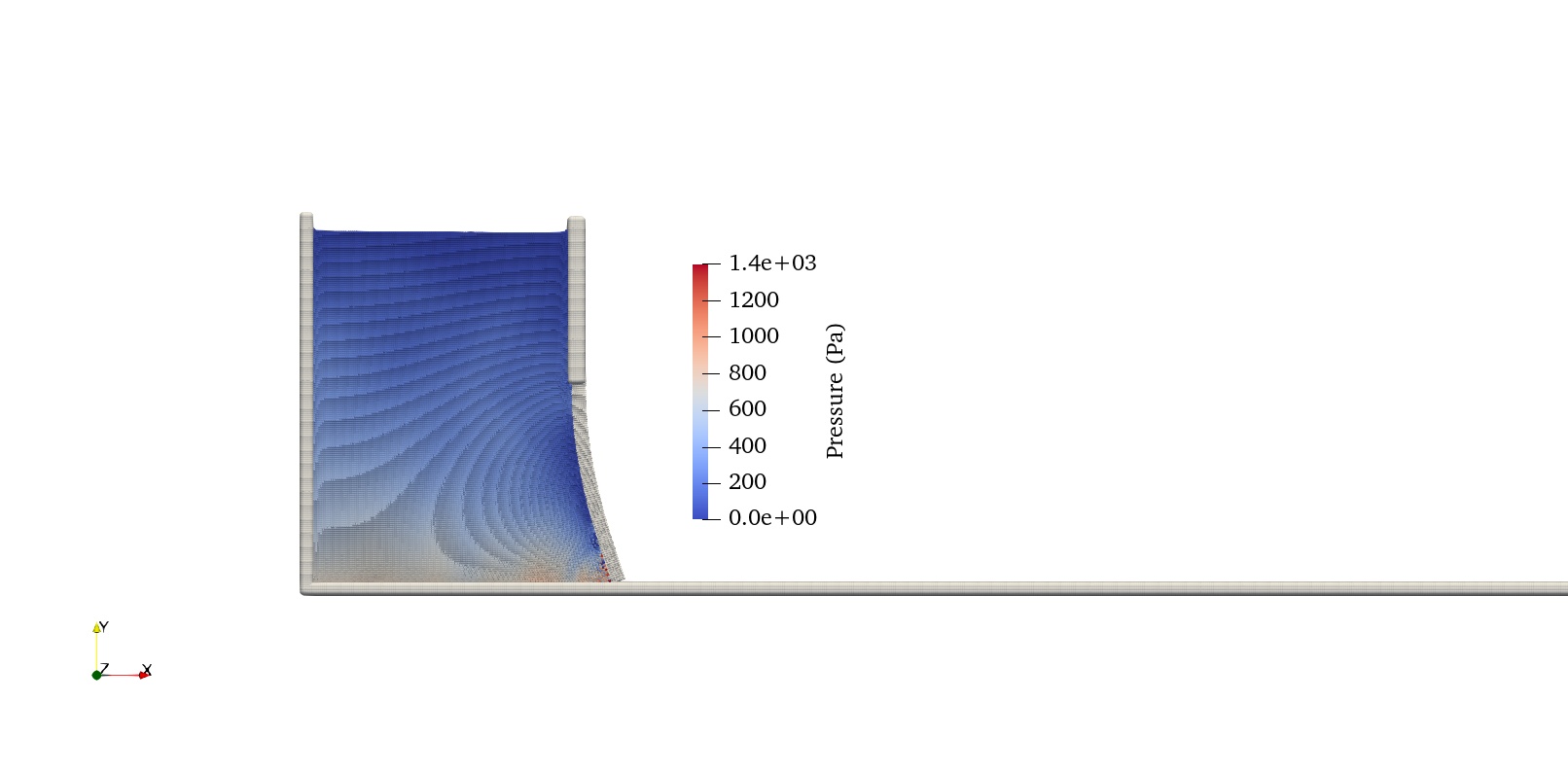}\caption{Time = 0.04 s}
\end{subfigure}
\begin{subfigure}[t]{0.35\textwidth}
\includegraphics[width=\textwidth,trim={0 0 0 0}, clip]{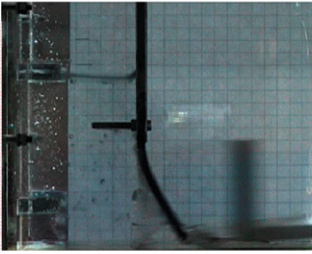}\caption{Time = 0.08 s}
\end{subfigure}
\begin{subfigure}[t]{0.37\textwidth}
\includegraphics[width=\textwidth,trim={250 180 700 125}, clip]{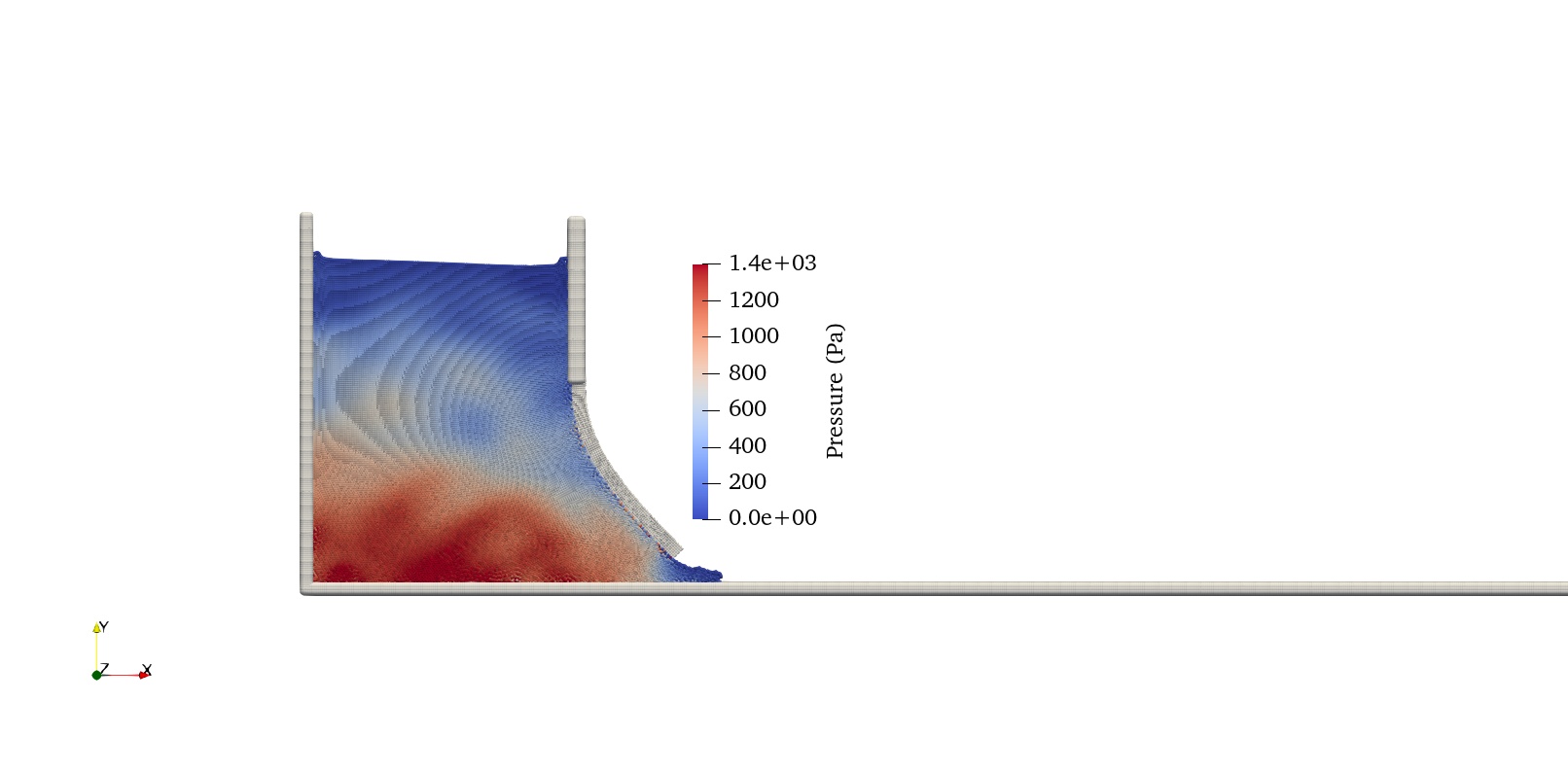}\caption{Time = 0.08 s}
\end{subfigure}
\begin{subfigure}[t]{0.35\textwidth}
\includegraphics[width=\textwidth,trim={0 0 0 0}, clip]{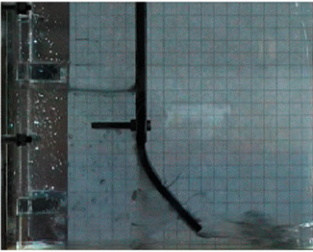}\caption{Time = 0.12 s}
\end{subfigure}
\begin{subfigure}[t]{0.37\textwidth}
\includegraphics[width=\textwidth,trim={250 180 700 125}, clip]{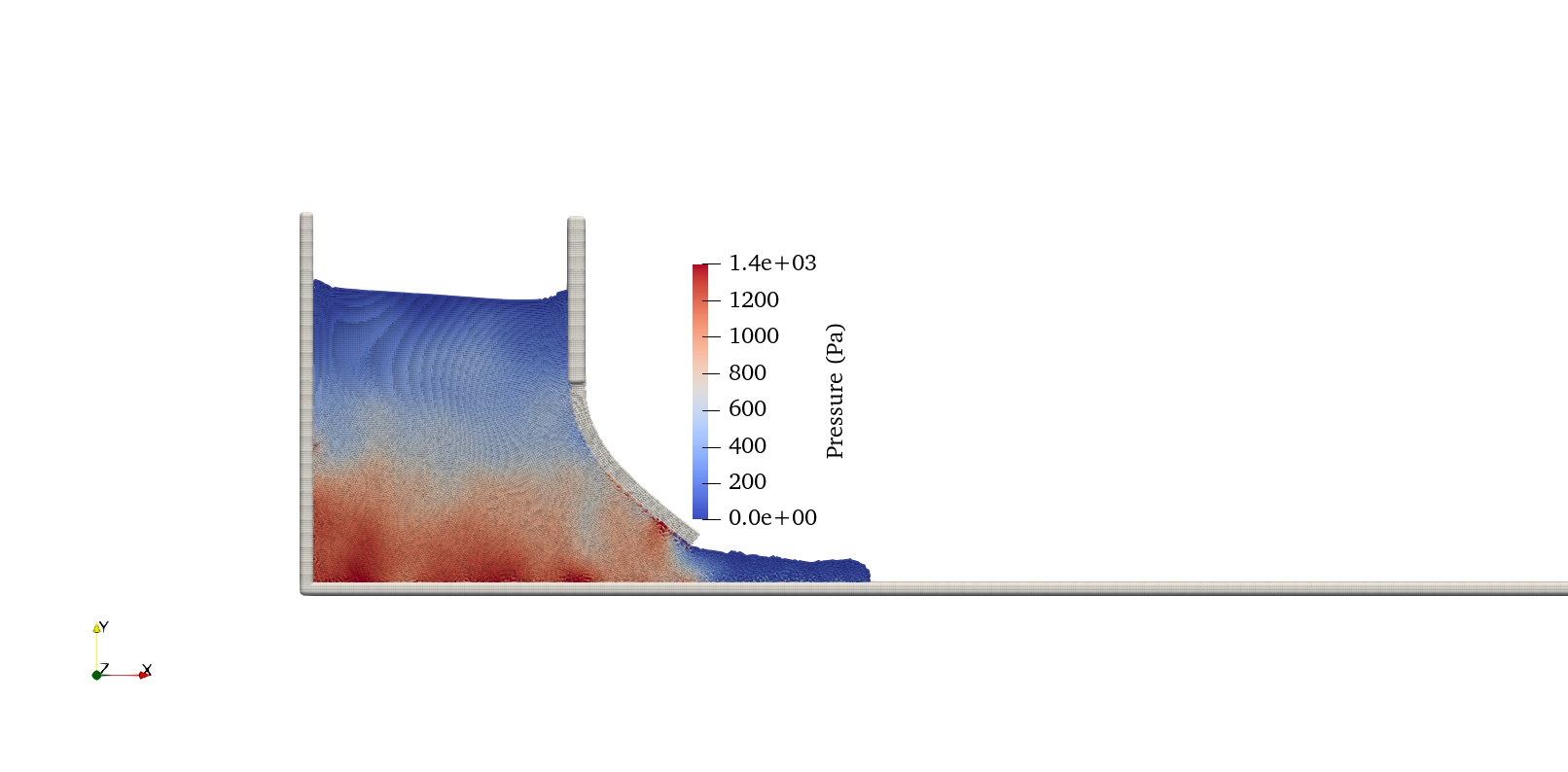}\caption{Time = 0.12 s}
\end{subfigure}
\begin{subfigure}[t]{0.35\textwidth}
\includegraphics[width=\textwidth,trim={0 0 0 0}, clip]{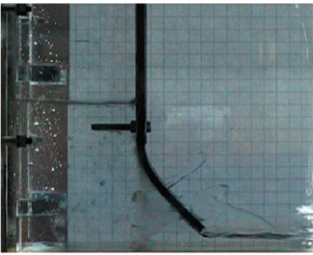}\caption{Time = 0.16 s}
\end{subfigure}
\begin{subfigure}[t]{0.37\textwidth}
\includegraphics[width=\textwidth,trim={240 180 700 125}, clip]{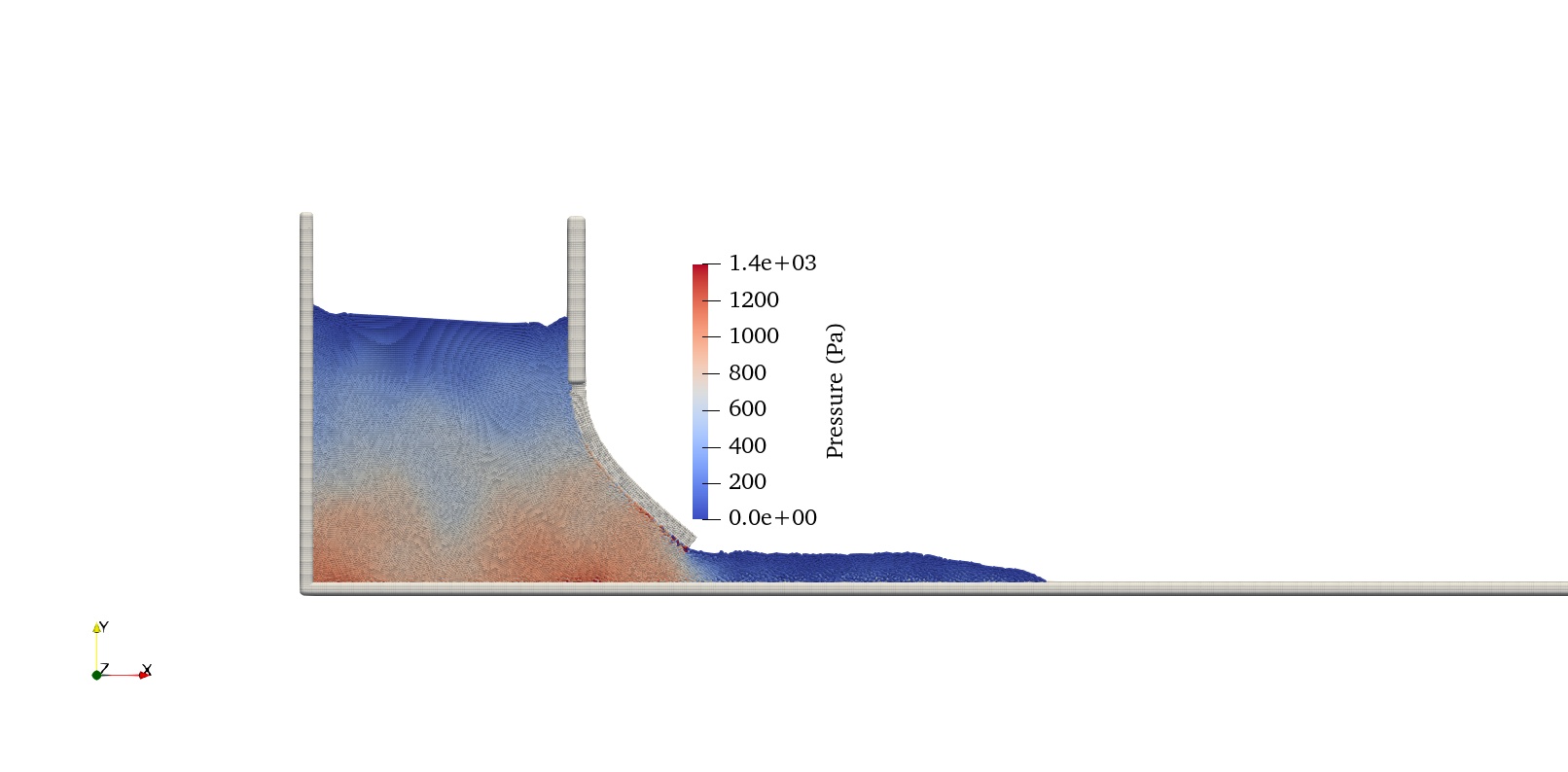}\caption{Time = 0.16 s}
\end{subfigure}
\caption{Qualitive comparison between experimental results \cite{antoci2007numerical} and present work at different time steps}\label{dam_break_contour}
\end{figure}

\subsection{Dam break flow impacting on flexible obstacle}
\begin{figure}[hbtp!] %trim={<left> <lower> <right> <upper>}
\centering
\begin{subfigure}[t]{0.8\textwidth}
\includegraphics[width=\textwidth,trim={100 70 100 100}, clip]{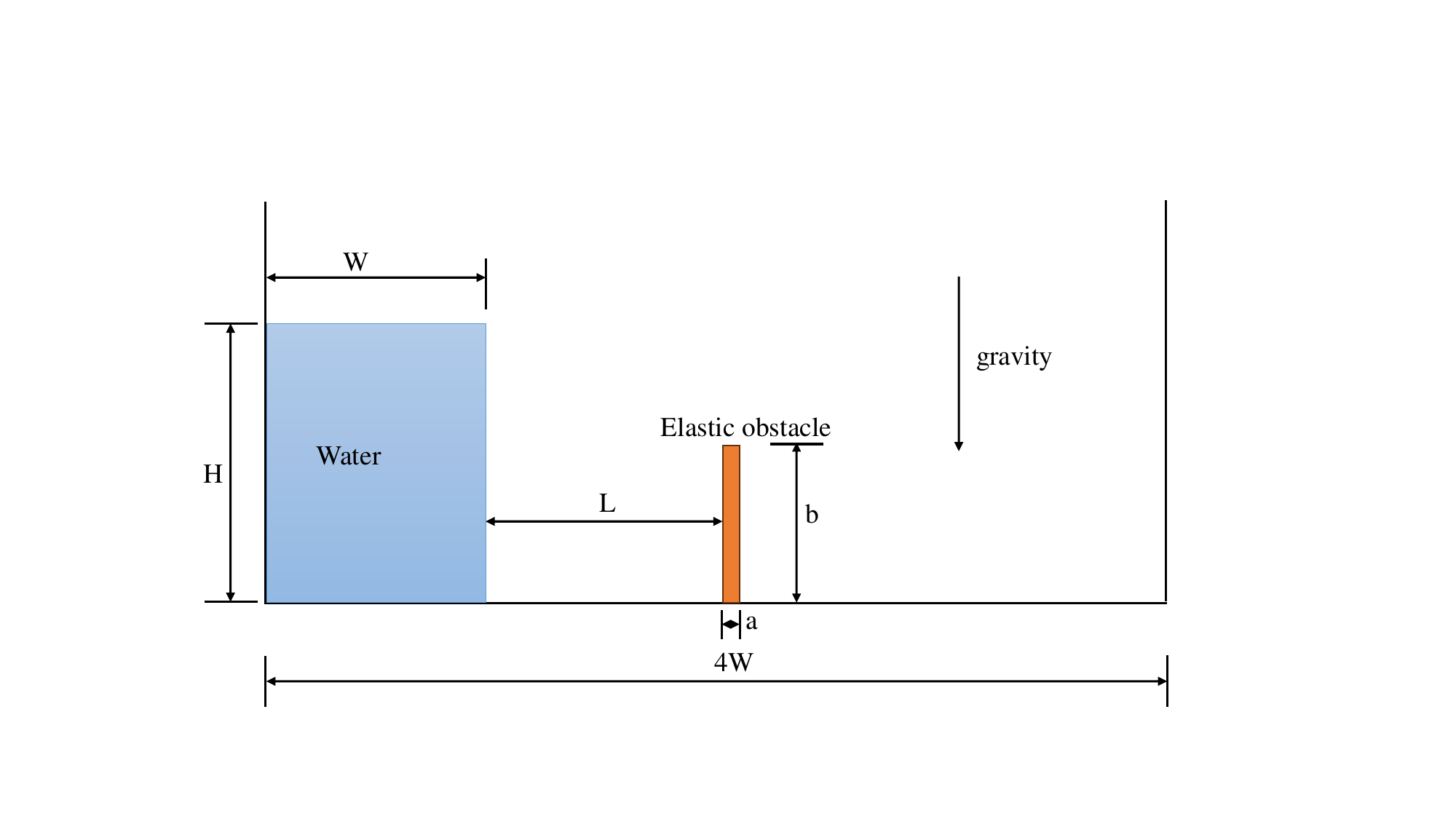}
\end{subfigure}
\caption{Setup for water impact on elastic obstacle}\label{dam_break_obstacle}
\end{figure}

In this segment, we undertake another numerical simulation to explore the dynamic interaction of a deformable structure in a fluid-structure interaction (FSI) scenario. Specifically, we simulate the scenario where a vertical water column collapses and strikes a flexible elastic wall, investigating the resulting complex dynamics. The system's configuration is shown in Fig. \ref{dam_break_obstacle}. 

The water column in this setup exhibits specific geometric parameters: its width, denoted as $W$, measures $0.146$ m, while its height, represented as $H$, measures $0.292$ m. The gap between the two vertical walls, which serves as the spatial constraint for the column, amounts to $4$ times the width, or $0.584$ m.
A deformable elastic plate occupies the central position within this confined space, fixed at its lower end. The plate is positioned at a horizontal distance of $L$, equivalent to $W$, from the approaching water column. The elastic plate has distinct dimensions, with a thickness denoted as $a$ measuring $0.012$ m and a height denoted as $b$ measuring $0.08$ m. The computational domain is discretized with $\Delta p = 0.0025$ m. The initial state of the experiment involves the water column being abruptly released, setting in motion its trajectory towards a collision with the stationary elastic plate. The density values for the water and the deformable elastic obstacle are initially set to be $1000~Kg/m^3$ and $2500~Kg/m^3$. Furthermore, the deformable elastic obstacle is characterized by an elastic modulus $E$ of $10^6~N/m^2$ and a Poisson ratio $\nu$ of $0$.

\begin{figure}[hbtp!] %trim={<left> <lower> <right> <upper>}
\centering
\begin{subfigure}[t]{0.49\textwidth}
\includegraphics[width=\textwidth,trim={375 275 300 200}, clip]{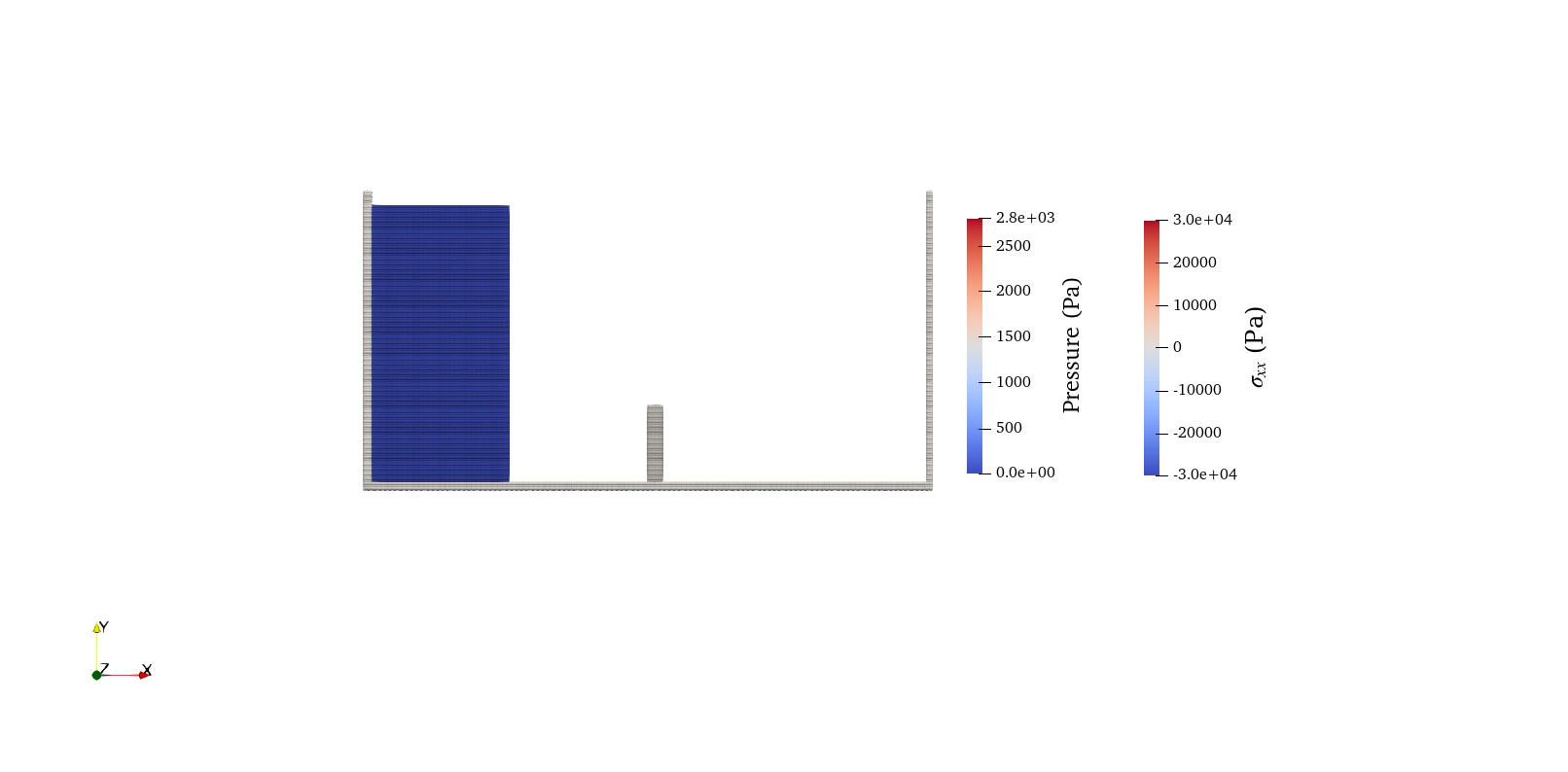}\caption{Time = 0.0 s}
\end{subfigure}
\begin{subfigure}[t]{0.49\textwidth}
\includegraphics[width=\textwidth,trim={375 275 300 200}, clip]{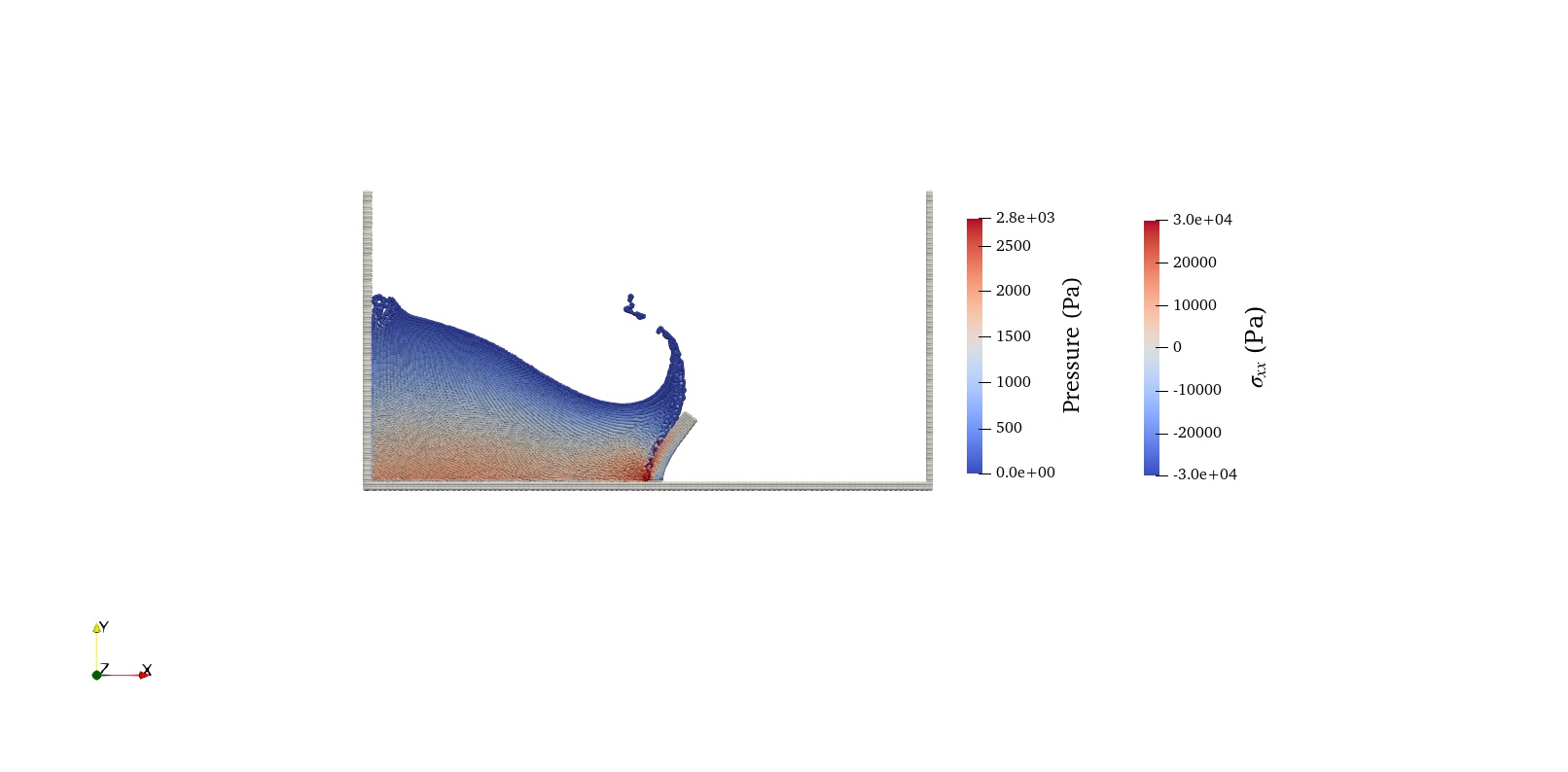}\caption{Time = 0.2 s}
\end{subfigure}
\begin{subfigure}[t]{0.49\textwidth}
\includegraphics[width=\textwidth,trim={375 275 300 200}, clip]{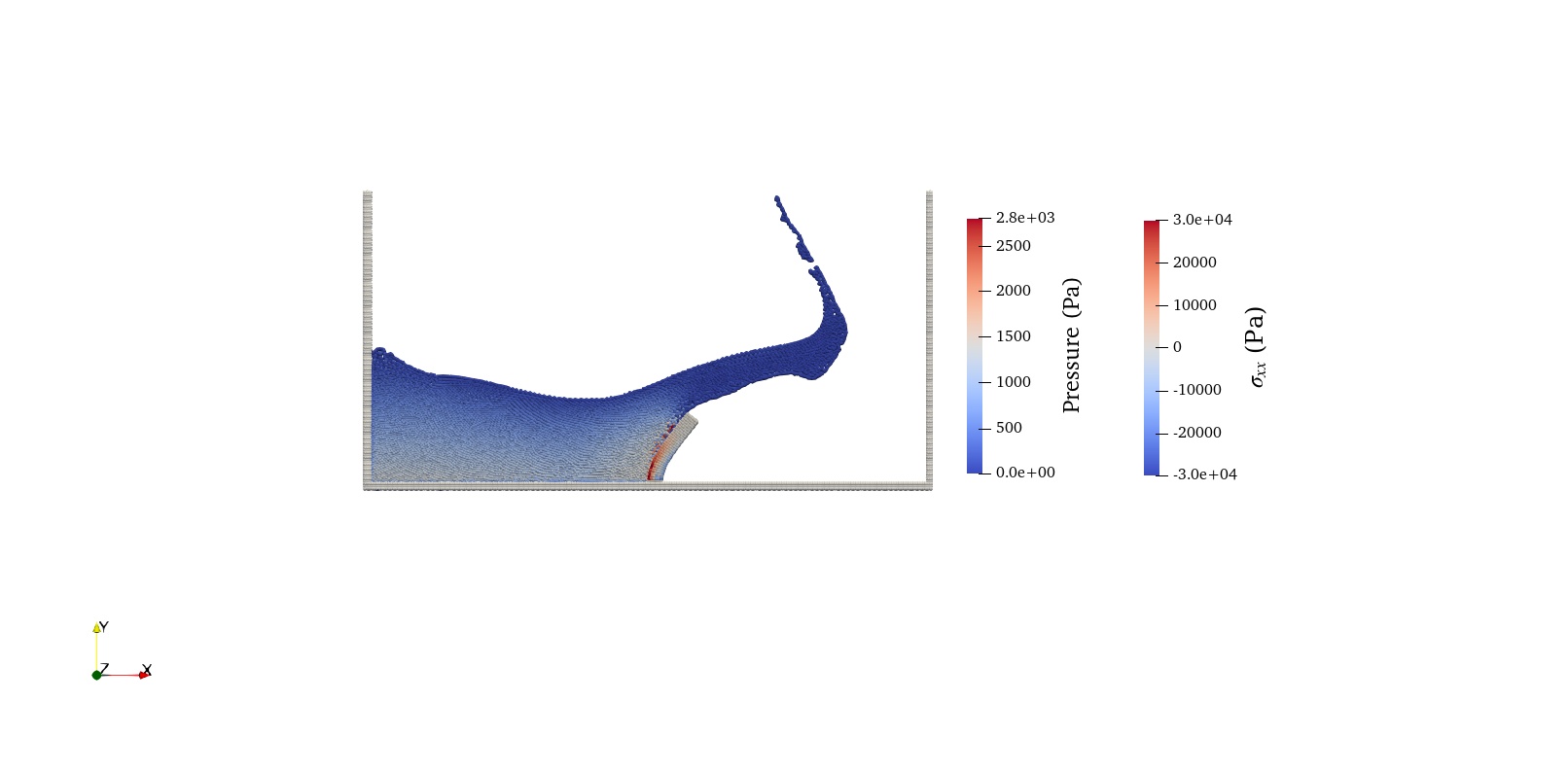}\caption{Time = 0.3 s}
\end{subfigure}
\begin{subfigure}[t]{0.49\textwidth}
\includegraphics[width=\textwidth,trim={375 275 300 200}, clip]{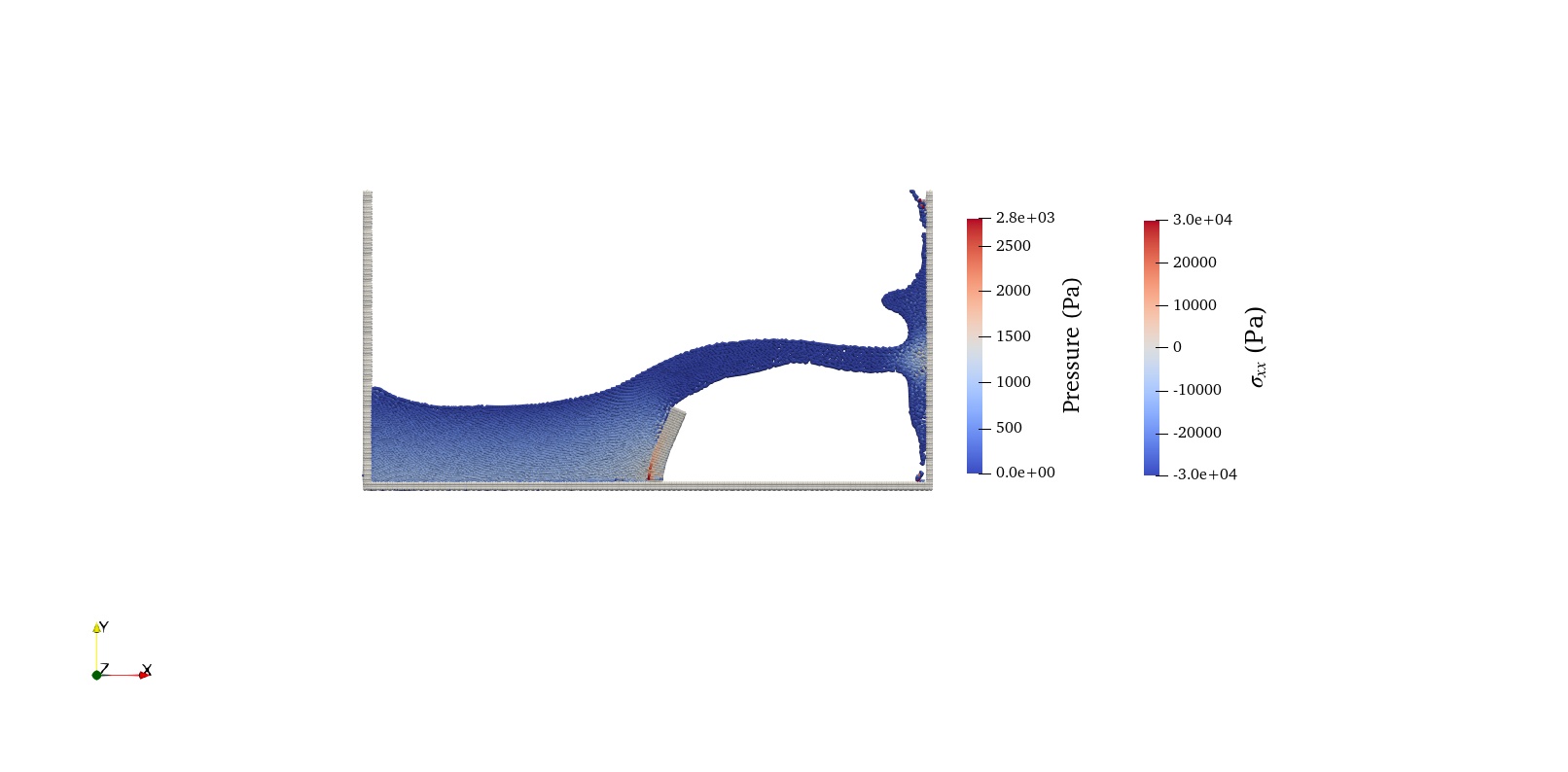}\caption{Time = 0.4 s}
\end{subfigure}
\begin{subfigure}[t]{0.49\textwidth}
\includegraphics[width=\textwidth,trim={375 275 300 200}, clip]{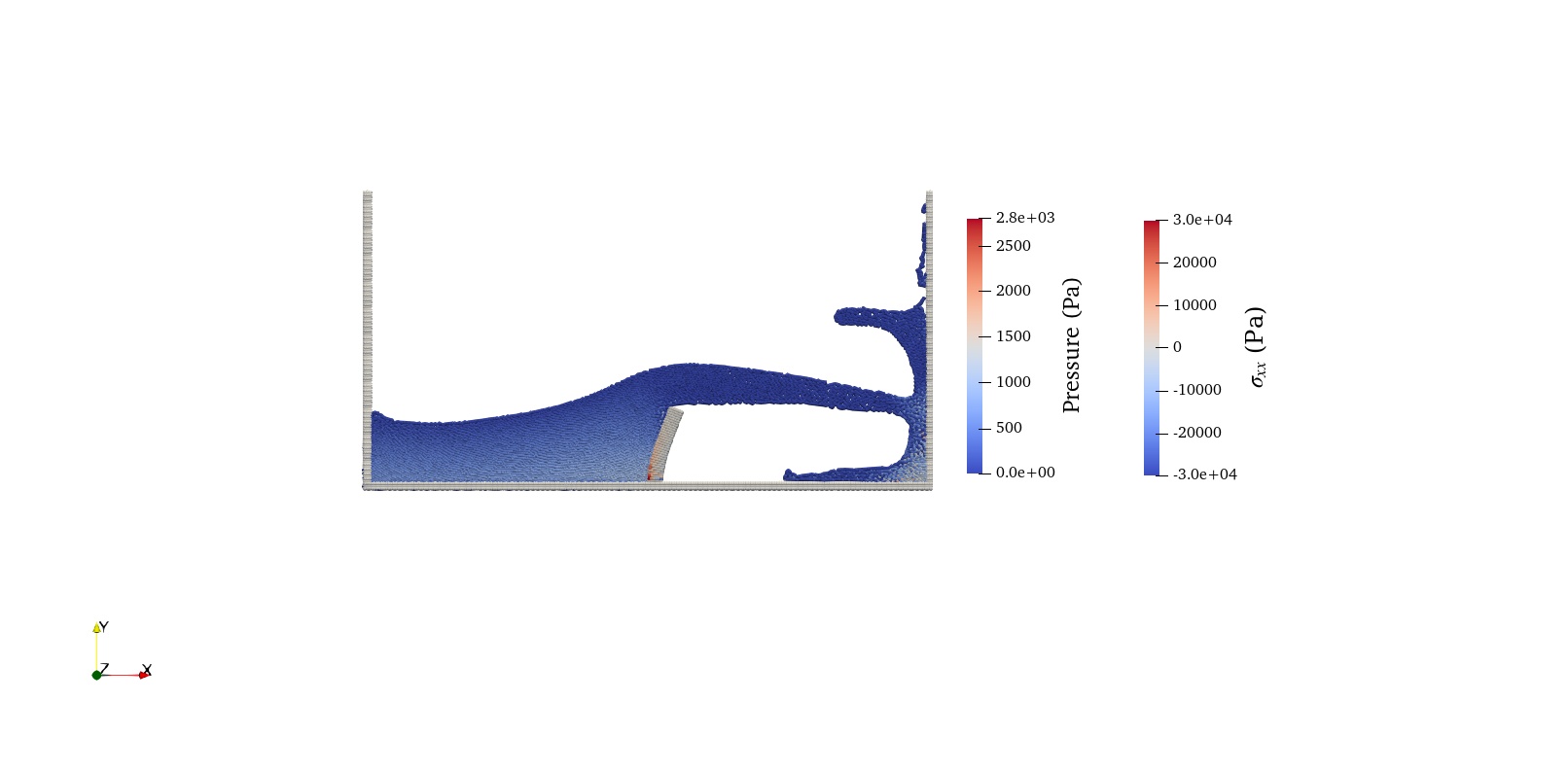}\caption{Time = 0.5 s}
\end{subfigure}
\begin{subfigure}[t]{0.49\textwidth}
\includegraphics[width=\textwidth,trim={375 275 300 200}, clip]{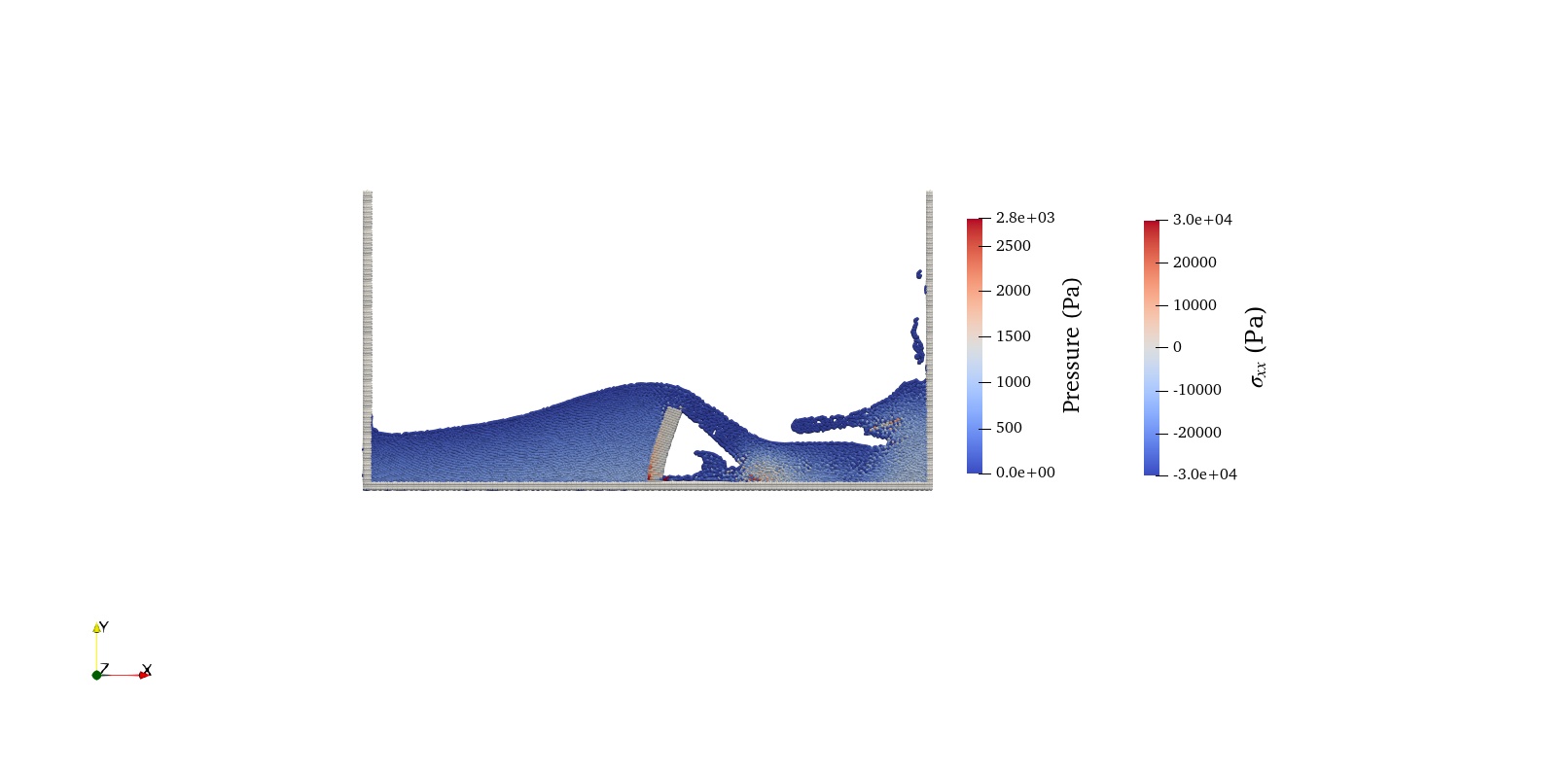}\caption{Time = 0.6 s}
\end{subfigure}
\begin{subfigure}[t]{0.49\textwidth}
\includegraphics[width=\textwidth,trim={375 275 300 200}, clip]{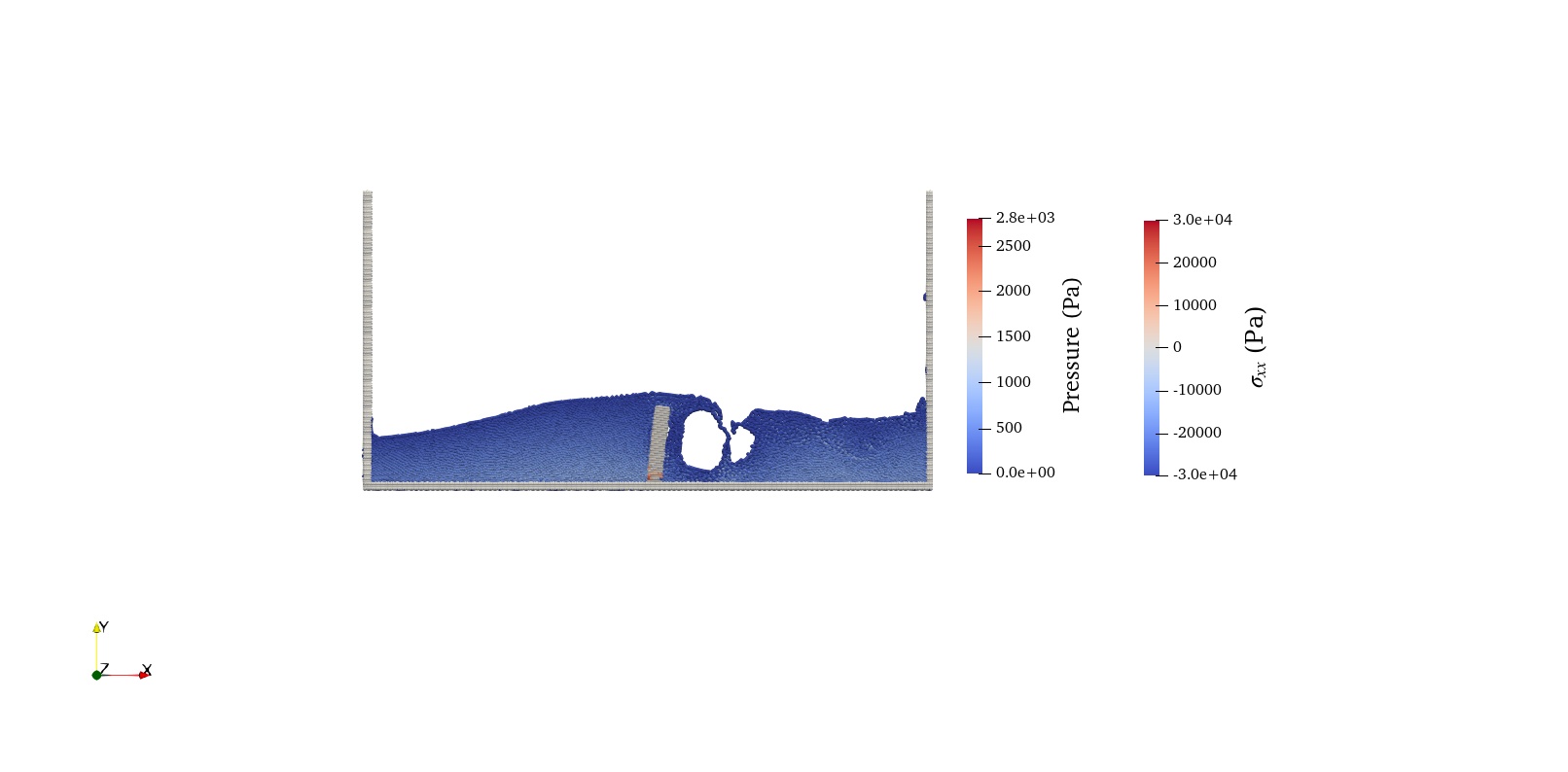}\caption{Time = 0.7 s}
\end{subfigure}
\begin{subfigure}[t]{0.49\textwidth}
\includegraphics[width=\textwidth,trim={375 275 300 200}, clip]{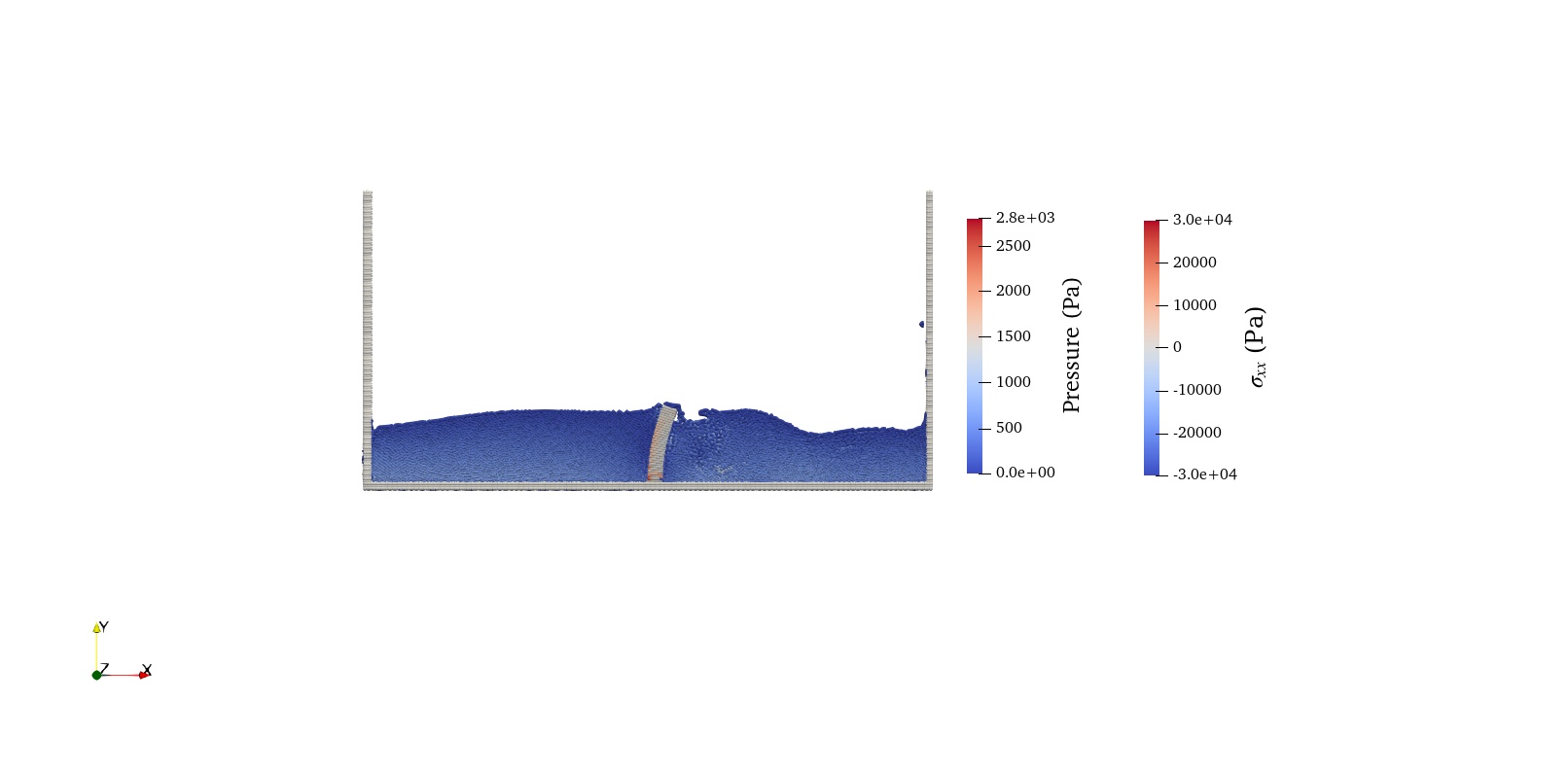}\caption{Time = 0.8 s}
\end{subfigure}
\begin{subfigure}[t]{0.49\textwidth}
\includegraphics[width=\textwidth,trim={375 275 300 200}, clip]{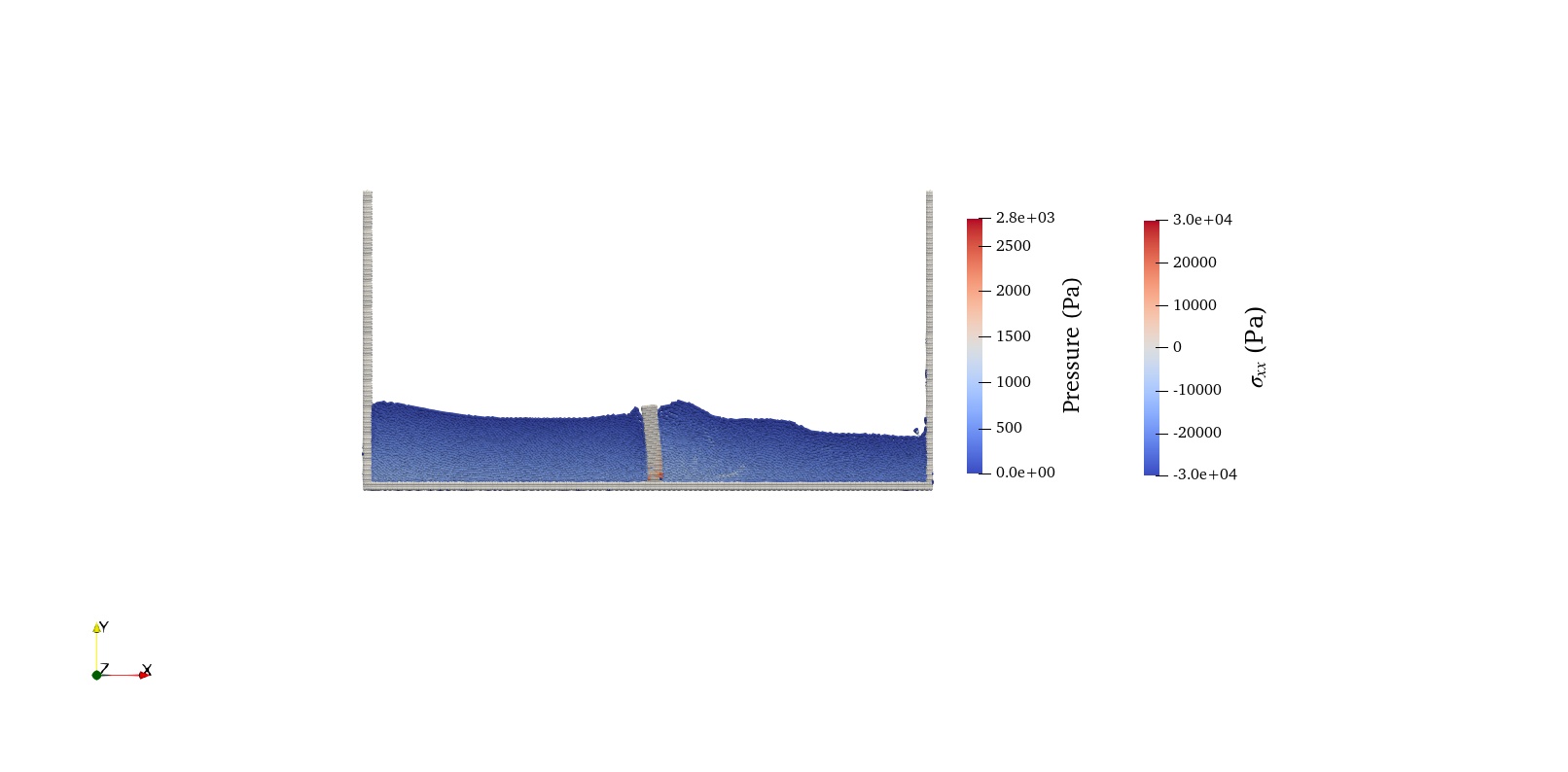}\caption{Time = 0.9 s}
\end{subfigure}
\begin{subfigure}[t]{0.49\textwidth}
\includegraphics[width=\textwidth,trim={375 275 300 200}, clip]{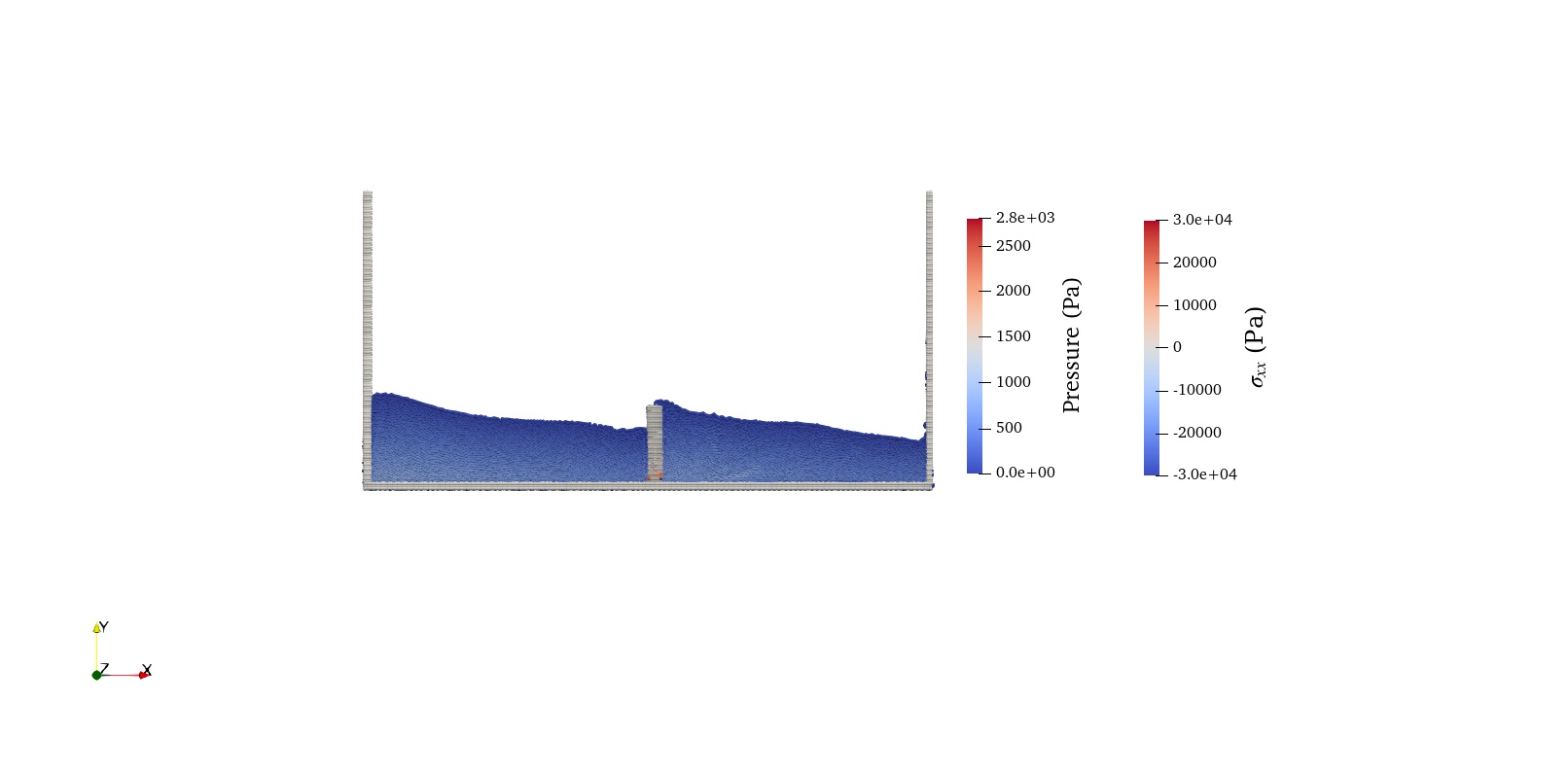}\caption{Time = 1.0 s}
\end{subfigure}
\caption{Pressure and stress distribution at different time steps for water impact on elastic obstacle}\label{dam_break_obstacle_contour}
\end{figure}

Fig. \ref{dam_break_obstacle_contour} comprehensively depicts various aspects of the scenario, showcasing the evolution of the water pressure, free-surface profile, and the deformation of the elastic obstacle at distinct time intervals. At the outset, the water flows freely, resembling a typical dam break scenario, with the flow front exhibiting low pressure. However, as the water collides with the elastic obstacle, a substantial surge in pressure becomes apparent, generating significant impact forces. This leads to a considerable deformation in the elastic obstacle, instigating notable changes in the dynamics of the flowing water. As the water progresses over the wall, the pressure gradually subsides, eventually reaching a state of hydrostatic equilibrium. Concurrently, the elastic wall rebounds from its deformed state. Notably, upon the initial impact of water on the tank's rigid wall, a localized high-pressure zone re-emerges. This method effectively captures the intricate interplay between the fluid and the deformable structure, resulting in a qualitative agreement between the numerical present simulations and the other observations from literature \cite{rafiee2009sph, zhan2019stabilized, rahimi2023sph}. The stress distribution in the deformable obstacle is also shown in Fig. \ref{dam_break_obstacle_contour}. Following the impact around $0.2$ s, the upstream face of the wall experiences tension, while the opposite face is subjected to compression. The maximum stress can be found near the fixed support, consistent with other literature findings. Also, there is no instability observed in the obstacle.

\begin{figure}[hbtp!] %trim={<left> <lower> <right> <upper>}
\centering
\begin{subfigure}[t]{0.75\textwidth}
\includegraphics[width=\textwidth]{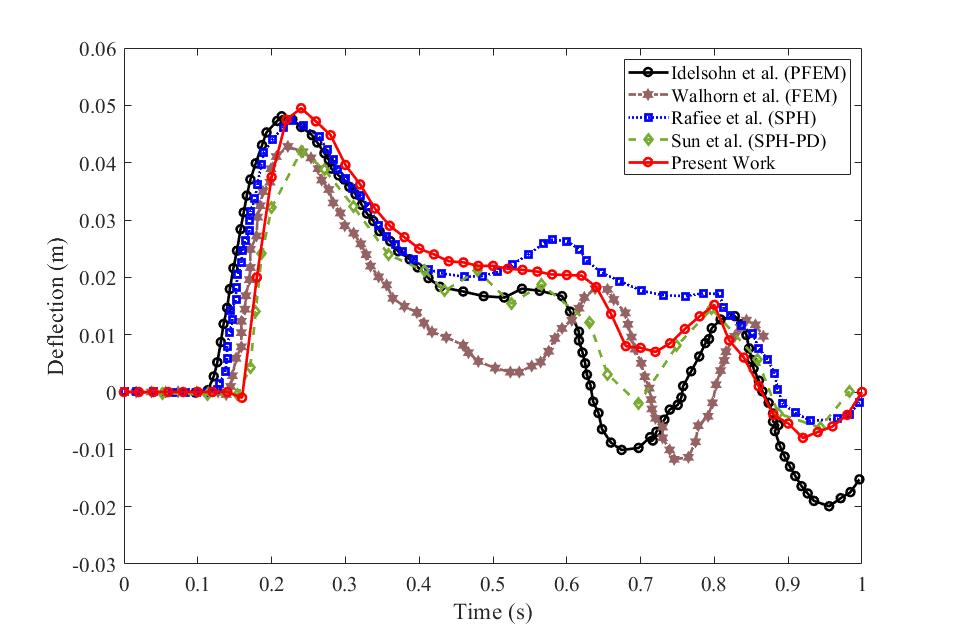}
\end{subfigure}
\caption{Comparison of time histories of the deflection of the free end of the elastic obstacle}\label{dam_break_obstacle_time_his}
\end{figure}

To further confirm the accuracy of the SPH–PD method, we analyzed the deflection process occurring at the upper-left corner of the elastic plate over time. In Fig. \ref{dam_break_obstacle_time_his}, we present the variations in horizontal displacement observed at the upper-left corner of the elastic wall. Furthermore, we include numerical results from previous literature \cite{rafiee2009sph, sun2020smoothed, walhorn2005fluid, idelsohn2008unified}, to enable a comparative assessment. It is evident from the results that the current approach effectively anticipates and replicates the overall response of the obstacle when subjected to hydrodynamic forces induced by the collapsing water column.

\subsection{Damage and fracture of an elastic obstacle due to water impact}

\begin{figure}[hbtp!] %trim={<left> <lower> <right> <upper>}
\centering
\begin{subfigure}[t]{0.75\textwidth}
\includegraphics[width=\textwidth,trim={100 70 100 100}, clip]{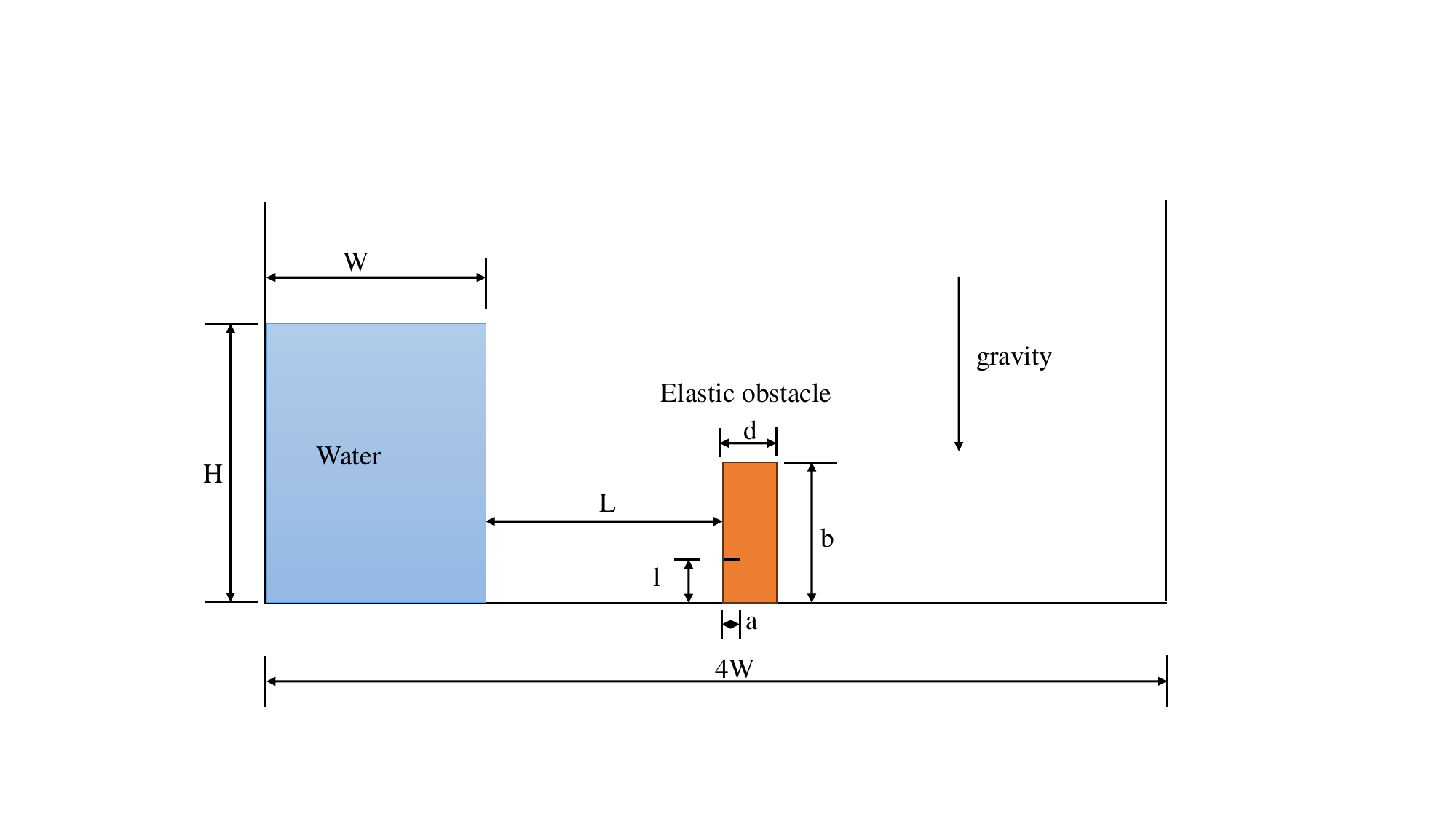}
\end{subfigure}
\caption{Setup for water impact on an elastic obstacle with an initial notch}\label{dam_break_obstacle_notch}
\end{figure}

We examine the interaction between a brittle obstacle and water. The specific geometric arrangement for this case is detailed in Fig. \ref{dam_break_obstacle_notch}. Here, the initial crack/ notch is made by deleting the particles of a row. The obstacle shown in the diagram has an initial crack measuring $a=0.008$ m in length, positioned and $l=0.025$ m above the ground. Simultaneously, water is in a state of descent due to the gravitational force. The other dimensions are as follows: $H=0.3$ m, $W=0.15$ m, $L=W$, $b=0.09$ m and $d=0.03$ m. The experiment begins with the sudden release of the water column, initiating its path towards a collision with the elastic obstacle. Initially, the density of the water and the flexible elastic barrier are established at $1000~Kg/m^3$ and $2500~Kg/m^3$, respectively. Additionally, the deformable elastic obstacle exhibits an elastic modulus denoted as $E$ with a value of $10^6~N/m^2$, and a Poisson ratio represented as $\nu$ with a value of $0$. The fracture strain, $\epsilon_f$, is set at $0.05$. Therefore, the interaction between a pair of particles $i$ and $j$ is stopped when the strain in the connecting pseudo-spring exceeds the value of $0.05$ ($f_{ij}=0~if~\epsilon_f>0.05$). The failure process is assumed to be permanent in this simulation. 

\begin{figure}[hbtp!] %trim={<left> <lower> <right> <upper>}
\centering
\begin{subfigure}[t]{0.4\textwidth}
\includegraphics[width=\textwidth,trim={375 150 500 200}, clip]{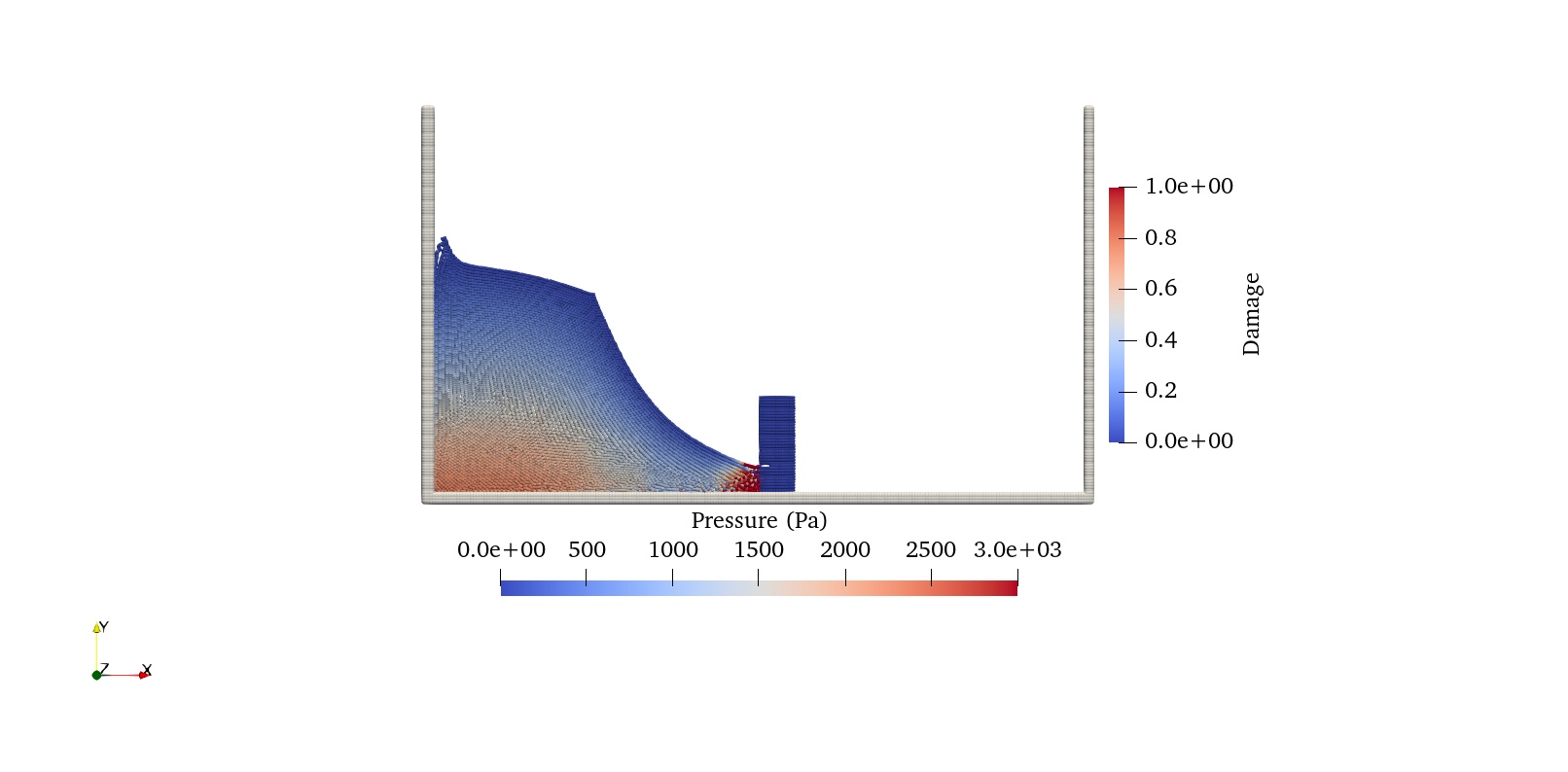}\caption{Time = 0.15 s}
\end{subfigure}
\begin{subfigure}[t]{0.4\textwidth}
\includegraphics[width=\textwidth,trim={550 250 300 150}, clip]{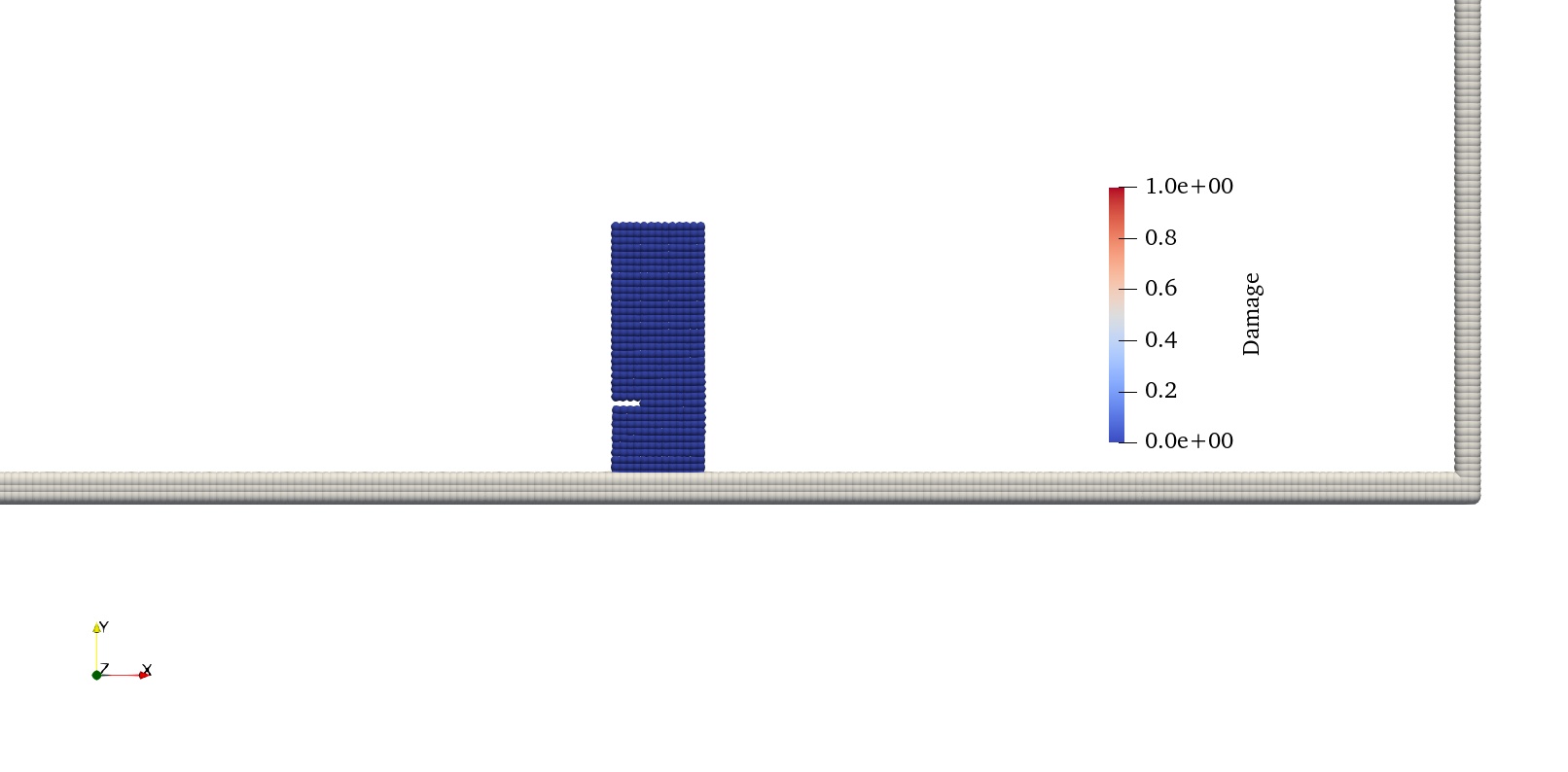}\caption{Time = 0.15 s}
\end{subfigure}
\begin{subfigure}[t]{0.4\textwidth}
\includegraphics[width=\textwidth,trim={375 150 500 200}, clip]{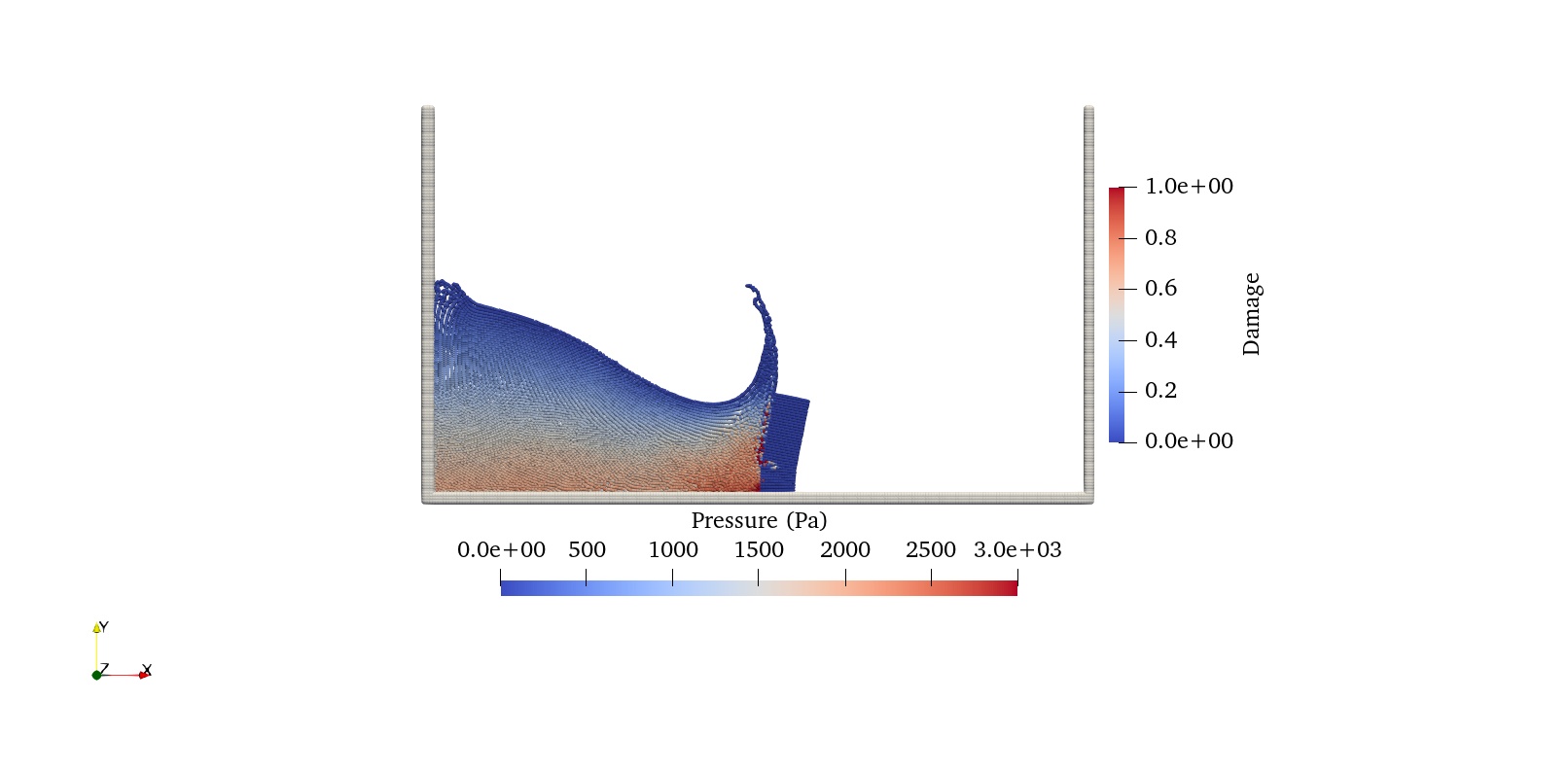}\caption{Time = 0.20 s}
\end{subfigure}
\begin{subfigure}[t]{0.4\textwidth}
\includegraphics[width=\textwidth,trim={550 250 300 150}, clip]{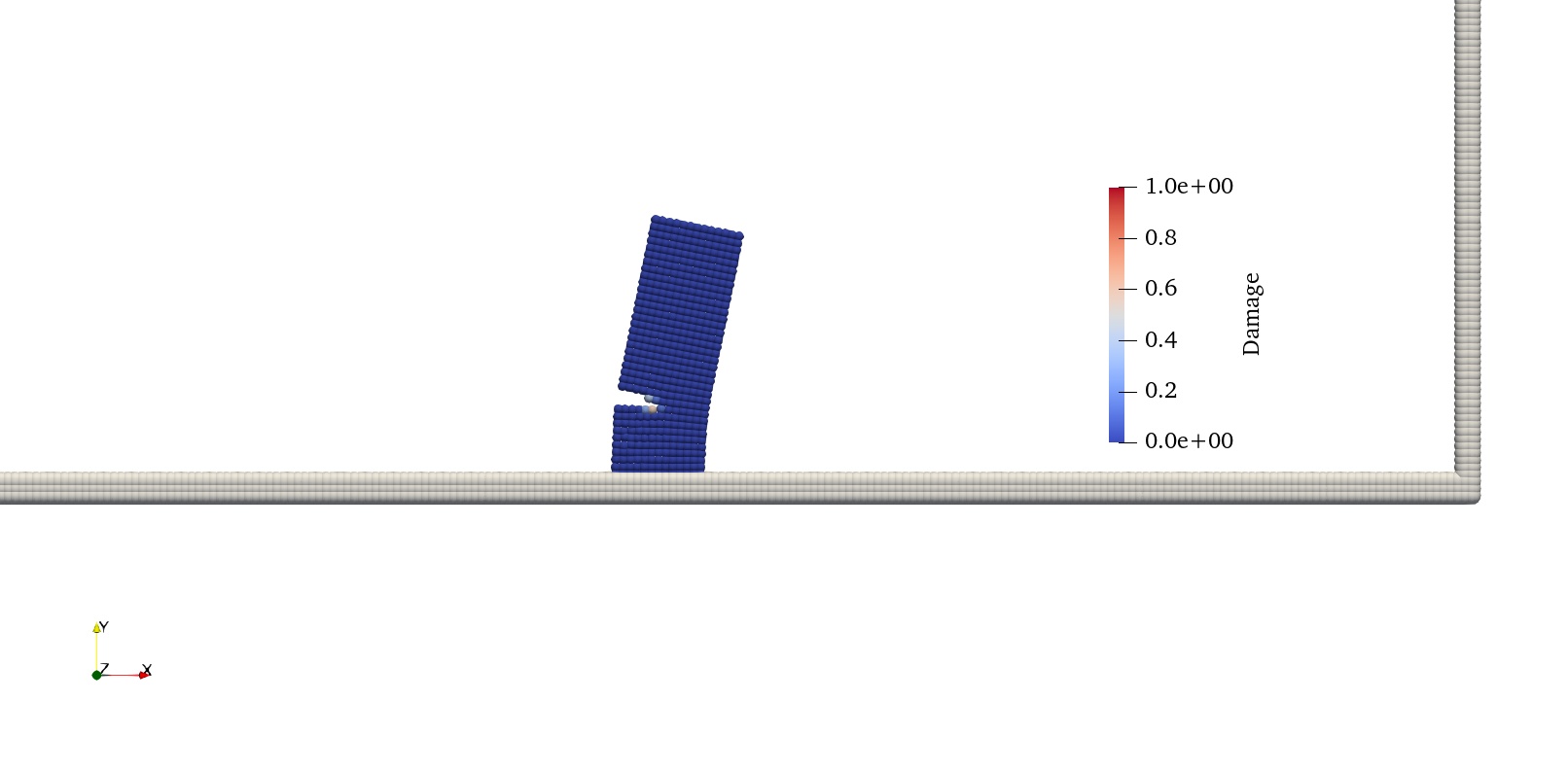}\caption{Time = 0.20 s}
\end{subfigure}
\begin{subfigure}[t]{0.4\textwidth}
\includegraphics[width=\textwidth,trim={375 150 500 200}, clip]{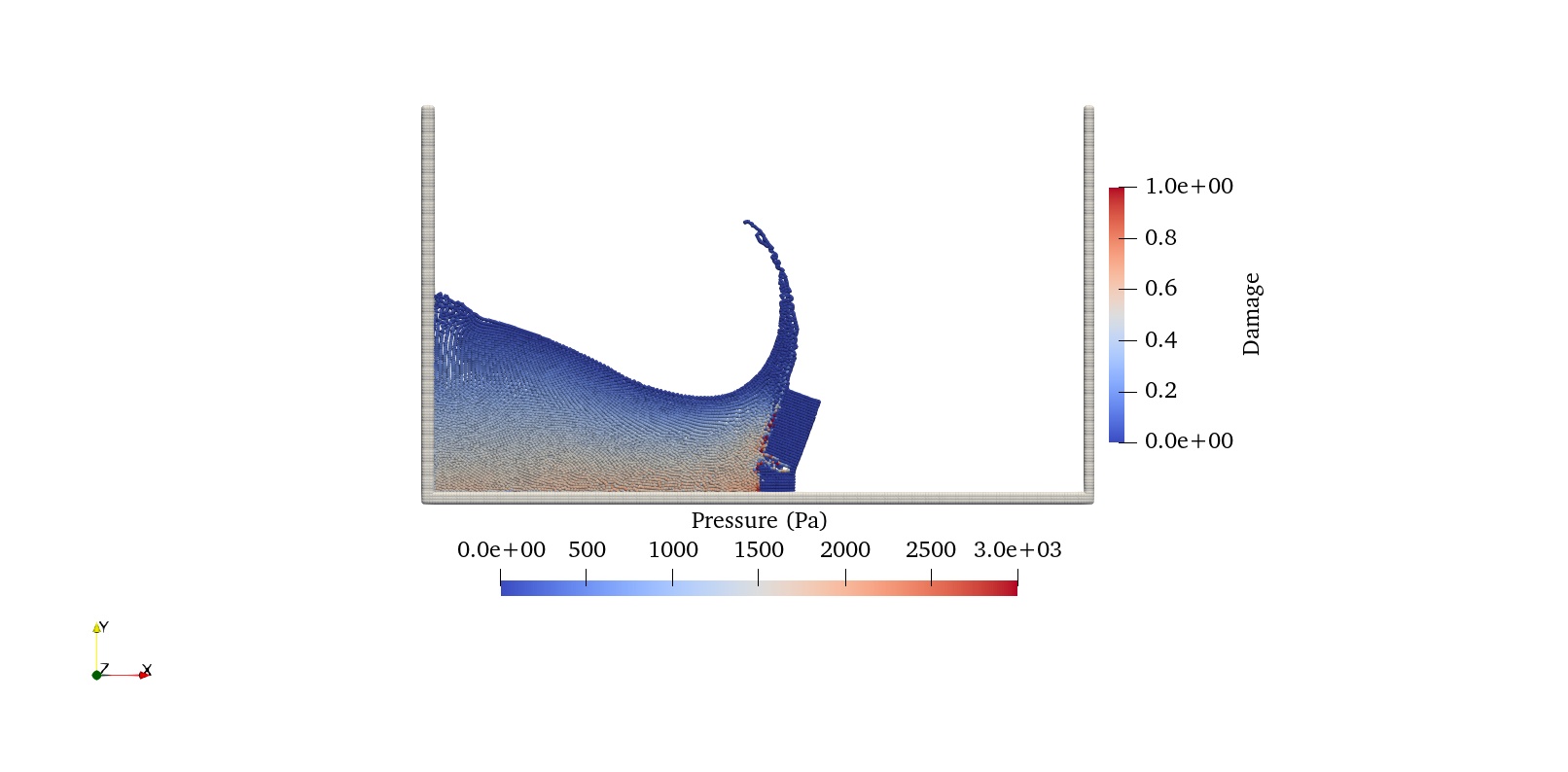}\caption{Time = 0.22 s}
\end{subfigure}
\begin{subfigure}[t]{0.4\textwidth}
\includegraphics[width=\textwidth,trim={550 250 300 150}, clip]{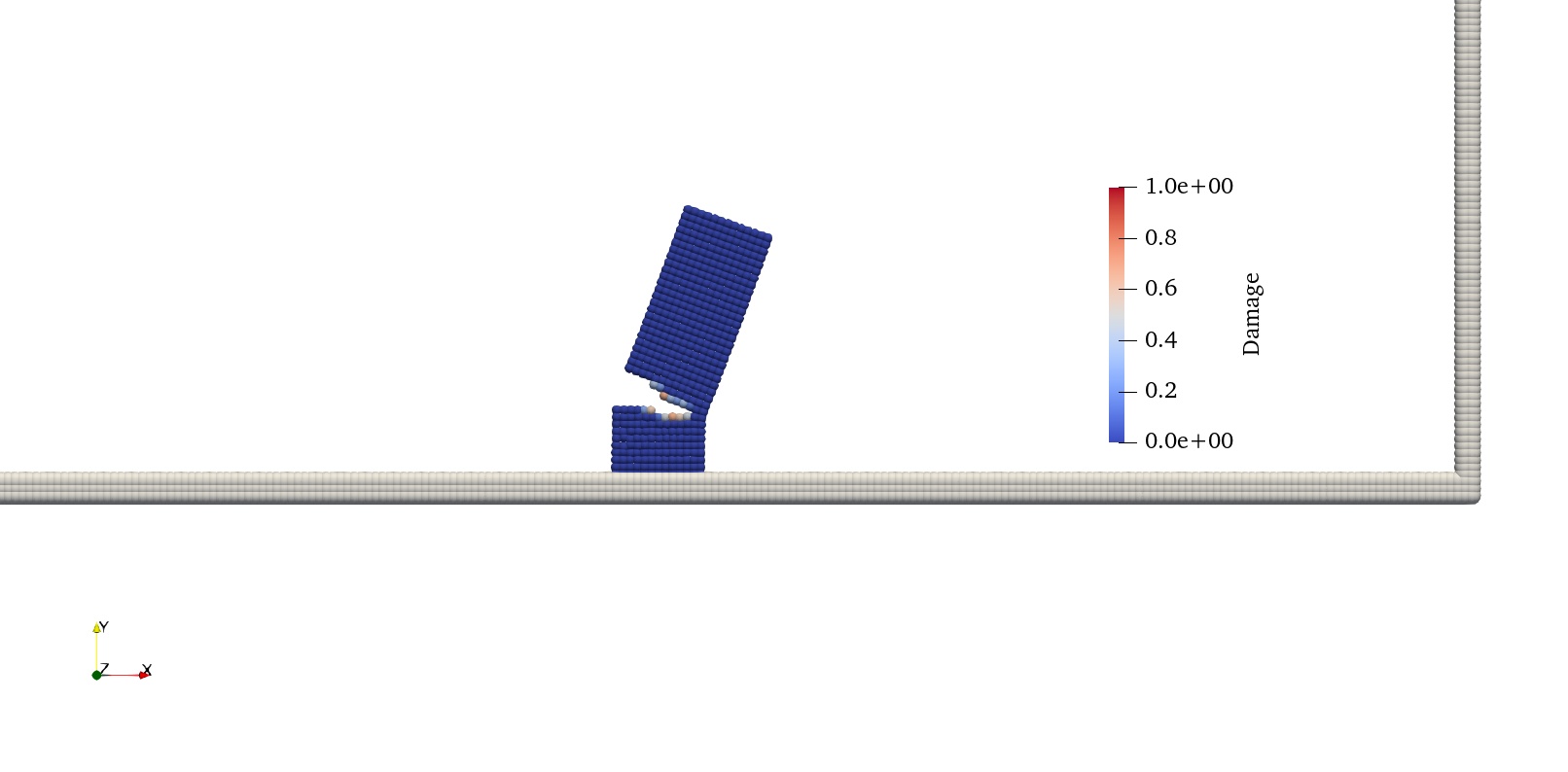}\caption{Time = 0.22 s}
\end{subfigure}
%\begin{subfigure}[t]{0.4\textwidth}
%\includegraphics[width=\textwidth,trim={375 150 500 200}, clip]{dam_break_fracture_0_24s.jpeg}\caption{Time = 0.24 s}
%\end{subfigure}
%\begin{subfigure}[t]{0.4\textwidth}
%\includegraphics[width=\textwidth,trim={550 250 300 150}, clip]{dam_break_fracture_0_24s_zoom_obs.jpeg}\caption{Time = 0.24 s}
%\end{subfigure}
\begin{subfigure}[t]{0.4\textwidth}
\includegraphics[width=\textwidth,trim={375 150 500 200}, clip]{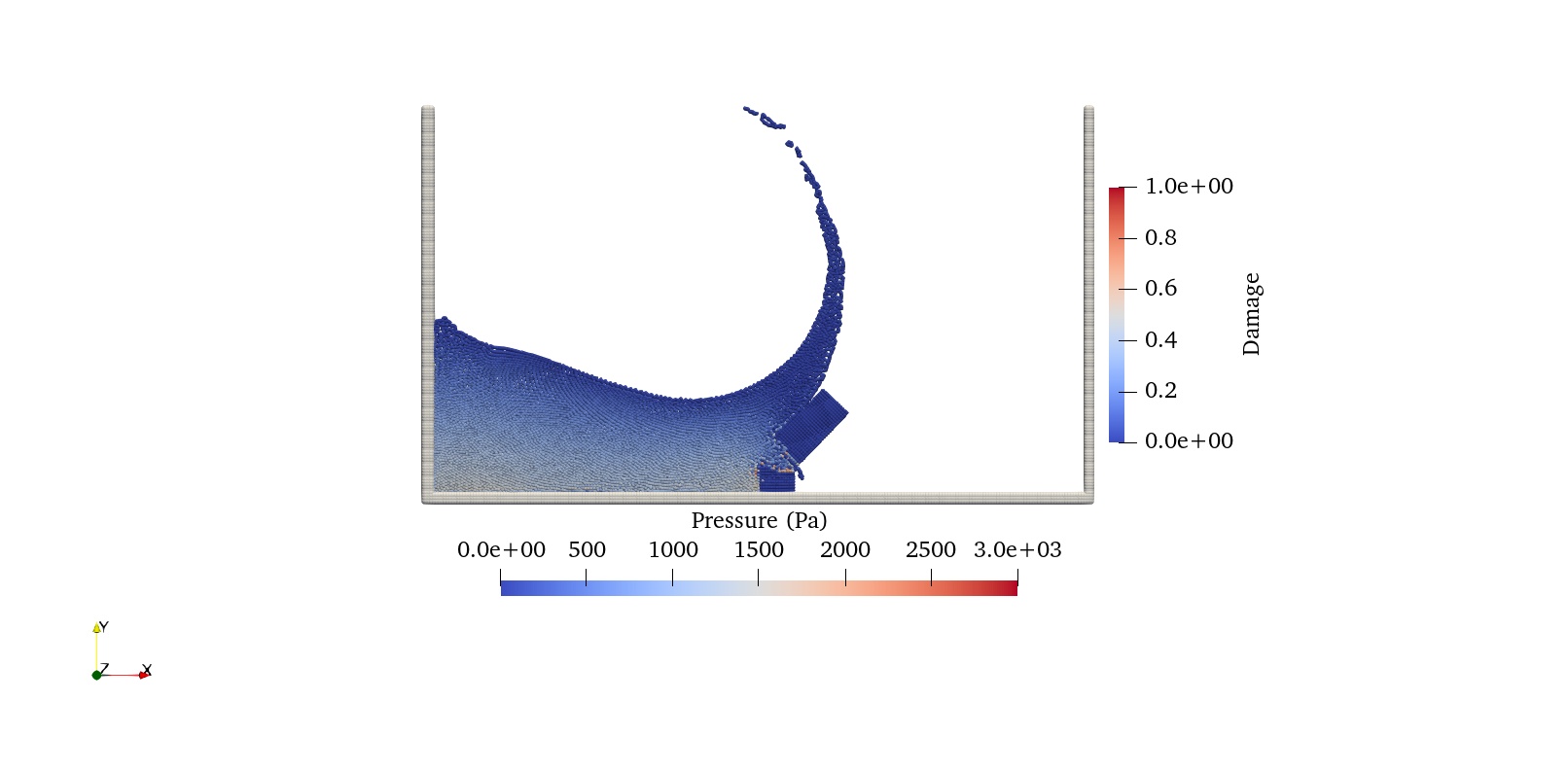}\caption{Time = 0.26 s}
\end{subfigure}
\begin{subfigure}[t]{0.4\textwidth}
\includegraphics[width=\textwidth,trim={550 250 300 150}, clip]{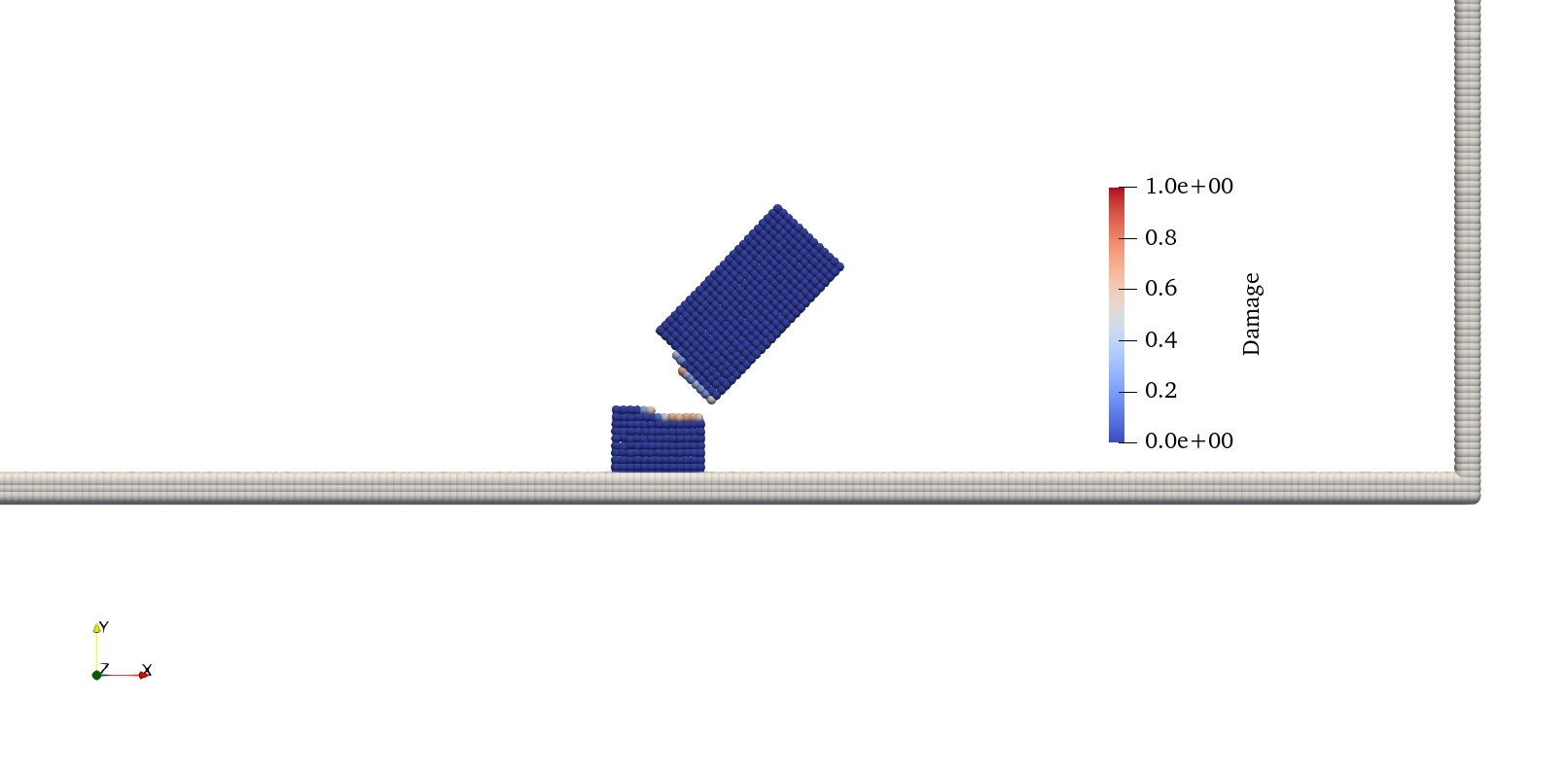}\caption{Time = 0.26 s}
\end{subfigure}
\caption{Pressure and Damage distribution at different time steps for water impact on an elastic obstacle ($\Delta p = 0.0025$ m)}\label{dam_break_fracture_contour}
\end{figure}

\begin{figure}[hbtp!] %trim={<left> <lower> <right> <upper>}
\centering
\begin{subfigure}[t]{0.4\textwidth}
\includegraphics[width=\textwidth,trim={375 150 500 200}, clip]{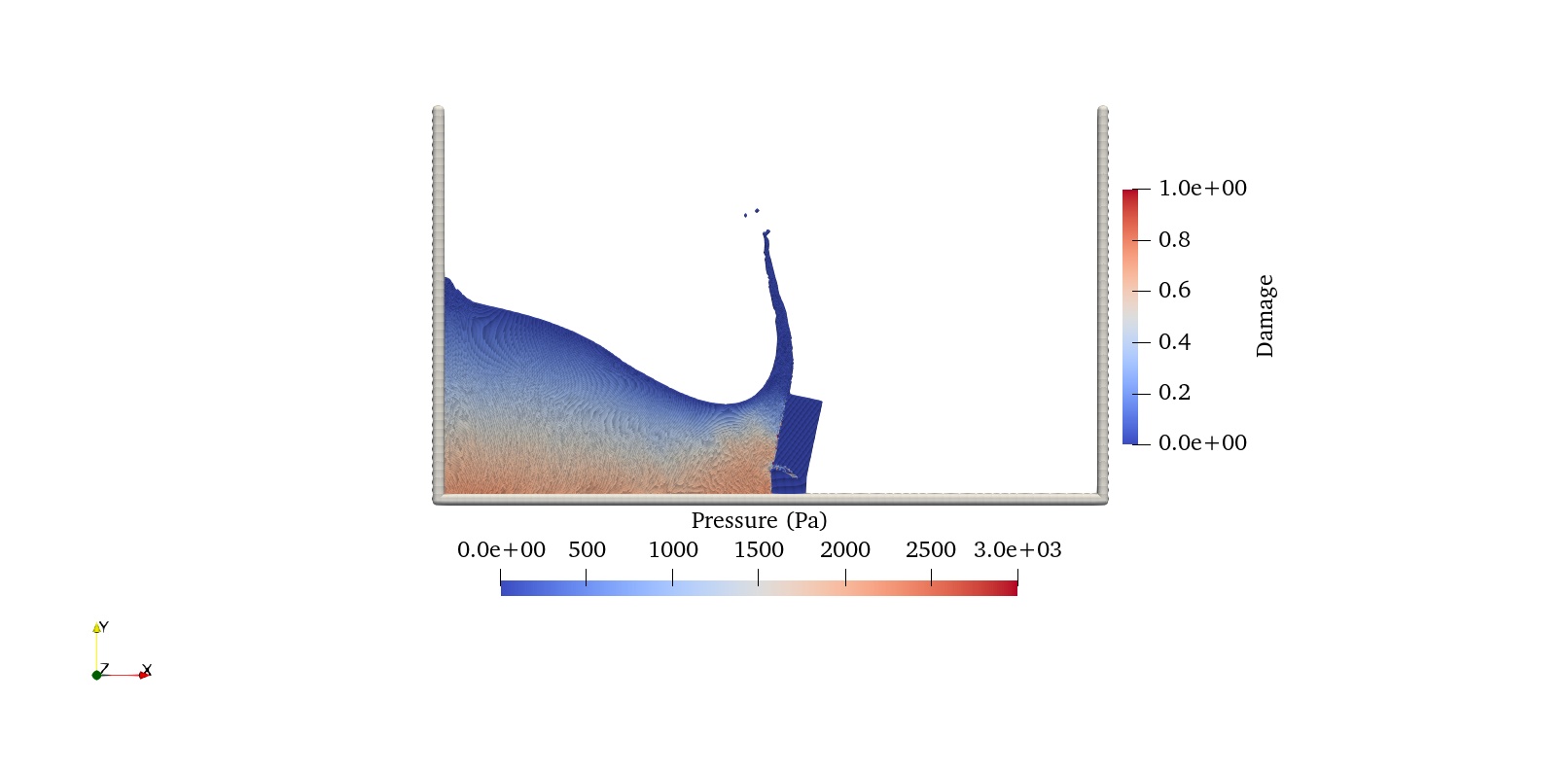}\caption{Time = 0.2 s}
\end{subfigure}
\begin{subfigure}[t]{0.4\textwidth}
\includegraphics[width=\textwidth,trim={550 250 300 150}, clip]{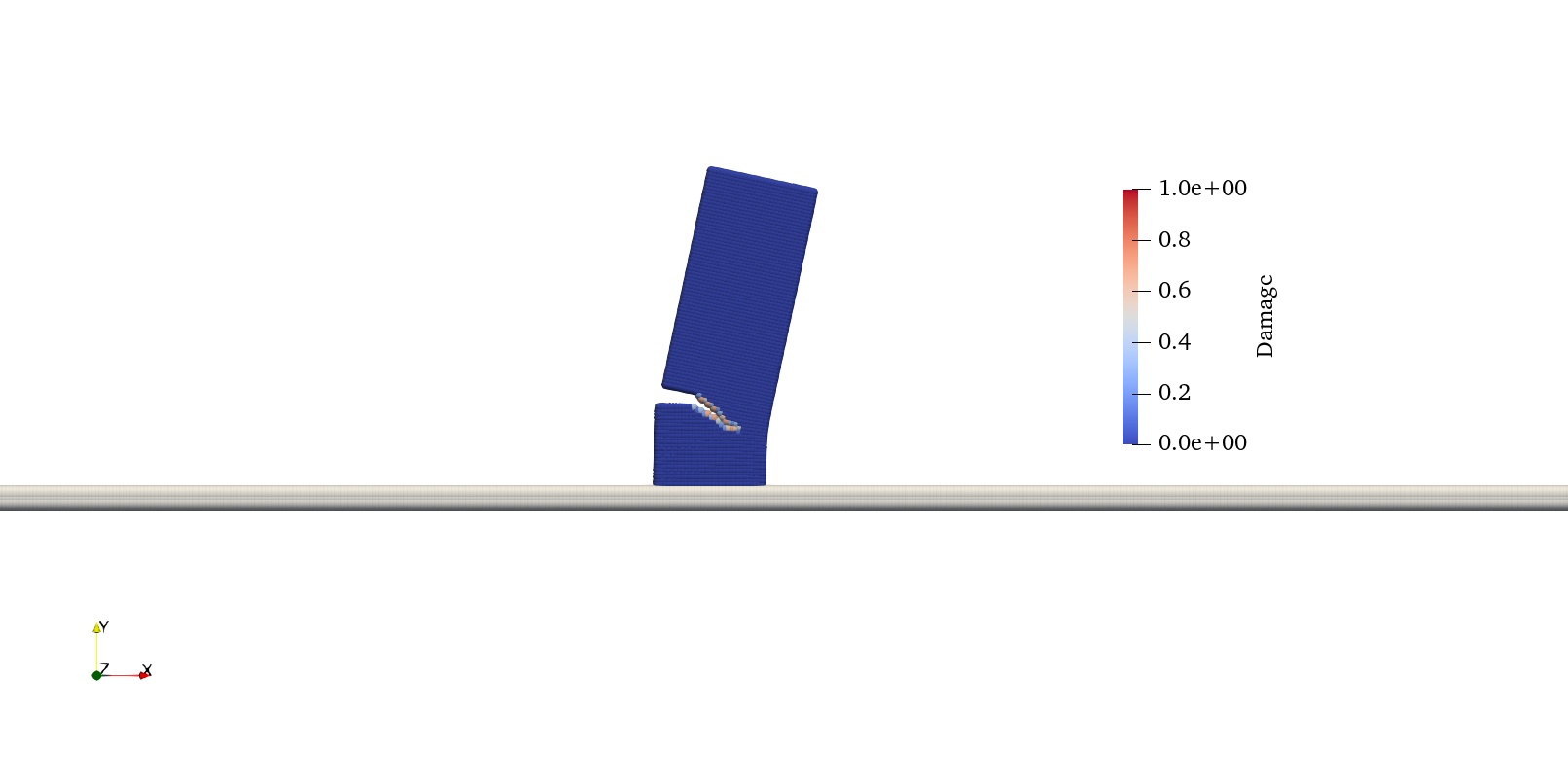}\caption{Time = 0.2 s}
\end{subfigure}
\begin{subfigure}[t]{0.4\textwidth}
\includegraphics[width=\textwidth,trim={375 150 500 150}, clip]{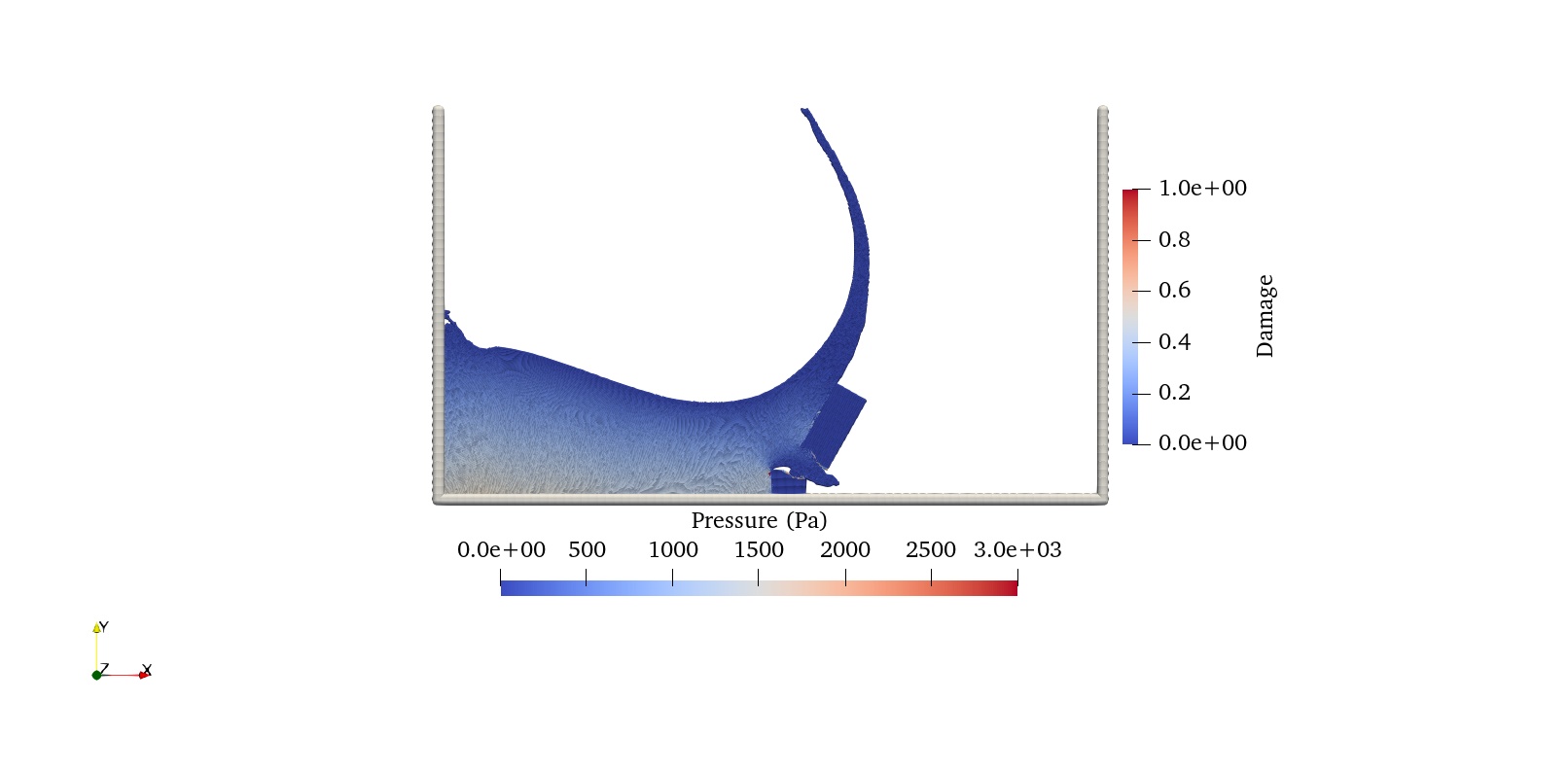}\caption{Time = 0.26 s}
\end{subfigure}
\begin{subfigure}[t]{0.4\textwidth}
\includegraphics[width=\textwidth,trim={550 250 300 125}, clip]{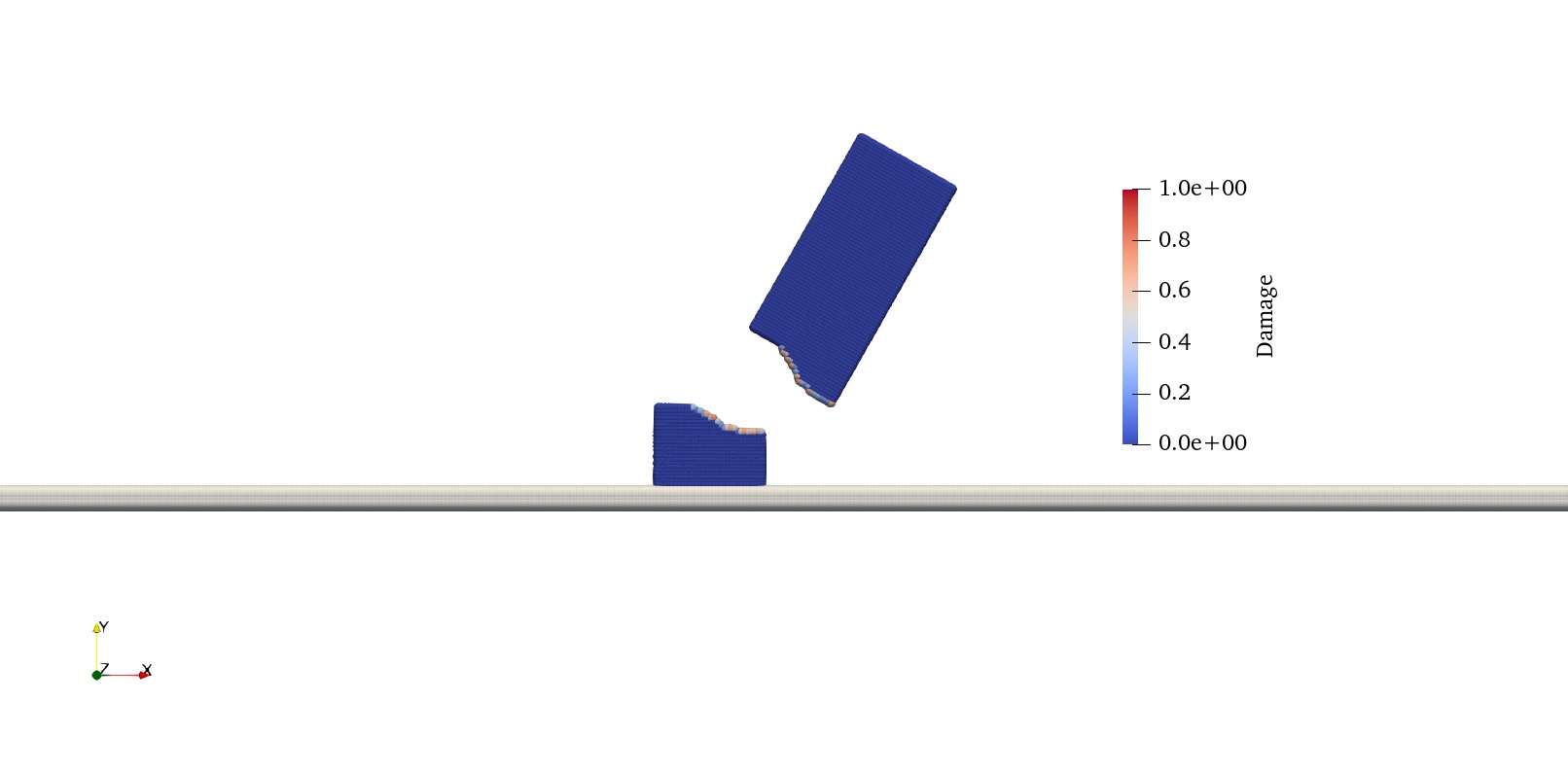}\caption{Time = 0.26 s}
\end{subfigure}
\caption{Pressure and Damage distribution at different time steps for water impact on an elastic obstacle ($\Delta p = 0.001$ m)}\label{dam_break_fracture_contour_fine}
\end{figure}

Fig. \ref{dam_break_fracture_contour} illustrates the progressive changes in the water's free surface, pressure contour patterns, and crack propagation throughout the simulation. As observed, when the fluid initially interacts with the obstacle, there is a significant surge in pressure within the FSI zone. Subsequently, due to this heightened pressure, the obstacle deforms, and the strains in the particles increase. After a specific duration ($t \approx 17.5$ s), the accumulated strain surpasses the material's fracture strain threshold $\epsilon_f = 0.05$, leading to the initiation of crack propagation of the obstacle and its eventual detachment from the weak area, i.e., the preexisting crack tip. It's worth noting that following the complete detachment of the upper section of the obstacle, the lower part exerts a jet effect on the fluid, causing the water to follow a predefined path. This phenomenon is depicted in the final snapshots of Fig. \ref{dam_break_fracture_contour}. Similar observations were made in \cite{rahimi2023sph}. The overall process of Fluid-Structure Interaction (FSI) and fracture is consistent when a fine discretization, i.e., inter-particle spacing ($\Delta p$), is used. It may be observed from Fig. \ref{dam_break_fracture_contour_fine} that fine resolution yields a better representation of crack initiation, propagation and fracturing process. 

\begin{figure}[hbtp!] %trim={<left> <lower> <right> <upper>}
\centering
\begin{subfigure}[t]{0.49\textwidth}
\includegraphics[width=\textwidth,trim={375 150 500 200}, clip]{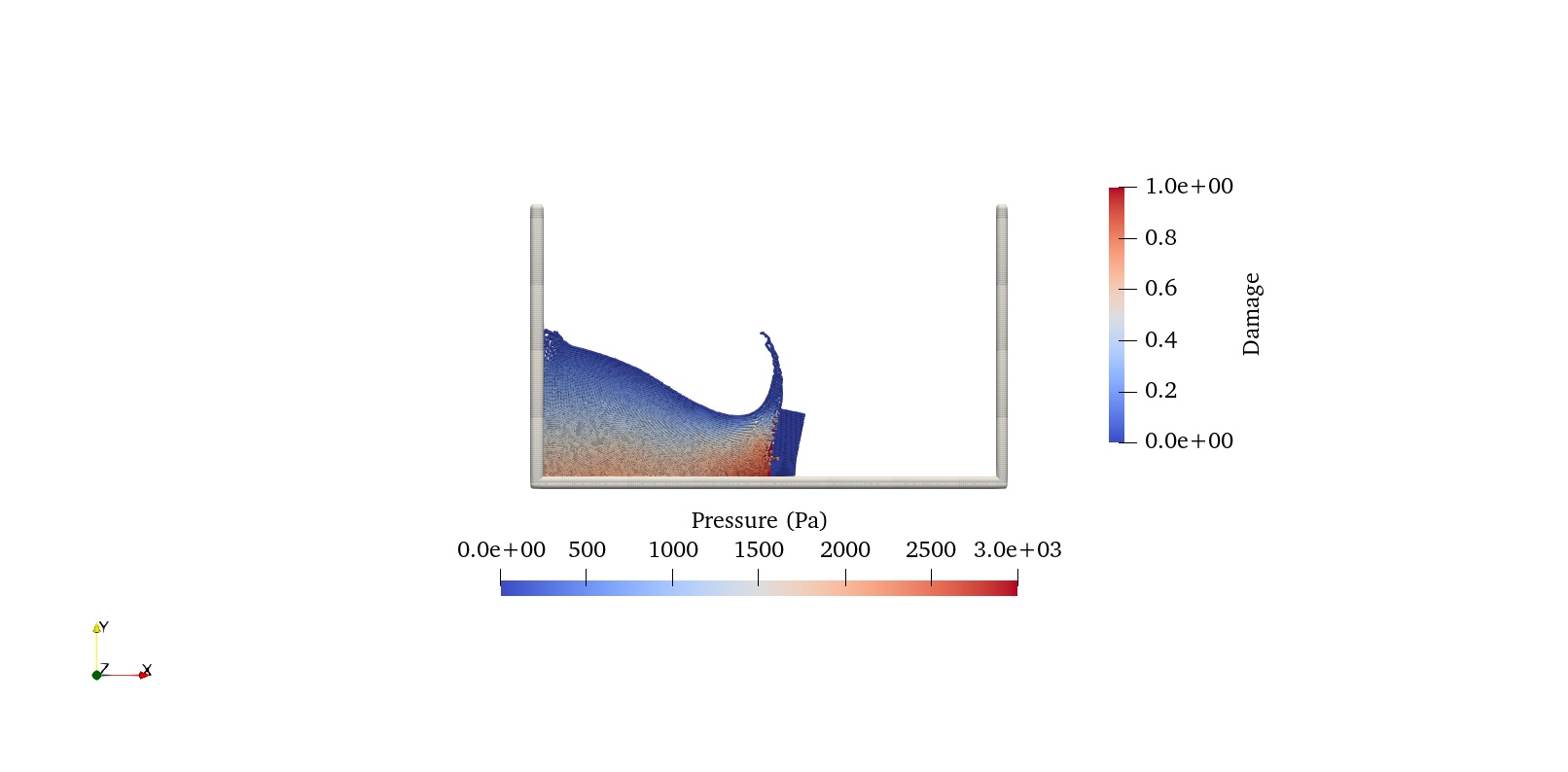}\caption{Time = 0.2 s}
\end{subfigure}
\begin{subfigure}[t]{0.49\textwidth}
\includegraphics[width=\textwidth,trim={375 150 500 200}, clip]{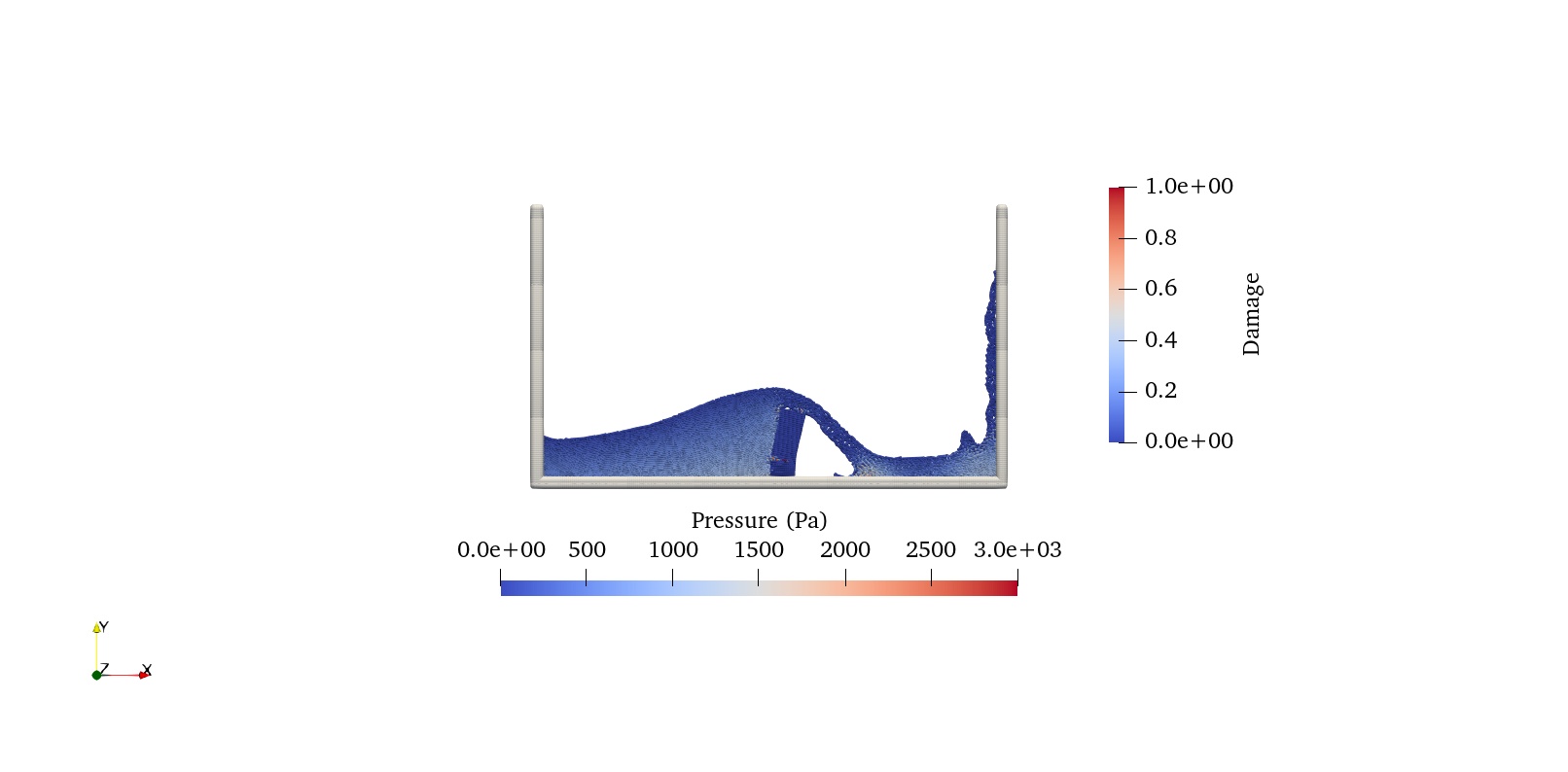}\caption{Time = 0.6 s}
\end{subfigure}
\caption{Pressure and Damage distribution at different time steps for water impact on an elastic obstacle without considering damage and fracturing ($\Delta p = 0.0025$ m)}\label{dam_break_fracture_contour_no_dam}
\end{figure}

In order to highlight the efficacy of the pseudo-spring analogy in modelling material damage and subsequent cracking, we perform a simulation of the same set-up without the failure strain (i.e., even if the strains in the pseudo-springs are greater than $\epsilon_f$, the interaction coefficient is kept same $f_{ij}=1$). It can be seen from Fig. \ref{dam_break_fracture_contour_no_dam} that the crack does not initiate, and the elastic obstacle remains undamaged, i.e., does not suffer any failure. 

\section{Conclusion}\label{conclu}
A computational framework for modelling large deformation and material damage and failure is proposed for fluid-structure interaction problems. In the integrated numerical approach, we utilize a two-pronged strategy. Firstly, the fluid phase is simulated by employing the WCSPH method, which includes a density diffusion term to enhance accuracy. The interaction between the fluid phase and the rigid walls is modelled through specialized boundary particles designed to extrapolate relevant variables. Secondly, for the solid phase, we implement a pseudo-spring analogy in SPH, where the immediate neighbour particles are used for approximation. The pseudo-springs help in modelling the material damage and subsequent crack propagation without requiring any computationally intensive processes such as visibility criteria, particle splitting, etc. The interaction between the moving fluid phase and the deformable solid structure/ obstacle is modelled by a soft repulsive particle contact model. This approach results in the establishment of a cohesive framework for effectively managing rigid wall boundaries and fluid-structure interactions.

The numerical results obtained in our study have been subjected to thorough comparisons with analytical solutions, experimental data, and other existing numerical findings from the literature. Our findings demonstrate the capability of accurately modelling free surface flow and dynamic elastic problems without encountering instability. The validation of the approach was carried out through the examination of different FSI scenarios involving deformable structures. Our numerical outcomes exhibit good agreement with existing experimental, numerical and analytical data from the literature, reaffirming the reliability of our method. We have also demonstrated in the last numerical example that the proposed framework is capable of modelling material damage and subsequent fracture under extreme hydrodynamic events. While our proposed method has shown promising accuracy in preliminary assessments, it is important to acknowledge the limited availability of experiments in the existing literature that specifically address FSI problems involving deformable structures exhibiting material damage and fracture. To ensure a comprehensive evaluation of the precision and reliability of our approach, further investigations are necessary. In light of this, we plan to conduct dedicated laboratory experiments as part of our validation process in future works. These experiments will provide valuable real-world data and insights that can help us refine and enhance the accuracy of our method in modelling FSI scenarios involving deformable structures experiencing material damage and fracture. However, the proposed framework has shown great stability and efficiency (when compared with the existing numerical schemes) and holds the potential to become a widely employed method for modelling finite deformation and material damage and failure in FSI problems.

\section{Acknowledgments}
The author acknowledges the computational support provided as a part of the IIT Delhi NFS grant, on which the simulations have been run.

\FloatBarrier

%\bibliographystyle{cas-model2-names}
\bibliographystyle{elsarticle-num}

\begin{thebibliography}{10}
\expandafter\ifx\csname url\endcsname\relax
  \def\url#1{\texttt{#1}}\fi
\expandafter\ifx\csname urlprefix\endcsname\relax\def\urlprefix{URL }\fi
\expandafter\ifx\csname href\endcsname\relax
  \def\href#1#2{#2} \def\path#1{#1}\fi

\bibitem{hu2009two}
C.~Hu, M.~Kashiwagi, Two-dimensional numerical simulation and experiment on
  strongly nonlinear wave--body interactions, Journal of marine science and
  technology 14 (2009) 200--213.

\bibitem{slone2002dynamic}
A.~Slone, K.~Pericleous, C.~Bailey, M.~Cross, Dynamic fluid--structure
  interaction using finite volume unstructured mesh procedures, Computers \&
  structures 80~(5-6) (2002) 371--390.

\bibitem{heil2004efficient}
M.~Heil, An efficient solver for the fully coupled solution of
  large-displacement fluid--structure interaction problems, Computer Methods in
  Applied Mechanics and Engineering 193~(1-2) (2004) 1--23.

\bibitem{belytschko1996difficulty}
T.~Belytschko, On difficulty levels in non linear finite element analysis of
  solids, Iacm Expressions 2 (1996) 6--8.

\bibitem{fries2010extended}
T.-P. Fries, T.~Belytschko, The extended/generalized finite element method: an
  overview of the method and its applications, International journal for
  numerical methods in engineering 84~(3) (2010) 253--304.

\bibitem{gingold1977smoothed}
R.~A. Gingold, J.~J. Monaghan, Smoothed particle hydrodynamics: theory and
  application to non-spherical stars, Monthly notices of the royal astronomical
  society 181~(3) (1977) 375--389.

\bibitem{lucy1977numerical}
L.~B. Lucy, A numerical approach to the testing of the fission hypothesis,
  Astronomical Journal, vol. 82, Dec. 1977, p. 1013-1024. 82 (1977) 1013--1024.

\bibitem{libersky2005smooth}
L.~D. Libersky, A.~G. Petschek, Smooth particle hydrodynamics with strength of
  materials, in: Advances in the Free-Lagrange Method Including Contributions
  on Adaptive Gridding and the Smooth Particle Hydrodynamics Method:
  Proceedings of the Next Free-Lagrange Conference Held at Jackson Lake Lodge,
  Moran, WY, USA 3--7 June 1990, Springer, 2005, pp. 248--257.

\bibitem{liu2010smoothed}
M.~Liu, G.~Liu, Smoothed particle hydrodynamics (sph): an overview and recent
  developments, Archives of computational methods in engineering 17 (2010)
  25--76.

\bibitem{monaghan1994simulating}
J.~J. Monaghan, Simulating free surface flows with sph, Journal of
  computational physics 110~(2) (1994) 399--406.

\bibitem{adami2012generalized}
S.~Adami, X.~Y. Hu, N.~A. Adams, A generalized wall boundary condition for
  smoothed particle hydrodynamics, Journal of Computational Physics 231~(21)
  (2012) 7057--7075.

\bibitem{fraga2019implementation}
C.~A.~D. Fraga~Filho, C.~Peng, M.~R. Ibne~Islam, C.~McCabe, S.~Baig, G.~V.
  Durga~Prasad, Implementation of three-dimensional physical reflective
  boundary conditions in mesh-free particle methods for continuum fluid
  dynamics: Validation tests and case studies, Physics of Fluids 31~(10)
  (2019).

\bibitem{bui2008lagrangian}
H.~H. Bui, R.~Fukagawa, K.~Sako, S.~Ohno, Lagrangian meshfree particles method
  (sph) for large deformation and failure flows of geomaterial using
  elastic--plastic soil constitutive model, International journal for numerical
  and analytical methods in geomechanics 32~(12) (2008) 1537--1570.

\bibitem{chen2011practical}
J.~Chen, Y.~Xin, Practical method of conical cam outline expansion, Chinese
  Journal of Mechanical Engineering-English Edition 24~(1) (2011) 127.

\bibitem{peng2015sph}
C.~Peng, W.~Wu, H.-s. Yu, C.~Wang, A sph approach for large deformation
  analysis with hypoplastic constitutive model, Acta Geotechnica 10 (2015)
  703--717.

\bibitem{peng2016unified}
C.~Peng, X.~Guo, W.~Wu, Y.~Wang, Unified modelling of granular media with
  smoothed particle hydrodynamics, Acta Geotechnica 11 (2016) 1231--1247.

\bibitem{liu2003smoothed}
M.~Liu, G.~Liu, K.~Lam, Z.~Zong, Smoothed particle hydrodynamics for numerical
  simulation of underwater explosion, Computational mechanics 30 (2003)
  106--118.

\bibitem{islam2017computational}
M.~R.~I. Islam, S.~Chakraborty, A.~Shaw, S.~Reid, A computational model for
  failure of ductile material under impact, International Journal of Impact
  Engineering 108 (2017) 334--347.

\bibitem{islam2020pseudo}
M.~R.~I. Islam, A.~Shaw, Pseudo-spring sph simulations on the perforation of
  metal targets with different damage models, Engineering Analysis with
  Boundary Elements 111 (2020) 55--77.

\bibitem{islam2023comparison}
M.~R.~I. Islam, C.~Peng, P.~K. Patra, A comparison of numerical stability for
  esph and tlsph for dynamic brittle fracture, Theoretical and Applied Fracture
  Mechanics (2023) 104052.

\bibitem{antoci2007numerical}
C.~Antoci, M.~Gallati, S.~Sibilla, Numerical simulation of fluid--structure
  interaction by sph, Computers \& structures 85~(11-14) (2007) 879--890.

\bibitem{rafiee2009sph}
A.~Rafiee, K.~P. Thiagarajan, An sph projection method for simulating
  fluid-hypoelastic structure interaction, Computer Methods in Applied
  Mechanics and Engineering 198~(33-36) (2009) 2785--2795.

\bibitem{salehizadeh2022coupled}
A.~Salehizadeh, A.~Shafiei, A coupled isph-tlsph method for simulating
  fluid-elastic structure interaction problems, Journal of Marine Science and
  Application 21~(1) (2022) 15--36.

\bibitem{marrone2015prediction}
S.~Marrone, A.~Colagrossi, A.~Di~Mascio, D.~Le~Touz{\'e}, Prediction of energy
  losses in water impacts using incompressible and weakly compressible models,
  Journal of Fluids and Structures 54 (2015) 802--822.

\bibitem{meringolo2017filtering}
D.~Meringolo, A.~Colagrossi, S.~Marrone, F.~Aristodemo, On the filtering of
  acoustic components in weakly-compressible sph simulations, Journal of Fluids
  and Structures 70 (2017) 1--23.

\bibitem{pahar2016modeling}
G.~Pahar, A.~Dhar, Modeling free-surface flow in porous media with modified
  incompressible sph, Engineering Analysis with Boundary Elements 68 (2016)
  75--85.

\bibitem{pahar2017coupled}
G.~Pahar, A.~Dhar, Coupled incompressible smoothed particle hydrodynamics model
  for continuum-based modelling sediment transport, Advances in water resources
  102 (2017) 84--98.

\bibitem{trask2015highly}
N.~Trask, K.~Kim, A.~Tartakovsky, M.~Perego, M.~L. Parks, A highly-scalable
  implicit sph code for simulating single-and multi-phase flows in
  geometrically complex bounded domains., Tech. rep., Sandia National
  Lab.(SNL-NM), Albuquerque, NM (United States) (2015).

\bibitem{guoa2015developing}
X.~Guoa, B.~D. Rogersb, Developing highly scalable 3-d incompressible sph,
  ARCHER Community, ISPH embedded CSE Report (2015).

\bibitem{guo2018new}
X.~Guo, B.~D. Rogers, S.~Lind, P.~K. Stansby, New massively parallel scheme for
  incompressible smoothed particle hydrodynamics (isph) for highly nonlinear
  and distorted flow, Computer Physics Communications 233 (2018) 16--28.

\bibitem{molteni2009simple}
D.~Molteni, A.~Colagrossi, A simple procedure to improve the pressure
  evaluation in hydrodynamic context using the sph, Computer Physics
  Communications 180~(6) (2009) 861--872.

\bibitem{marrone2011delta}
S.~Marrone, M.~Antuono, A.~Colagrossi, G.~Colicchio, D.~Le~Touz{\'e},
  G.~Graziani, $\delta$-sph model for simulating violent impact flows, Computer
  Methods in Applied Mechanics and Engineering 200~(13-16) (2011) 1526--1542.

\bibitem{ferrari2009new}
A.~Ferrari, M.~Dumbser, E.~F. Toro, A.~Armanini, A new 3d parallel sph scheme
  for free surface flows, Computers \& Fluids 38~(6) (2009) 1203--1217.

\bibitem{swegle1995smoothed}
J.~W. Swegle, D.~L. Hicks, S.~W. Attaway, Smoothed particle hydrodynamics
  stability analysis, Journal of computational physics 116~(1) (1995) 123--134.

\bibitem{monaghan2000sph}
J.~J. Monaghan, Sph without a tensile instability, Journal of computational
  physics 159~(2) (2000) 290--311.

\bibitem{gray2001sph}
J.~P. Gray, J.~J. Monaghan, R.~Swift, Sph elastic dynamics, Computer methods in
  applied mechanics and engineering 190~(49-50) (2001) 6641--6662.

\bibitem{vignjevic2006sph}
R.~Vignjevic, J.~R. Reveles, J.~Campbell, Sph in a total lagrangian formalism,
  CMC-Tech Science Press- 4~(3) (2006) 181.

\bibitem{belytschko2000unified}
T.~Belytschko, Y.~Guo, W.~Kam~Liu, S.~Ping~Xiao, A unified stability analysis
  of meshless particle methods, International Journal for Numerical Methods in
  Engineering 48~(9) (2000) 1359--1400.

\bibitem{vidal2007stabilized}
Y.~Vidal, J.~Bonet, A.~Huerta, Stabilized updated lagrangian corrected sph for
  explicit dynamic problems, International journal for numerical methods in
  engineering 69~(13) (2007) 2687--2710.

\bibitem{islam2022equivalence}
M.~R.~I. Islam, K.~V. Ganesh, P.~K. Patra, On the equivalence of eulerian
  smoothed particle hydrodynamics, total lagrangian smoothed particle
  hydrodynamics and molecular dynamics simulations for solids, Computer Methods
  in Applied Mechanics and Engineering 391 (2022) 114591.

\bibitem{islam2022large}
M.~R.~I. Islam, W.~Zhang, C.~Peng, Large deformation analysis of geomaterials
  using stabilized total lagrangian smoothed particle hydrodynamics,
  Engineering Analysis with Boundary Elements 136 (2022) 252--265.

\bibitem{rahimi2023sph}
M.~N. Rahimi, G.~Moutsanidis, An sph-based fsi framework for phase-field
  modeling of brittle fracture under extreme hydrodynamic events, Engineering
  with Computers (2023) 1--35.

\bibitem{rahimi2022modeling}
M.~N. Rahimi, G.~Moutsanidis, Modeling dynamic brittle fracture in functionally
  graded materials using hyperbolic phase field and smoothed particle
  hydrodynamics, Computer Methods in Applied Mechanics and Engineering 401
  (2022) 115642.

\bibitem{zhao2023simulation}
Y.~Zhao, Z.~Zhou, J.~Bi, C.~Wang, Simulation of brittle fractures using
  energy-bond-based smoothed particle hydrodynamics, International Journal of
  Mechanical Sciences 248 (2023) 108236.

\bibitem{zhou2015novel}
X.~Zhou, Y.~Zhao, Q.~Qian, A novel meshless numerical method for modeling
  progressive failure processes of slopes, Engineering Geology 192 (2015)
  139--153.

\bibitem{zhai2016effects}
S.~Zhai, X.~Zhou, J.~Bi, N.~Xiao, The effects of joints on rock fragmentation
  by tbm cutters using general particle dynamics, Tunnelling and Underground
  Space Technology 57 (2016) 162--172.

\bibitem{zhou20173d}
X.~Zhou, X.~Zhang, The 3d numerical simulation of damage localization of rocks
  using general particle dynamics, Engineering Geology 224 (2017) 29--42.

\bibitem{chakraborty2013pseudo}
S.~Chakraborty, A.~Shaw, A pseudo-spring based fracture model for sph
  simulation of impact dynamics, International Journal of Impact Engineering 58
  (2013) 84--95.

\bibitem{wendland1995piecewise}
H.~Wendland, Piecewise polynomial, positive definite and compactly supported
  radial functions of minimal degree, Advances in computational Mathematics 4
  (1995) 389--396.

\bibitem{monaghan1983shock}
J.~J. Monaghan, R.~A. Gingold, Shock simulation by the particle method sph,
  Journal of computational physics 52~(2) (1983) 374--389.

\bibitem{eliezer1986introduction}
S.~Eliezer, A.~K. Ghatak, H.~Hora, E.~Teller, An introduction to equations of
  state: theory and applications, Cambridge University Press Cambridge, 1986.

\bibitem{chen1999corrective}
J.~Chen, J.~Beraun, T.~Carney, A corrective smoothed particle method for
  boundary value problems in heat conduction, International Journal for
  Numerical Methods in Engineering 46~(2) (1999) 231--252.

\bibitem{zhang2018meshfree}
Z.~Zhang, K.~Walayat, J.~Chang, M.~Liu, Meshfree modeling of a fluid-particle
  two-phase flow with an improved sph method, International Journal for
  Numerical Methods in Engineering 116~(8) (2018) 530--569.

\bibitem{liu2019smoothed}
M.~Liu, Z.~Zhang, Smoothed particle hydrodynamics (sph) for modeling
  fluid-structure interactions, Science China Physics, Mechanics \& Astronomy
  62 (2019) 1--38.

\bibitem{martin1952part}
J.~Martin, W.~Moyce, Part iv. an experimental study of the collapse of liquid
  columns on a rigid horizontal plane, Philosophical Transactions of the Royal
  Society of London Series A 244~(882) (1952) 312--324.

\bibitem{ritter1892reproduction}
A.~Knight, The propagation of water waves, Journal of the Association of German
  Engineers 36~(33) (1892) 947--954.

\bibitem{colagrossi2005meshless}
A.~Colagrossi, A meshless lagrangian method for free-surface and interface
  flows with fragmentation, These, Universita di Roma (2005).

\bibitem{sun2020smoothed}
W.-K. Sun, L.-W. Zhang, K.~Liew, A smoothed particle
  hydrodynamics--peridynamics coupling strategy for modeling fluid--structure
  interaction problems, Computer Methods in Applied Mechanics and Engineering
  371 (2020) 113298.

\bibitem{zhan2019stabilized}
L.~Zhan, C.~Peng, B.~Zhang, W.~Wu, A stabilized tl--wc sph approach with gpu
  acceleration for three-dimensional fluid--structure interaction, Journal of
  Fluids and Structures 86 (2019) 329--353.

\bibitem{walhorn2005fluid}
E.~Walhorn, A.~K{\"o}lke, B.~H{\"u}bner, D.~Dinkler, Fluid--structure coupling
  within a monolithic model involving free surface flows, Computers \&
  structures 83~(25-26) (2005) 2100--2111.

\bibitem{idelsohn2008unified}
S.~R. Idelsohn, J.~Marti, A.~Limache, E.~O{\~n}ate, Unified lagrangian
  formulation for elastic solids and incompressible fluids: application to
  fluid--structure interaction problems via the pfem, Computer Methods in
  Applied Mechanics and Engineering 197~(19-20) (2008) 1762--1776.

\end{thebibliography}


\end{document}